\documentclass[prd,aps,floatfix,superscriptaddress,twocolumn]{revtex4-2}
\usepackage{amsfonts,amsmath,amsthm,amssymb}
\usepackage{bm}
\usepackage{booktabs}
\usepackage{braket} 
\usepackage[colorlinks,citecolor=blue,anchorcolor=red,menucolor=red,linkcolor=red,filecolor=red,runcolor=red,urlcolor=blue,frenchlinks=red]{hyperref}
\usepackage{cases}
\usepackage{color}
\usepackage{dcolumn}
\usepackage{enumerate}
\usepackage{epsfig}
\usepackage{epstopdf}
\usepackage{graphicx}
\usepackage{indentfirst}
\usepackage{latexsym}
\usepackage{longtable}
\usepackage{multirow}
\usepackage{placeins}
\usepackage{rotating}
\usepackage{slashed}  
\usepackage{subfigure}
\usepackage{youngtab}
\newcommand{\etal}{\textit{et~al}.}

\allowdisplaybreaks[4]
\begin{document}
\title{Heavy baryons in the relativized quark model with chromodynamics}
\author{Xin-Zhen Weng}
\email{xinzhenweng@mail.tau.ac.il}
\affiliation{School of Physics and Astronomy, Tel Aviv University, Tel Aviv 6997801, Israel}
\author{Wei-Zhen Deng}
\email{dwz@pku.edu.cn}
\affiliation{School of Physics, Peking University, Beijing 100871, China}
\author{Shi-Lin Zhu}
\email{zhusl@pku.edu.cn}
\affiliation{School of Physics, Peking University, Beijing 100871, China}
\affiliation{Center of High Energy Physics, Peking University, Beijing 100871, China}
\date{\today}

\begin{abstract}

Following the work of Capstick and Isgur [\href{https://doi.org/10.1103/PhysRevD.34.2809}{Phys.~Rev.~D~34,~2809~(1986)}],
we systematically study the mass spectrum of the heavy baryons in the relativized quark potential model with chromodynamics.
Besides the original Godfrey-Isgur (GI) model, we also adopt a modified GI model which replaces the linear confinement by a screened one.
The two models give similar results in our work.
All heavy baryons observed so far can be explained as three-quark states.
In particular, we identify the $\Omega_{c}(3000)$/$\Omega_{b}(6316)$, $\Omega_{c}(3050)$/$\Omega_{b}(6330)$, $\Omega_{c}(3065)$/$\Omega_{b}(6340)$, and $\Omega_{c}(3090)$/$\Omega_{b}(6350)$ states as the $p_{\lambda}$ excitations with quantum numbers $1/2^{-}$, $3/2^{-}$, $3/2^{-}$, and $5/2^{-}$.
The $\Omega_{c}(3120)$ is a $3/2^{-}$ state with the $p_{\rho}$ excitation, whose bottom partner is predicted to be $\Omega_{b}(6446/6457,3/2^{-})$.
The higher state $\Omega_{c}(3188)$ is the $2s_{\lambda}$ excitation with quantum numbers $1/2^{+}$, and $\Omega_{c}(3327)$ is a $d_{\lambda}$ excitation with quantum numbers $3/2^{+}$ or $5/2^{+}$.
%
%

\end{abstract}

\maketitle
\thispagestyle{empty} 

\section{Introduction}
\label{Sec:Introduction}

In the past few decades, we have seen large experimental progress in searching for the heavy baryons.
Lots of charmed baryons were observed in the last few years.
In 2017, the LHCb collaboration observed five narrow states in the $\Xi_{c}^{+}K^{-}$ mass spectrum: $\Omega_{c}(3000)^{0}$, $\Omega_{c}(3050)^{0}$, $\Omega_{c}(3066)^{0}$, $\Omega_{c}(3090)^{0}$, and $\Omega_{c}(3119)^{0}$~\cite{LHCb:2017uwr}, where the first four of them were confirmed by the Belle collaboration~\cite{Belle:2017ext}.
Later LHCb observed $\Lambda_{c}(2860)^{+}$ with quantum numbers $J^{P}=3/2^{+}$ in the $D^{0}p$ channel~\cite{LHCb:2017jym}.
In 2020, LHCb further studied the $\Lambda_{c}^{+}K^{-}$ mass spectrum and observed $\Xi_{c}(2923)^{0}$, $\Xi_{c}(2939)^{0}$ and $\Xi_{c}(2965)^{0}$~\cite{LHCb:2020iby}, where the first two are new states.
In 2022, the Belle collaboration observed $\Lambda_{c}(2910)^{+}$ decaying into $\Sigma_{c}\pi$~\cite{Belle:2022hnm}.
Recently, the LHCb collaboration observed two new states $\Omega_{c}(3185)^{0}$ and $\Omega_{c}(3227)^{0}$ in the $\Xi_{c}^{+}K^{-}$ decay modes~\cite{LHCb:2023sxp}.
In addition, many bottom baryons were also found in experiments.
In 2018, the LHCb collaboration reported the observation of $\Sigma_{b}(6097)^{\pm}$ in the $\Lambda_{b}^{0}\pi^{\pm}$ final states~\cite{LHCb:2018haf}.
The same year, they found the $\Xi_{b}(6227)^{-}$ as a peak in both the $\Lambda_{b}^{0}K^{-}$ and $\Xi_{b}^{0}\pi^{-}$ invariant mass spectra~\cite{LHCb:2018vuc}.
The isospin partner $\Xi_{b}(6227)^{0}$ was soon found~\cite{LHCb:2020xpu}.
In 2019, the LHCb collaboration observed the $\Lambda_{b}(6146)^{0}$ and $\Lambda_{b}(6152)^{0}$ in the $\Lambda_{b}^{0}\pi^{+}\pi^{-}$ channel~\cite{LHCb:2019soc}.
Another state $\Lambda_{b}(6072)^{0}$ was also found in the same channel by the CMS collaboration\cite{CMS:2020zzv}.
In 2020, the LHCb collaboration observed four narrow peak, $\Omega_{b}(6316)^{-}$, $\Omega_{b}(6330)^{-}$, $\Omega_{b}(6340)^{-}$, and $\Omega_{b}(6350)^{-}$, in the $\Xi_{b}^{0}K^{-}$ channel~\cite{LHCb:2020tqd}.
Later in 2021, the CMS collaboration found $\Xi_{b}(6100)^{-}$ in the $\Xi_{b}^{-}\pi^{+}\pi^{-}$ channel~\cite{CMS:2021rvl}.
Recently, the LHCb collaboration confirmed the existence of $\Xi_{b}(6227)^{0}$ and observed another state $\Xi_{b}(6333)^{0}$~\cite{LHCb:2021ssn}.
To date, there are 42 singly-charmed and 25 singly-bottomed baryons collected in PDG~\cite{ParticleDataGroup:2022pth}, plus 3 singly-charmed and 1 singly-bottomed baryons observed more recently.

Inspired by the experimental progress, physicists used various methods to study heavy baryons, such as
the quark models~\cite{Isgur:1977ef,Isgur:1978wd,Martin:1995vk,Ebert:2005xj,Ebert:2007nw,Roberts:2007ni,Karliner:2008sv,Valcarce:2008dr,Yang:2008zzi,Vijande:2014uma,Karliner:2015ema,Karliner:2017kfm,Wang:2017kfr,Yang:2017qan,Karliner:2018bms,Shi:2019tji,Karliner:2020fqe,Wang:2020gkn,Xiao:2020gjo,Chen:2021eyk,Wang:2021bmz,Garcia-Tecocoatzi:2022zrf,Ma:2022vqf,Wang:2022dmw,Karliner:2023okv,Ortiz-Pacheco:2023bns},
lattice QCD~\cite{Liu:2009jc,Briceno:2012wt,Namekawa:2013vu,Brown:2014ena,Bali:2015lka,Bahtiyar:2020uuj,Zhang:2021oja},
QCD sum rules~\cite{Bagan:1992tp,Wang:2009cr,Zhang:2009iya,Chen:2015kpa,Mao:2015gya,Agaev:2017lip,Yang:2022oog},
heavy quark effective theory (HQET)~\cite{Vishwakarma:2022vzy},
and the Regge phenomenology~\cite{Wei:2016jyk,Jia:2019bkr,Oudichhya:2023awb}.
For extra references, see recent review Refs.~\cite{Klempt:2009pi,Crede:2013sze,Chen:2016spr,Eichmann:2016yit,Cheng:2021qpd,Chen:2022asf}.

A well-known quark potential model is the relativized quark model proposed by Godfrey and Isgur (GI) in 1985~\cite{Godfrey:1985xj}, which was used to investigate the meson systems extensively~\cite{Godfrey:1985xj,Godfrey:1986wj,Godfrey:2004ya,Barnes:2005pb,Godfrey:2015dia,Godfrey:2016nwn,Godfrey:2015dva}.
In 1986, Capstick and Isgur extended this model to study the baryon systems~\cite{Capstick:1986ter}.
Due to the lack of experimental data of heavy baryons, their work mainly focus on the light baryons and $nnQ$ baryons [Here, we assume $\mathrm{SU}(2)$ isospin symmetry and denote $n=\{u,d\}$].
In Refs.~\cite{Lu:2016ctt,Lu:2017meb}, L\"u~\etal~used this model, together with the diquark approximation, to study the $\Lambda_{c}$ and doubly heavy baryons.
Recently, ~Yu~\etal~studied the singly and doubly heavy baryons within this model~\cite{Yu:2022ymb,Li:2022xtj,Li:2022oth,Yu:2022lel,Li:2022ywz}.
However, they assumed $\{l_{\rho}l_{\lambda}LSj\}IJ^{P}$ ($\bm{j}=\bm{l}_{\rho}+\bm{S}$) to be good quantum numbers, while the Hamiltonian only ascertains that the $IJ^{P}$ are good quantum numbers.
Furthermore, they ignored the $l_{\rho}\neq0$ ($l_{\lambda}\neq0$) components for the singly (doubly) heavy baryons.
In this series, we will investigate the spectra of heavy baryons in the GI model, without using these approximations.

When going to the higher excited states, the coupled-channel effects may be of importance.
For instance, the physical $D_{s0}^{*}(2317)$ and $D_{s1}^{*}(2460)$ are the mixture of the $c\bar{s}$ core and the $D^{(*)}K$ component~\cite{Yang:2021tvc}.
Another example would be the $X(3872)$ of $J^{PC}=1^{++}$.
Its mass is about 80 MeV lower than the theoretical prediction in the GI~\cite{Godfrey:1985xj} and similar quark potential models.
Theoretical studies suggested that the coupled-channel effect with $D\bar{D}^{*}$ is very important~\cite{Kalashnikova:2005ui,Ortega:2009hj,Danilkin:2010cc,Padmanath:2015era}.
In the baryon sector, the $\Lambda_{c}(2910)$ and $\Lambda_{c}(2940)$ are argued to be conventional $\Lambda_{c}$ excitations dressed with the $D^{*}N$ channel~\cite{Luo:2019qkm,Zhang:2022pxc}.
In Ref.~\cite{Li:2009ad}, Li~\etal~found that the coupled-channel model and the screened potential model have similar global features in describing the charmonium spectrum (below 4.0~GeV) since they approximately embody the same effect of the vacuum polarization of dynamical light quark pairs.
Based on this observation, Song~\etal~implemented the GI model by replacing the linear confining potential $br$ by a screened one $V^{\text{scr}}=b(1-\mathrm{e}^{-{\mu}r})/\mu$ to reflect the coupled-channel effect~\cite{Song:2015nia}.
They found that the modified GI (MGI) model is more suitable to describe the experimental data, especially the higher excited mesons~\cite{Song:2015nia,Song:2015fha,Pang:2017dlw,Wang:2018rjg}.

In this work, we use the two relativized quark models, both GI and MGI models, to study the mass spectra of the singly heavy baryons.
In Sec~\ref{Sec:Model}, we briefly introduce the models.
Then we present the numerical results and discussions in Sec.~\ref{Sec:Result}.
Finally, we conclude in Sec.~\ref{Sec:Summary}.
%

\section{The Relativized Quark Model}
\label{Sec:Model}

\subsection{Godfrey-Isgur (GI) model}
\label{sec:GI}

In the relativized quark model, the Hamiltonian of a baryon reads~\cite{Godfrey:1985xj,Capstick:1986ter}
\begin{equation}
H=H_{0}+V_{\text{OGE}}+V_{\text{conf}}\,,
\end{equation}
where
\begin{equation}
H_{0}=\sum_{i=1}^{3}\left(\mathbf{p}_{i}^2+m_{i}^2\right)^{1/2}
\end{equation}
is the relativistic kinetic energy, in which $m_{i}$ and $\mathbf{p}_{i}$ are the $i$th quark's mass and momentum.
$V_{\text{OGE}}$ is a one-gluon-exchange (OGE) potential, and $V_{\text{conf}}$ consists of a string potential and a spin-orbit term arising via Thomas precession.
%

\begin{figure}[htbp]
\includegraphics[width=200pt]{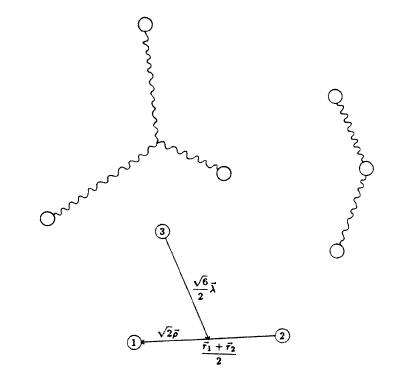}
\caption{The string confinement and the relative coordinates $\bm{\rho}$ and $\bm{\lambda}$~\cite{Capstick:1986ter}.}
\label{fig:V_string}
\end{figure}
%
In the nonrelativistic limit, the potentials go to the usual Breit-Fermi interaction
\begin{equation}
V_{\text{OGE}}
\rightarrow
\sum_{i<j}
\left(
{V}_{ij}^{\text{Coul}}
+
{V}_{ij}^{\text{hyp}}
+
{V}_{ij}^{\text{so}(v)}
\right)\,,
\end{equation}
and
\begin{equation}
V_{\text{conf}}
\rightarrow
V_{\text{string}}
+
\sum_{i<j}
{V}_{ij}^{\text{so}(v)}\,.
\end{equation}
Here, ${V}_{ij}^{\text{Coul}}$ is the spin-independent Coulomb-type interaction;
${V}_{ij}^{\text{hyp}}$ is the color-hyperfine interaction which consists of the contact and tensor interaction;
${V}_{ij}^{\text{so}(v)}+{V}_{ij}^{\text{so}(v)}$ is the spin-orbit interaction,
and $V_{\text{string}}$ is the three-body $Y$-shape confinement interaction (see Fig.~\ref{fig:V_string})
\begin{equation}
V_{\text{string}}
=
C_{qqq}
+
b
\sum_{i=1}^{3}
\left|\mathbf{r}_{i}-\mathbf{r}_{\text{junction}}\right|\,.
\end{equation}
The $V_{\text{string}}$ interaction can be broken into an effective two-body piece and a three-body piece
\begin{equation}
V_{\text{string}}
=
C_{qqq}
+
fb
\sum_{i<j}r_{ij}
+
V_{3b}
\end{equation}
where
\begin{equation}\label{eqn:Vconf:V3b}
V_{3b}
=
b
\left(
\sum_{i=1}^{3}
\left|\mathbf{r}_{i}-\mathbf{r}_{\text{junction}}\right|
-
f\sum_{i<j}r_{ij}
\right)
\end{equation}
with $f=0.5493$~\cite{Dosch:1975gf,Carlson:1982xi,Capstick:1986ter} chosen to minimize the size of the expectation value of $V_{3b}$.
Then the $V_{3b}$ can be treated perturbatively.
Note that the contribution of $V_{3b}$ is \emph{always small}~\cite{Capstick:1986ter} and we will not consider it in the present work for simplification.

As pointed out in Refs.~\cite{Godfrey:1985xj,Capstick:1986ter}, the Breit-Fermi interaction should be modified for many reasons:
(a) the proceeding potentials are constructed by reproducing the \emph{on-shell} $qq$ scattering, which will be modified by the off-shell properties of quark;
(b) the constituent quarks are not pointlike, but rather will have a
graininess appropriate to some finite scale $\mu$;
(c) the higher Fock space such as $\ket{qqqg}$ is not considered in the present work, which will introduce additional momentum-dependence in the potential.
To incorporate these effects, Godfrey and Isgur further built a semiquantitative \emph{model}~\cite{Godfrey:1985xj}.
Firstly, by introducing the smearing function
\begin{equation}
\rho_{ij}\left(\mathbf{r}-\mathbf{r}'\right)
=
\frac{\sigma_{ij}^{3}}{\pi^{3/2}}
\mathrm{e}^{-\sigma_{ij}^2\left(\mathbf{r}-\mathbf{r}'\right)^2}\,,
\end{equation}
the potentials $V_{ij}^{\text{OGE}}$ and $V_{ij}^{\text{conf}}$ are smeared to $\tilde{V}_{ij}^{\text{OGE}}$ and $\tilde{V}_{ij}^{\text{conf}}$ via
\begin{equation}
\tilde{V}_{ij}(r)=\int\mathrm{d}^3{r}'\rho_{ij}\left(\mathbf{r}-\mathbf{r}'\right)V_{ij}(r'),
\end{equation}
with the prescription
\begin{equation}
\sigma_{ij}^2
=
\frac{\sigma_{0}^2}{2}
\left\{1+\left[\frac{4m_{i}m_{j}}{\left(m_{i}+m_{j}\right)^2}\right]^4\right\}
+
s^2\left(\frac{2m_{i}m_{j}}{m_{i}+m_{j}}\right)^2,
\end{equation}
where $\sigma_{0}$ and $s$ are free parameters.
Secondly, through the introduction of the momentum-dependent factors, the Coulomb term is modified according to
\begin{equation}
\tilde{V}_{ij}^{\text{Coul}}
{\to}
\left(\beta_{ij}\right)^{1/2+\epsilon_\text{Coul}}
\tilde{V}_{ij}^{\text{Coul}}
\left(\beta_{ij}\right)^{1/2+\epsilon_\text{Coul}},
\end{equation}
and the contact, tensor, vector spin-orbit, and scalar spin-orbit potentials are modified according to
\begin{equation}
\tilde{V}_{ij}^{\alpha}
{\to}
\left(\delta_{ij}^{[ij]}\right)^{1/2+\epsilon_{\alpha}}
\tilde{V}_{ij}^{\alpha}
\left(\delta_{ij}^{[ij]}\right)^{1/2+\epsilon_{\alpha}}
\end{equation}
where $\epsilon_{\alpha}$ corresponds to the contact (cont), tensor (tens), vector spin-orbit [so$(v)$], and scalar spin-orbit [so$(s)$].
The momentum-dependent factors are
\begin{equation}
\beta_{ij}
=
1+\frac{p_{ij}^2}{(p_{ij}^2+m_{i}^2)^{1/2}(p_{ij}^2+m_{j}^2)^{1/2}}
\end{equation}
and
\begin{equation}
\delta_{ij}^{[i'j']}
=
\frac{m_{i}m_{j}}{(p_{i'j'}^2+m_{i}^2)^{1/2}(p_{i'j'}^2+m_{j}^2)^{1/2}}\,,
\end{equation}
where $p_{ij}$ is the magnitude of the momentum of either of the quarks in the $ij$ center-of-mass frame.
%

\begin{figure}[htbp]
\includegraphics[width=200pt]{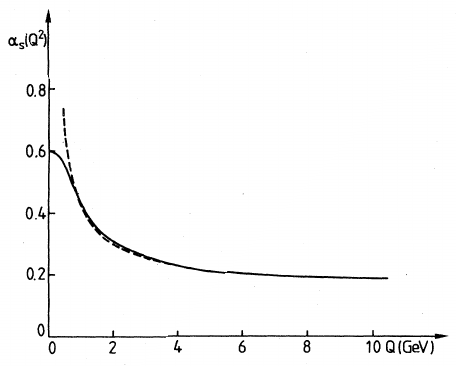}
\caption{The leading order QCD formula \eqref{eqn:alpha_s:QCD} for the effective coupling constant with $\Lambda_{\text{QCD}}=200~\text{MeV}$ and the fit $\alpha_{s}(Q^2)=0.25\mathrm{e}^{-Q^2}+0.15\mathrm{e}^{-Q^2/10}+0.20\mathrm{e}^{-Q^2/1000}$ ($Q^2$ in $\text{GeV}^{2}$)~\cite{Godfrey:1985xj,Capstick:1986ter}.}
\label{fig:alpha_s}
\end{figure}
%
The last piece of the model is the QCD coupling constant $\alpha_{s}$.
With $N_{f}$ flavors of quark with masses below $Q^2$, the lowest-order coupling constant reads
\begin{equation}\label{eqn:alpha_s:QCD}
\alpha_{s}(Q^2)
=
\frac{12\pi}{33-2N_{f}\ln(Q^2/\Lambda_{\text{QCD}}^{2})}\,.
\end{equation}
The $\alpha_{s}(Q^2)$ is small for large $Q^2$ but diverges as $Q^2$ approach $\Lambda_{\text{QCD}}\approx200~\text{MeV}$, which indicates the confinement.
Since we are interested in the soft regime, we cannot avoid this divergence;
rather we assume that $\alpha_{s}$ saturates at some value $\alpha_{s}^{\text{critical}}$ for low $Q^2$ as confinement emerges.
For convenience, the coupling constant $\alpha_{s}$ is parametrized as
\begin{equation}\label{eqn:alpha_s:param:formal}
\alpha_{s}(Q^2)
=
\sum_{k}
\alpha_{k}
\mathrm{e}^{-Q^2/4\gamma_{k}^{2}}\,,
\end{equation}
where
\begin{equation}
\alpha_{s}^{\text{critical}}
=
\sum_{k}
\alpha_{k}
\end{equation}
is a free parameter.
Fitting the QCD curve [Eq.~\eqref{eqn:alpha_s:QCD}, see Fig.~\ref{fig:alpha_s}] gives~\cite{Godfrey:1985xj}
\begin{equation}\label{eqn:alpha_s:param:numerical}
\alpha_{s}(Q^2)
=
0.25\mathrm{e}^{-Q^2}
+
0.15\mathrm{e}^{-Q^2/10}
+
0.20\mathrm{e}^{-Q^2/1000}
\end{equation}
with $Q^2$ in $\text{GeV}^{2}$.

Now we can finally present the effective potentials~\cite{Capstick:1986ter}.
The smeared OGE propagator is
\begin{equation}
\tilde{G}\left(r_{ij}\right)
=
-\sum_{k}\frac{2\alpha_{k}}{3r_{ij}}
\cdot
\text{Erf}\left(\tau_{kij}r_{ij}\right)
\end{equation}
where $\text{Erf}(x)$ is the Gauss error function and
\begin{equation}
\frac{1}{\tau_{kij}^2}
=
\frac{1}{\gamma_{k}^2}+\frac{1}{\sigma_{ij}^2}\,.
\end{equation}
All of the OGE potential can be expressed in terms of $\tilde{G}\left(r_{ij}\right)$.
More precisely, the OGE potentials
\begin{equation}
\tilde{V}_{\text{OGE}}
=
\sum_{i<j}\tilde{V}_{ij}^{\text{OGE}}
\end{equation}
with
\begin{equation}
\tilde{V}_{ij}^{\text{OGE}}
=
\tilde{V}_{ij}^{\text{Coul}}+\tilde{V}_{ij}^{\text{hyp}}+\tilde{V}_{ij}^{\text{so}(v)}
\end{equation}
where
\begin{equation}
\tilde{V}_{ij}^{\text{Coul}}
=
\left(\beta_{ij}\right)^{1/2+\epsilon_\text{Coul}}
\tilde{G}\left(r_{ij}\right)
\left(\beta_{ij}\right)^{1/2+\epsilon_\text{Coul}}
\end{equation}
is the Coulomb-type interaction.
\begin{equation}
\tilde{V}_{ij}^{\text{hyp}}
=
\tilde{V}_{ij}^{\text{cont}}+\tilde{V}_{ij}^{\text{tens}}
\end{equation}
is the hyperfine interaction with contact term
\begin{widetext}
\begin{equation}
\tilde{V}_{ij}^{\text{cont}}
=
\left(\delta_{ij}^{[ij]}\right)^{1/2+\epsilon_{\text{cont}}}
\frac{2\mathbf{S}_{i}\cdot\mathbf{S}_{j}}{3m_{i}m_{j}}
\nabla_{r_{ij}}^2\tilde{G}\left(r_{ij}\right)
\left(\delta_{ij}^{[ij]}\right)^{1/2+\epsilon_{\text{cont}}}
\end{equation}
and tensor term
\begin{equation}
\tilde{V}_{ij}^{\text{tens}}
=
\left(\delta_{ij}^{[ij]}\right)^{1/2+\epsilon_{\text{tens}}}
\frac{1}{3m_{i}m_{j}}
\left(\frac{3\mathbf{S}_{i}\cdot\mathbf{r}_{ij}\mathbf{S}_{j}\cdot\mathbf{r}_{ij}}{r_{ij}^2}-\mathbf{S}_{i}\cdot\mathbf{S}_{j}\right)
\left(\frac{1}{r_{ij}}\frac{\mathrm{d}\tilde{G}\left(r_{ij}\right)}{\mathrm{d}r_{ij}}-\frac{\mathrm{d}^2\tilde{G}\left(r_{ij}\right)}{\mathrm{d}r_{ij}^2}\right)
\left(\delta_{ij}^{[ij]}\right)^{1/2+\epsilon_{\text{tens}}}
\,.
\end{equation}
%
%
\begin{align}
\tilde{V}_{ij}^{\text{so}(v)}
={}&
\frac{1}{r_{ij}}\frac{\mathrm{d}\tilde{G}\left(r_{ij}\right)}{\mathrm{d}r_{ij}}
\Bigg[
\left(\delta_{ii}^{[ij]}\right)^{1/2+\epsilon_{\text{so}(v)}}
\frac{\mathbf{r}_{ij}\times\mathbf{p}_{i}\cdot\mathbf{S}_{i}}{2m_{i}^2}
\left(\delta_{ii}^{[ij]}\right)^{1/2+\epsilon_{\text{so}(v)}}
-
\left(\delta_{jj}^{[ij]}\right)^{1/2+\epsilon_{\text{so}(v)}}
\frac{\mathbf{r}_{ij}\times\mathbf{p}_{j}\cdot\mathbf{S}_{j}}{2m_{j}^2}
\left(\delta_{jj}^{[ij]}\right)^{1/2+\epsilon_{\text{so}(v)}}
\notag\\
&\qquad\qquad-
\left(\delta_{ij}^{[ij]}\right)^{1/2+\epsilon_{\text{so}(v)}}
\frac{\mathbf{r}_{ij}\times\mathbf{p}_{j}\cdot\mathbf{S}_{i}-\mathbf{r}_{ij}\times\mathbf{p}_{i}\cdot\mathbf{S}_{j}}{m_{i}m_{j}}
\left(\delta_{ij}^{[ij]}\right)^{1/2+\epsilon_{\text{so}(v)}}
\Bigg]
\end{align}
is the spin-orbit interaction.

The confinement potential $V_{\text{conf}}$ consists of a string potential and a spin-orbit term arising from Thomas precession
\begin{equation}
\tilde{V}_{\text{conf}}
=
C_{qqq}
+
\sum_{i<j}
\tilde{V}_{ij}^{\text{string}}
+
\sum_{i<j}
\tilde{V}_{ij}^{\text{so}(s)}
\end{equation}
with
\begin{equation}
\tilde{V}_{ij}^{\text{string}}
=
fbr_{ij}
\left[
\frac{\mathrm{e}^{-\sigma_{ij}^2r_{ij}^2}}{\sqrt{\pi}\sigma_{ij}r_{ij}}
+
\left(1+\frac{1}{2\sigma_{ij}^2r_{ij}^2}\right)\text{Erf}\left(\sigma_{ij}r_{ij}\right)
\right]
\end{equation}
and
\begin{equation}
\tilde{V}_{ij}^{\text{so}(s)}
=
-\frac{1}{r_{ij}}\frac{\partial\tilde{V}_{ij}^{\text{string}}}{\partial{r}_{ij}}
\Bigg[
\left(\delta_{ii}^{[ij]}\right)^{1/2+\epsilon_{\text{so}(s)}}
\frac{\mathbf{r}_{ij}\times\mathbf{p}_{i}\cdot\mathbf{S}_{i}}{2m_{i}^2}
\left(\delta_{ii}^{[ij]}\right)^{1/2+\epsilon_{\text{so}(s)}}
-
\left(\delta_{jj}^{[ij]}\right)^{1/2+\epsilon_{\text{so}(s)}}
\frac{\mathbf{r}_{ij}\times\mathbf{p}_{j}\cdot\mathbf{S}_{j}}{2m_{j}^2}
\left(\delta_{jj}^{[ij]}\right)^{1/2+\epsilon_{\text{so}(s)}}
\Bigg]
\end{equation}
where we have ignored the small perturbation term $V_{3b}$ [Eq.~\eqref{eqn:Vconf:V3b}].
%

\subsection{Modified Godfrey-Isgur (MGI) model with the screening effect}
\label{sec:MGI}

In order to take into account the screening effect, Refs.~\cite{Song:2015nia,Song:2015fha} proposed the following replacement for the linear confinement interaction
\begin{equation}
br
\rightarrow
V^{\text{scr}}\left(r\right)
=
\frac{b\left(1-\mathrm{e}^{-\mu{r}}\right)}{\mu}\,.
\end{equation}
where $V^{\text{scr}}\left(r\right)$ behaves like the linear confinement potential $br$ at short distance and approaches constant $b/\mu$ at long distance.
After smearing, we have
\begin{equation}
\tilde{V}_{\text{conf}}
\to
\tilde{V}_{\text{conf}}^{\text{scr}}
=
C_{qqq}
+
\sum_{i<j}
\tilde{V}_{ij}^{\text{scr}}
+
\sum_{i<j}
\tilde{V}_{ij}^{\text{so(scr)}}
\end{equation}
where
%
%
\begin{align}
\tilde{V}_{ij}^{\text{scr}}
={}&
\frac{fb}{\mu}
\Bigg\{
1
-
\frac{1}{2\sigma_{ij}r_{ij}}
\Bigg[
\left({\frac{\mu}{2\sigma_{ij}}}+\sigma_{ij}r_{ij}\right)
\exp\left(\frac{\mu^{2}}{4\sigma_{ij}^{2}}+\mu{r}_{ij}\right)
\cdot
\text{Erfc}\left({\frac{\mu}{2\sigma_{ij}}}+\sigma_{ij}r_{ij}\right)
\notag\\
&\qquad\qquad\qquad~~
-
\left({\frac{\mu}{2\sigma_{ij}}}-\sigma_{ij}r_{ij}\right)
\exp\left(\frac{\mu^{2}}{4\sigma_{ij}^{2}}-\mu{r}_{ij}\right)
\cdot
\text{Erfc}\left({\frac{\mu}{2\sigma_{ij}}}-\sigma_{ij}r_{ij}\right)
\Bigg]
\Bigg\}\,,
\end{align}
and
\begin{equation}
\tilde{V}_{ij}^{\text{so(scr)}}
=
-\frac{1}{r_{ij}}\frac{\partial\tilde{V}_{ij}^{\text{scr}}}{\partial{r}_{ij}}
\Bigg[
\left(\delta_{ii}^{[ij]}\right)^{1/2+\epsilon_{\text{so}(s)}}
\frac{\mathbf{r}_{ij}\times\mathbf{p}_{i}\cdot\mathbf{S}_{i}}{2m_{i}^2}
\left(\delta_{ii}^{[ij]}\right)^{1/2+\epsilon_{\text{so}(s)}}
-
\left(\delta_{jj}^{[ij]}\right)^{1/2+\epsilon_{\text{so}(s)}}
\frac{\mathbf{r}_{ij}\times\mathbf{p}_{j}\cdot\mathbf{S}_{j}}{2m_{j}^2}
\left(\delta_{jj}^{[ij]}\right)^{1/2+\epsilon_{\text{so}(s)}}
\Bigg]\,.
\end{equation}
where $\text{Erfc}(x)$ is the complementary error function.
\end{widetext}
%

\subsection{Baryon wave function}
\label{Sec:WaveFunc}

To investigate the mass spectra of the baryons, we need to construct the wave functions.
The total wave function is a direct product of the flavor, color, spin, and spatial wave functions.

In this work, we ignore the isospin violation and abbreviate $n=\{u,d\}$.
The flavor wave functions of heavy baryons can be easily written down directly.
Note that for the $nnQ$ systems we have $\{nn\}Q$ for $\Sigma_{Q}$ with isospin $I=1$ and $[nn]Q$ for $\Lambda_{Q}$ with $I=0$, where we use the brace $\{\ldots\}$ to symmetrize the quark flavors and the bracket $[\ldots]$ to antisymmetrize the flavors.

The baryons should be color-singlet, thus we have only one totally antisymmetric color wave function
\begin{equation}
\varphi^{c}=\ket{(q_{1}q_{2})^{3_{c}}q_{3}}^{1_{c}}\,.
\end{equation}
where the superscripts are color representations.

Next we consider the spin wave functions. The total spin of the baryons can be either of $1/2$ or $3/2$.
A complete set of the spin wave functions are listed as follows
\begin{equation}
\left\{
\begin{split}
&\chi_{1}^{s}=\ket{(q_{1}q_{2})_{1}q_{3}}_{3/2}\,,\\
&\chi_{2}^{s}=\ket{(q_{1}q_{2})_{1}q_{3}}_{1/2}\,,\\
&\chi_{3}^{s}=\ket{(q_{1}q_{2})_{0}q_{3}}_{1/2}\,,
\end{split}
\right.
\end{equation}
where the subscripts are spins $\{S_{12},S\}$.
The third component $M_{S}$ is not shown for simplicity.

Finally for spatial wave functions we adopt the harmonic-oscillator wave function
\begin{align}
&\Psi_{LM_{L}n_{\rho}l_{\rho}n_{\lambda}l_{\lambda}}\left(\bm{\rho},\bm{\lambda}\right)
=
\sum_{m_{l_{\rho}}m_{l_{\lambda}}}
\Braket{l_{\rho}m_{l_{\rho}};l_{\lambda}m_{l_{\lambda}}|LM_{L}}
\notag\\
&\qquad\qquad\qquad\times
\psi_{n_{\rho}l_{\rho}m_{l_{\rho}}}\left(\alpha,\bm{\rho}\right)
\psi_{n_{\lambda}l_{\lambda}m_{l_{\lambda}}}\left(\alpha,\bm{\lambda}\right)
\end{align}
in the two relative coordinates (see Fig.~\ref{fig:V_string})
\begin{equation}
\left\{
\begin{split}
&\bm{\rho}=\frac{\mathbf{r}_{1}-\mathbf{r}_{2}}{\sqrt{2}}\,,\\
&\bm{\lambda}=\frac{\mathbf{r}_{1}+\mathbf{r}_{2}-2\mathbf{r}_{3}}{\sqrt{6}}\,,
\end{split}
\right.
\end{equation}
and $\psi_{nlm}(\alpha,\bm{r})$ is the harmonic oscillator wave function with variational parameter $\alpha$.

We then expand the wave function of quantum numbers $J^{P}$ in states of the form
\begin{align}
\Ket{\alpha}
={}&
\chi^{f}
\varphi^{c}
\sum_{M_{L}M_{S}}
\Braket{LM_{L};SM_{S}|JM}
\notag\\
&\times
\Psi_{LM_{L}n_{\rho}l_{\rho}n_{\lambda}l_{\lambda}}
\chi_{SM_{S}}^{s}\,,
\end{align}
where $P=(-1)^{l_{\rho}+l_{\lambda}}$.
Note that there are infinite number of bases for any quantum numbers $IJ^{P}$.
We truncate the bases by $N\leq10$ with $N=2(n_{\rho}+n_{\lambda})+l_{\rho}+l_{\lambda}$.
Diagonalizing the Hamiltonian in these bases and minimizing each eigenvalues by varying $\alpha$, we obtain the mass spectra of the heavy baryons.
More details can be found in Ref.~\cite{Capstick:1986ter}.
%

\section{Numerical results}
\label{Sec:Result}

\subsection{Parameters}
\label{sec:Parameter}

\begin{table}[htbp]
\centering
\caption{The parameters of the relativized quark potential model. Note that we ignore isospin violation.}
\label{table:parameter}
\begin{tabular}{cccccccccccc}
\toprule[1pt]
\toprule[1pt]
\multirow{2}{*}{Parameters}&\multirow{2}{*}{GI85~\cite{Godfrey:1985xj}}&\multirow{2}{*}{CI86~\cite{Capstick:1986ter}}&\multicolumn{2}{c}{This~work}\\
\cmidrule(lr){4-5}
&&&GI&MGI\\
\midrule[1pt]
\midrule[1pt]
\multicolumn{5}{c}{Masses~(MeV)}\\
$m_{n}$&$220$&$220$&$247$&$244$\\
$m_{s}$&$419$&$419$&$455$&$452$\\
$m_{c}$&$1628$&$1628$&$1652$&$1644$\\
$m_{b}$&$4977$&$4977$&$5000$&$4991$\\
\midrule[1pt]
\multicolumn{5}{c}{Potentials}\\
$b$~($\text{GeV}^2$)&$0.18$&$0.15$&$0.14$&$0.17$\\
$\alpha_{s}^{\text{critical}}$&$0.60$&Same&Same&Same\\
$\Lambda_{\text{QCD}}$~(MeV)&$220$&Same&Same&Same\\
$C_{q\bar{q}}$~(MeV)&$\frac{4}{3}(-253)=-340$&---&---&---\\
$C_{qqq}$~(MeV)&---&$-615$&$-547$&$-656$\\
$\mu$~(MeV)&---&---&---&$0.06$\\
\midrule[1pt]
\multicolumn{5}{c}{Relativistic effects}\\
$\sigma_{0}$~(GeV)&$1.80$&$1.80$&$1.60$&$1.60$\\
$s$&$1.55$&$1.55$&$1.75$&$1.67$\\
$\epsilon_{\text{Coul}}$&---&$0$&$+0.060$&$+0.071$\\
$\epsilon_{\text{cont}}$&$-0.168$&$-0.168$&$-0.179$&$-0.173$\\
$\epsilon_{\text{tens}}$&$+0.025$&$-0.168$&$-0.015$&$-0.125$\\
$\epsilon_{\text{so}(v)}$&$-0.035$&$0$&$-0.220$&$-0.237$\\
$\epsilon_{\text{so}(s)}$&$+0.055$&$+0.300$&$+0.039$&$-0.062$\\
\bottomrule[1pt]
\bottomrule[1pt]
\end{tabular}
\end{table}
%
\begin{table*}[htbp]
\centering
\caption{Masses of light and singly-heavy baryons used for fitting parameters. All masses are in units of MeV.}
\label{table:mass:fit}
\begin{tabular}{cccccccccccc}
\toprule[1pt]
\toprule[1pt]
\multirow{2}{*}{State,~$J^{P}$}&\multirow{2}{*}{CI86~\cite{Capstick:1986ter}}&\multicolumn{2}{c}{This~work}&\multirow{2}{*}{Exp.~\cite{ParticleDataGroup:2022pth}}&
\multirow{2}{*}{State,~$J^{P}$}&\multirow{2}{*}{CI86~\cite{Capstick:1986ter}}&\multicolumn{2}{c}{This~work}&\multirow{2}{*}{Exp.~\cite{ParticleDataGroup:2022pth}}\\
\cmidrule(lr){3-4}
\cmidrule(lr){8-9}
&&GI&MGI&&&
&GI&MGI\\
\midrule[1pt]
$N\frac12^{+}$&960&948&946&$938.9187544\pm0.0000003${\footnote{The baryon masses are taken by their isospin averages.}}&
$\Delta\frac32^{+}$&1230&1233&1232&$1232\pm2$\\
$\Lambda\frac12^{+}$&1115&1122&1122&$1115.683\pm0.006$\\
$\Sigma\frac12^{+}$&1190&1179&1179&$1193.15\pm0.03$&
$\Sigma\frac32^{+}$&1370&1387&1388&$1384.6\pm0.4$\\
$\Xi\frac12^{+}$&1305&1318&1318&$1318.28\pm0.11$&
$\Xi\frac32^{+}$&1505&1532&1535&$1533.4\pm0.3$\\
&&&&&$\Omega\frac32^{+}$&1635&1668&1670&$1672.45\pm0.29$\\
\midrule[1pt]
$\Lambda_{c}\frac12^{+}$&2265&2290&2288&$2286.46\pm0.14$&
$\Lambda_{c}^{*}\frac32^{+}$&\underline{2910}&2870&2871&$2856.1_{-6.0}^{+2.3}$\\
$\Lambda_{c}^{*}\frac12^{-}$&\underline{2630}{\footnote{Most heavy baryons are observed after the publication of Ref.~\cite{Capstick:1986ter}. These states are indicated by an underline.}}&2594&2592&$2592.25\pm0.28$&
$\Lambda_{c}^{*}\frac32^{-}$&\underline{2640}&2624&2627&$2628.11\pm0.19$\\
$\Sigma_{c}\frac12^{+}$&2440&2439&2437&$2453.45\pm0.10$&
$\Sigma_{c}\frac32^{+}$&\underline{2495}&2519&2518&$2518.1\pm0.3$\\
$\Xi_{c}\frac12^{+}$&---&2476&2474&$2469.08\pm0.18$&
$\Xi_{c}\frac32^{+}$&---&2649&2649&$2645.63\pm0.20$\\
&---&2572&2571&$2578.5\pm0.4$\\
$\Xi_{c}^{*}\frac12^{-}$&---&2787&2793&$2792.9\pm0.4$&
$\Xi_{c}^{*}\frac32^{-}$&---&2814&2818&$2818.15\pm0.20$\\
$\Omega_{c}\frac12^{+}$&---&2695&2692&$2695.2\pm1.7$&
$\Omega_{c}\frac32^{+}$&---&2769&2768&$2765.9\pm2.0$\\
\midrule[1pt]
$\Lambda_{b}\frac12^{+}$&\underline{5585}&5625&5623&$5619.60\pm0.17$\\
$\Lambda_{b}^{*}\frac12^{-}$&\underline{5912}&5894&5896&$5912.19\pm0.17$&
$\Lambda_{b}^{*}\frac32^{-}$&\underline{5920}&5906&5908&$5920.09\pm0.17$\\
$\Sigma_{b}\frac12^{+}$&\underline{5795}&5809&5809&$5813.10\pm0.18$
&$\Sigma_{b}\frac32^{+}$&\underline{5805}&5838&5838&$5832.53\pm0.20$\\
$\Xi_{b}\frac12^{+}$&---&5805&5803&$5794.5\pm0.4$&
$\Xi_{b}\frac32^{+}$&---&5963&5964&$5953.8\pm0.3$\\
&---&5935&5934&$5935.02\pm0.05$\\
$\Omega_{b}\frac12^{+}$&---&6049&6047&$6045.2\pm1.2$\\
\bottomrule[1pt]
\bottomrule[1pt]
\end{tabular}
\end{table*}

The GI model is successful in studying the mesons and baryons.
In 1985, Godfrey and Isgur fitted the parameters for mesons~\cite{Godfrey:1985xj}.
Later Capstick and Isgur refitted the parameters for the baryons~\cite{Capstick:1986ter}.
The concrete values of these parameters are collected in Table~\ref{table:parameter}.
After their works, lots of heavy baryons had been observed in experiments~\cite{ParticleDataGroup:2022pth}.
To appreciate the experimental progress in the past three decades, we perform the $\chi^{2}$ fits to obtain the 13/14 parameters of the GI/MGI models.
The $\chi^{2}$ is defined as
\begin{equation}
\chi^{2}
=
\sum_{i}
\frac{\left(m_{i}^{\text{exp.}}-m_{i}^{\text{th.}}\right)^{2}}{\sigma_{i}^{2}}\,,
\end{equation}
where the $m_{i}^{\text{exp.}}$ and $m_{i}^{\text{th.}}$ are experimental and theoretical masses of the baryons.
The baryons used for fits include 24 ground states and 6 excited states (see Table~\ref{table:mass:fit}).
The $\sigma_{i}$ is usually chosen to be the uncertainty of the experimental mass ${\Delta}m_{i}^{\text{exp.}}$.
However, the ${\Delta}m_{N}^{\text{exp.}}\sim10^{-7}~\text{MeV}$ is too small compares to the uncertainties of other baryons, which will deteriorate the fits.
Thus we choose $\sigma_{i}$ to be a constant
\begin{equation}
\sigma_{i}
=
k~\text{MeV}\,.
\end{equation}
The results are independent of the concrete value of $k$.
Our fits give $\chi^{2}=\{1776,1558\}/k^{2}$ for the GI and MGI models, respectively.
The parameters obtained are listed in Table~\ref{table:parameter}.
The comparison of the fitted masses with experimental data is listed in Table~\ref{table:mass:fit}.
With parameters obtained, we can calculate the heavy baryon masses.
The spectroscopic results are given in Tables~\ref{table:mass:nnc}--\ref{table:mass:ssQ} and Figs.~\ref{fig:mass:nnc:MS}--\ref{fig:mass:ssb}.
Within uncertainty of the quark model, the GI and MGI models give similar results.
%

\subsection{The $\Sigma_{Q}$ and $\Lambda_{Q}$ baryons}
\label{sec:nnQ}

\begin{figure*}
\includegraphics[width=500pt]{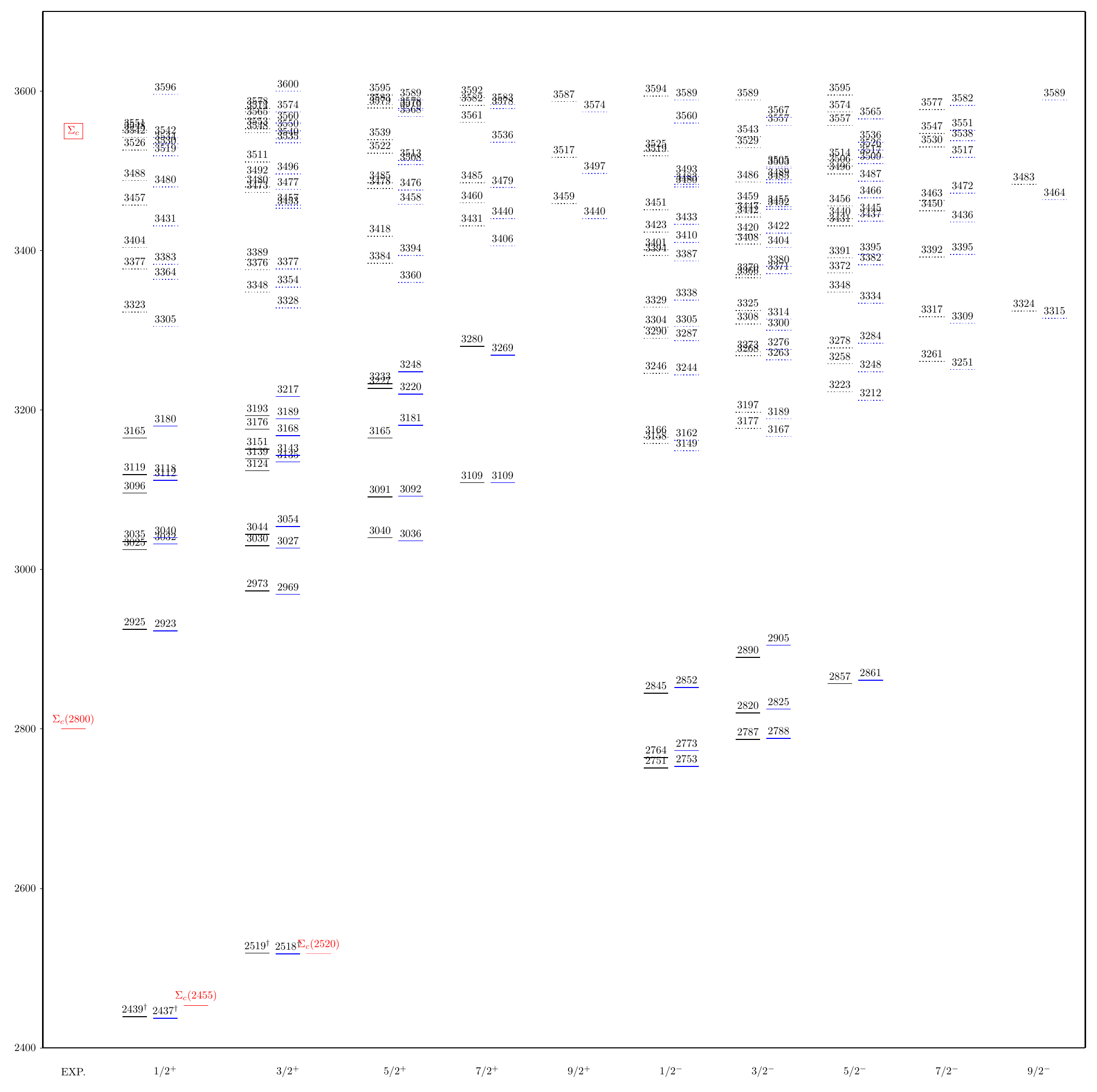}
\caption{Masses of the $\Sigma_{c}$ baryons (in units of MeV). Here, the black/blue lines stand for the numerical results obtained by the GI/MGI models. The red lines represent the experimental results taken from PDG~\cite{ParticleDataGroup:2022pth}. In the first column, we list the experimental observed states whose quantum numbers are not determined yet. The states used for fittings are indicated by daggers ($\dagger$).}
\label{fig:mass:nnc:MS}
\end{figure*}

\begin{figure*}
\includegraphics[width=500pt]{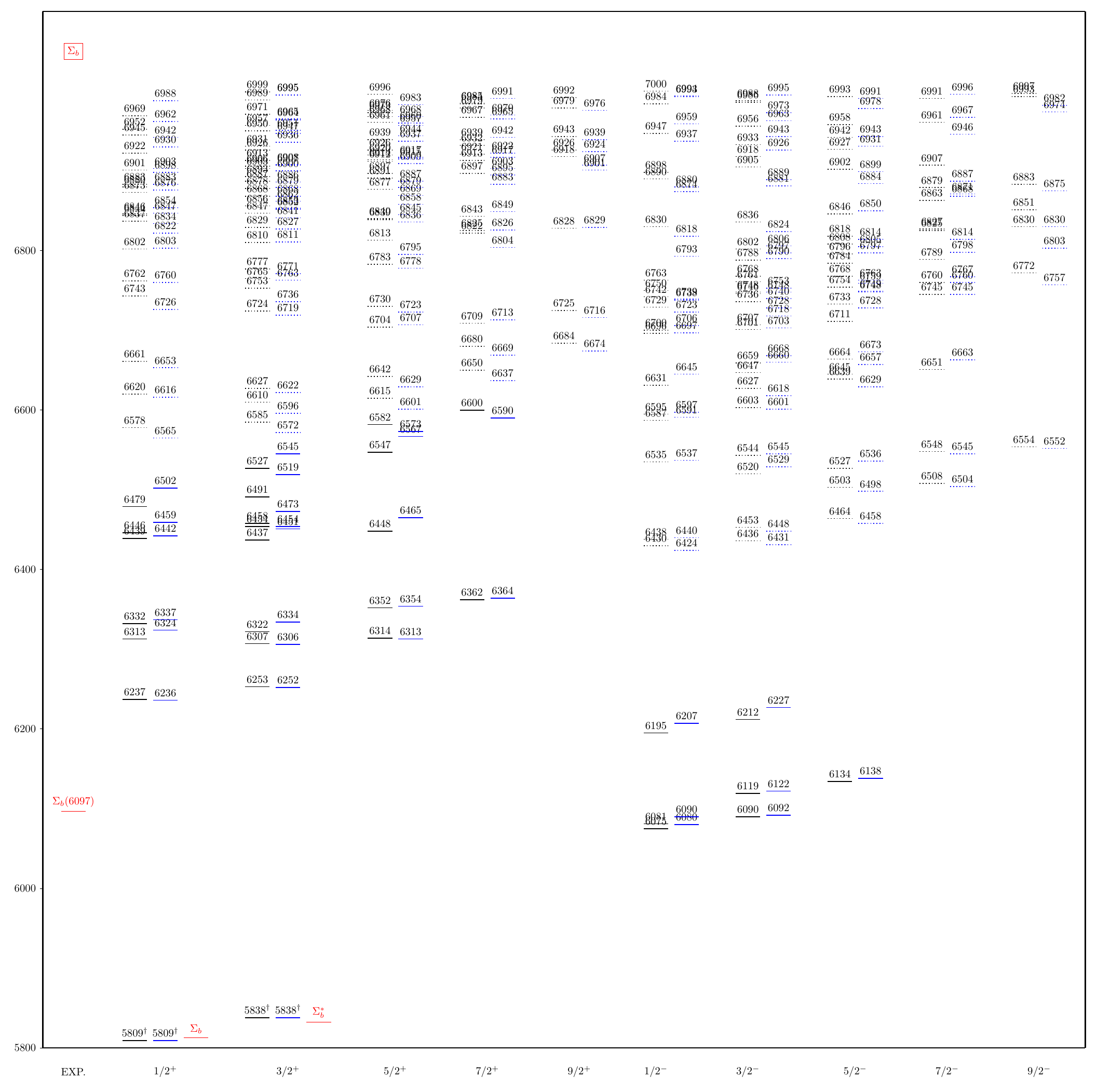}
\caption{Masses of the $\Sigma_{b}$ baryons (in units of MeV). Here, the black/blue lines stand for the numerical results obtained by the GI/MGI models. The red lines represent the experimental results taken from PDG~\cite{ParticleDataGroup:2022pth}. In the first column, we list the experimental observed states whose quantum numbers are not determined yet. The states used for fittings are indicated by daggers ($\dagger$).}
\label{fig:mass:nnb:MS}
\end{figure*}

\begin{figure*}
\includegraphics[width=500pt]{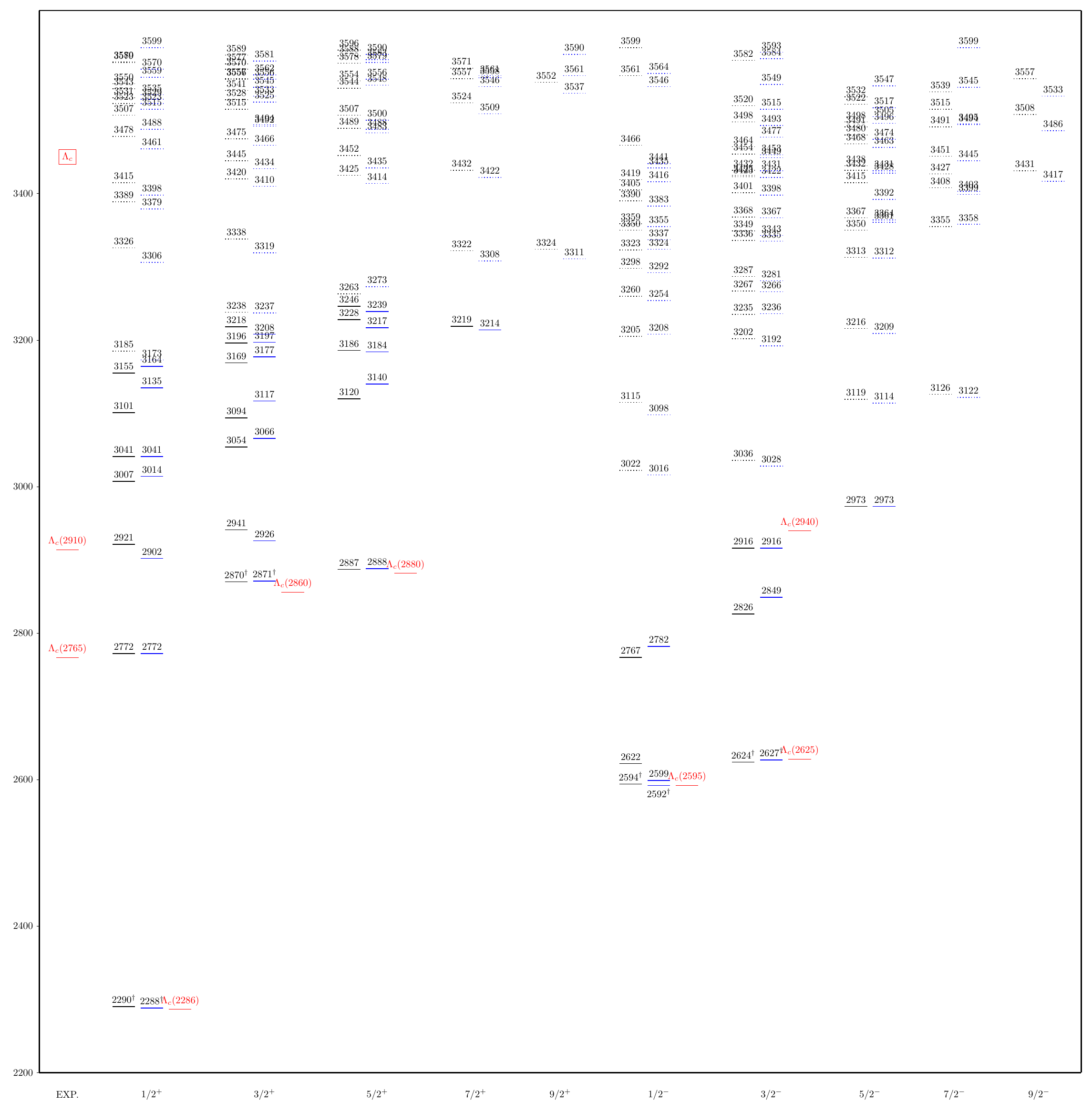}
\caption{Masses of the $\Lambda_{c}$ baryons (in units of MeV). Here, the black/blue lines stand for the numerical results obtained by the GI/MGI models. The red lines represent the experimental results taken from PDG~\cite{ParticleDataGroup:2022pth} and Ref.~\cite{Belle:2022hnm}. In the first column, we list the experimental observed states whose quantum numbers are not determined yet. The states used for fittings are indicated by daggers ($\dagger$).}
\label{fig:mass:nnc:MA}
\end{figure*}

\begin{figure*}
\includegraphics[width=500pt]{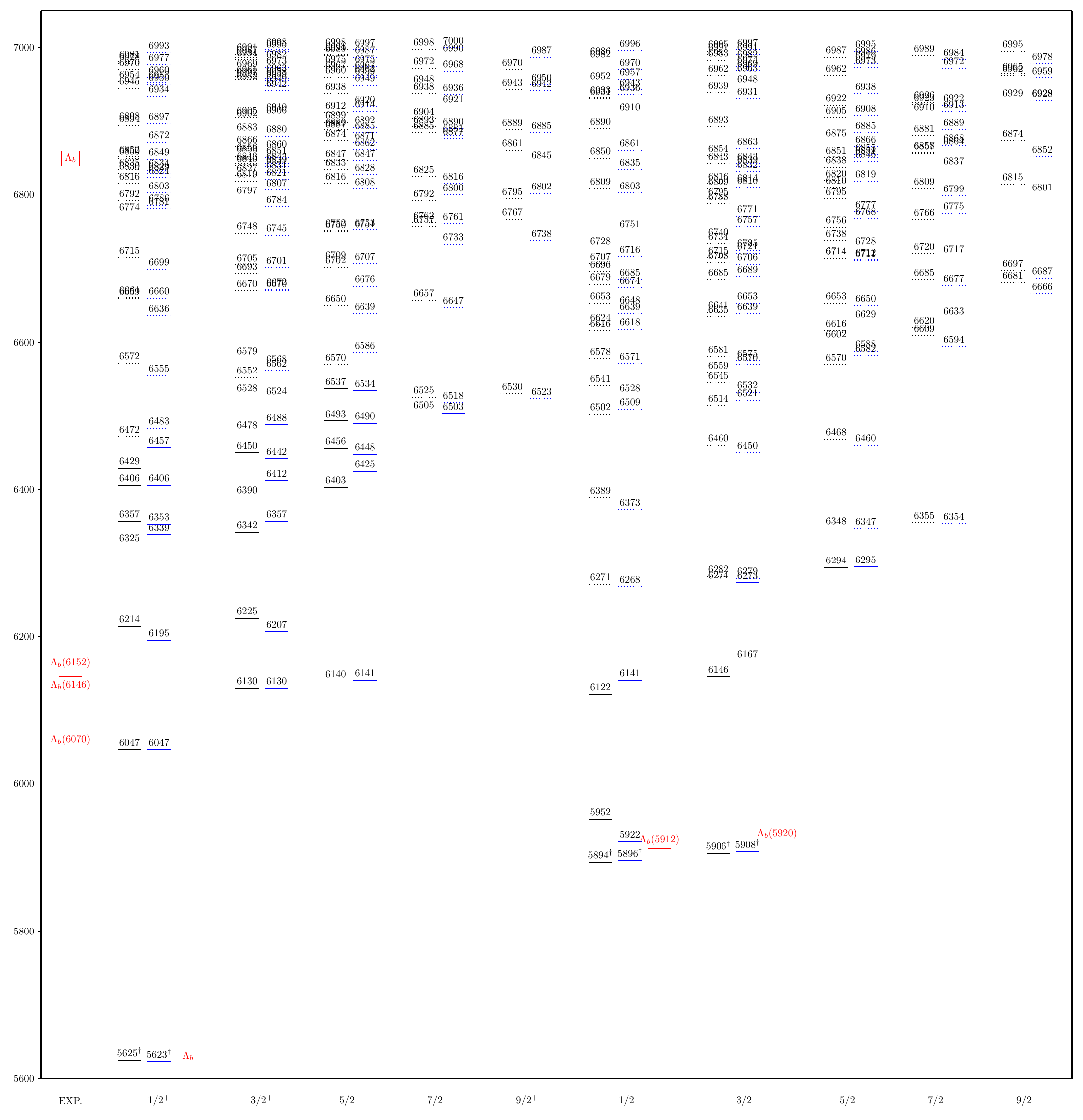}
\caption{Masses of the $\Lambda_{b}$ baryons (in units of MeV). Here, the black/blue lines stand for the numerical results obtained by the GI/MGI models. The red lines represent the experimental results taken from PDG~\cite{ParticleDataGroup:2022pth}. In the first column, we list the experimental observed states whose quantum numbers are not determined yet. The states used for fittings are indicated by daggers ($\dagger$).}
\label{fig:mass:nnb:MA}
\end{figure*}

The $\Sigma_{Q}$ and $\Lambda_{Q}$ baryon masses are listed in Tables~\ref{table:mass:nnc}--\ref{table:mass:nnb}.
Their relative positions are plotted in Figs.~\ref{fig:mass:nnc:MS}--\ref{fig:mass:nnb:MA}.

First we consider the $\Sigma_{c}$ baryons.
From Fig.~\ref{fig:mass:nnc:MS}, we can clearly see that the low-lying states form band structures.
The lowest band consists of the ground states of spin $1/2^{+}$ and $3/2^{+}$.
%
%
The next band consists of seven negative parity states dominated by $P$-wave excitation components.
Among them, the $1/2^{-}$ and $3/2^{-}$ states further form a two semi-band structure.
The lower two states are dominated by $p_{\lambda}$ modes, while the higher ones are dominated by $p_{\rho}$ excitations.
In 2004, the Belle Collaboration observed the $\Sigma_{c}(2800)$ state in the $\Lambda_{c}\pi$ channel~\cite{Belle:2004zjl}.
Its quantum numbers are not determined yet~\cite{ParticleDataGroup:2022pth}.
Our calculation suggests that it is most likely a $3/2^{-}$ state dominated by the $p_{\lambda}$ excitation.
The third band is dominated by the $N=2$ excitation (For convenience, we use the oscillator band to label the quark model states. When we refer to a state within $N=2$ band, we are implying that \emph{most} part of its wave functions belong to the $N=2$ band of oscillator).
More precisely, this band consists of $2s_{\rho}$, $2s_{\lambda}$, $d_{\rho}$, $d_{\lambda}$ and $[p_{\rho}p_{\lambda}]_{L}$ excitations.
We again find that the excitation over ${\lambda}$ mode is easier than over ${\rho}$ mode.
Taking $1/2^{+}$ states as an example, the $\Sigma_{c}(2925/2923)$ and $\Sigma_{c}(3025/3032)$ are dominated by the $2s_{\lambda}$ and $d_{\lambda}$ modes, the $\Sigma_{c}(3096/3112)$ and $\Sigma_{c}(3119/3118)$ are dominated by $d_{\rho}$ and $2s_{\rho}$ modes, while the $\Sigma_{c}(3025/3040)$ and $\Sigma_{c}(3165/3180)$ are both dominated by the $[p_{\rho}p_{\lambda}]_{L}$ mode.
We also calculate the heavier states which include higher excitations.
They are more packed with each other.
A more detailed study of their decay properties is needed.

The $\Sigma_{b}$ baryon spectrum is similar.
A main difference is that the splitting between the excitations over ${\lambda}$ and ${\rho}$ is larger.
For the $1/2^{-}$ states in the $N=1$ band, the $p_{\lambda}$-$p_{\rho}$ is about $110~\text{MeV}$ for the $\Sigma_{b}$ states and $80~\text{MeV}$ for the $\Sigma_{c}$ states.
The $3/2^{-}$ states are similar.
Another consequence is that the $N=2$ band begins to overlap with the $N\leq2$ bands.
Experimentally, there is one $\Sigma_{b}$ excited state observed by the LHCb Collaboration, namely the $\Sigma_{b}(6097)$~\cite{LHCb:2018haf}.
Our calculation suggests that it is a $1/2^{-}$ or $3/2^{-}$ state dominated by the $p_{\lambda}$ excitation.

Next we turn to the $\Lambda_{Q}$ baryons.
From Figs.~\ref{fig:mass:nnc:MA}--\ref{fig:mass:nnb:MA}, we see that our calculated results fit quite well with the experimental observations.
In Sec.~\ref{sec:Parameter}, we used the lowest $1/2^{\pm}$ and $3/2^{\pm}$ $\Lambda_{Q}$ states for parameter fits.
There remain four $\Lambda_{c}$ and three $\Lambda_{b}$ excitations.
Most of their quantum numbers are undetermined, except the $\Lambda_{c}(2880)$ and $\Lambda_{c}(2940)$ states.
The $\Lambda_{c}(2880)$ was observed in 2000 by the CLEO Collaboration~\cite{CLEO:2000mbh}.
The Belle Collaboration latter determined its quantum numbers to be $J^{P}=5/2^{+}$~\cite{Belle:2006xni}.
Our calculation shows that the mass of the lowest $5/2^{+}$ state is $2887/2888~\text{MeV}$, which lies very close to the experimental mass.
This state is dominated by the $d_{\lambda}$ excited component.
In 2006, the BABAR Collaboration observed the $\Lambda_{c}(2940)$ in the $D^{0}p$ decay mode~\cite{BaBar:2006itc}.
The LHCb Collaboration found that its quantum numbers is $3/2^{-}$~\cite{LHCb:2017jym}.
From Fig.~\ref{fig:mass:nnc:MA}, we see that it is consistent with the third $\Lambda_{c}$ state $\Lambda_{c}(2916/2916,3/2^{-})$ with quantum numbers $3/2^{-}$ within GI/MGI models.
Note that Capstick and Isgur also predicted this state back in 1986.
Their predicted mass was $2935~\text{MeV}$~\cite{Capstick:1986ter}.
The quantum numbers of $\Lambda_{c}(2765)$ and $\Lambda_{c}(2910)$ states are not determined.
Comparing to our numerical results, $\Lambda_{c}(2765)$ might be a $1/2^{+}$ or $1/2^{-}$ state.
And the $\Lambda_{c}(2910)$ is most likely a $1/2^{+}$ state, though it is also possibly a $3/2^{+}$ one.
Now we consider the $\Lambda_{b}$ baryons.
Their theoretical and experimental masses are plotted together in Fig.~\ref{fig:mass:nnb:MA}.
Our numerical results suggest that $\Lambda_{b}(6070)$ is a $1/2^{+}$ state dominated by the $2s_{\lambda}$ excitation.
The $\Lambda_{b}(6146)$ and $\Lambda_{b}(6152)$ states might belong to the $\{3/2^{+},5/2^{+}\}$ or $\{1/2^{-},3/2^{-}\}$ doublet.
A detailed study of their decay behavior is needed.
%

\subsection{The $\Omega_{Q}$ baryons}
\label{sec:ssQ}

\begin{figure*}
\includegraphics[width=500pt]{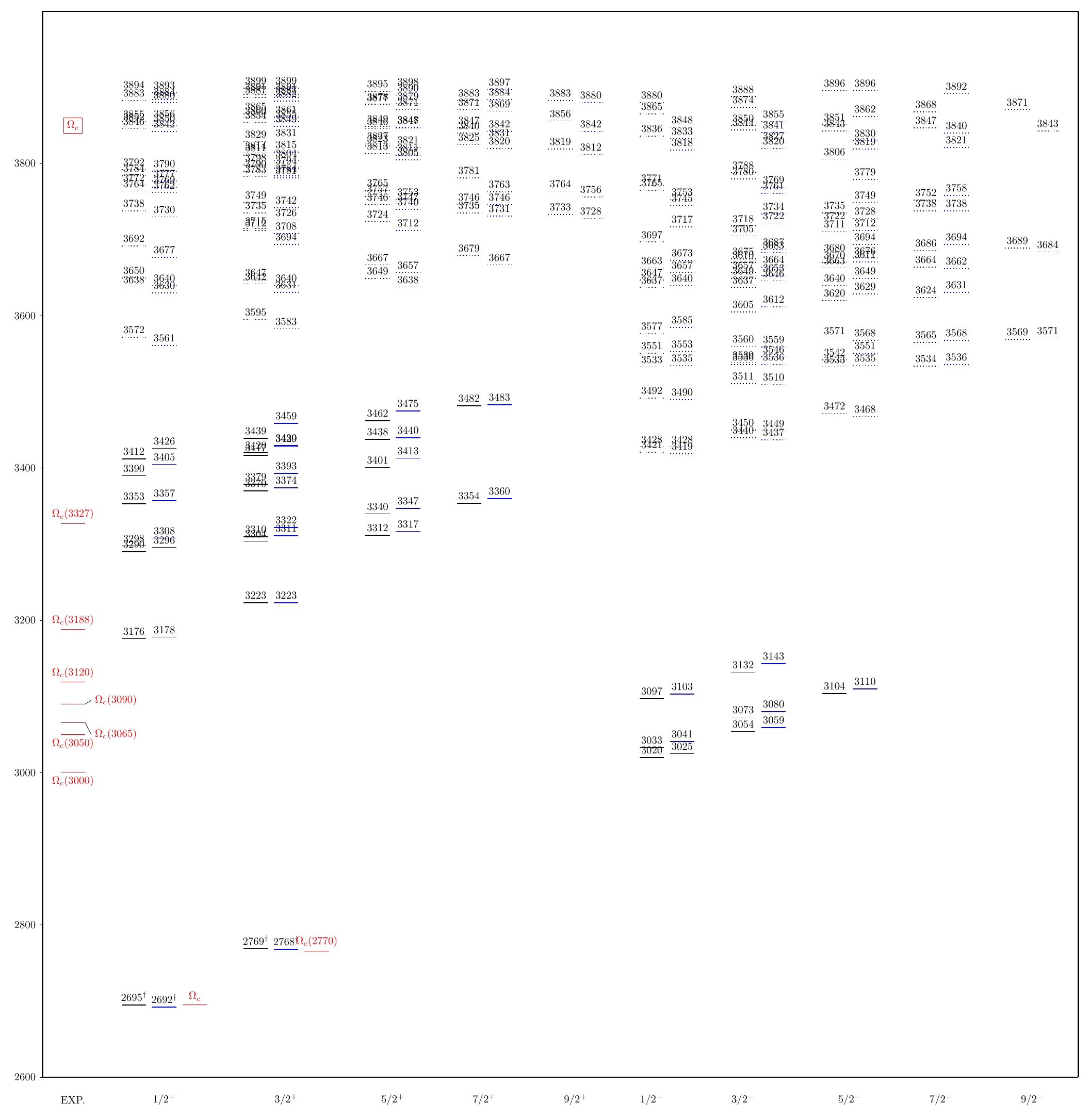}
\caption{Masses of the $\Omega_{c}$ baryons (in units of MeV). Here, the black/blue lines stand for the numerical results obtained by the GI/MGI models. The red lines represent the experimental results taken from PDG~\cite{ParticleDataGroup:2022pth}. In the first column, we list the experimental observed states whose quantum numbers are not determined yet. The states used for fittings are indicated by daggers ($\dagger$).}
\label{fig:mass:ssc}
\end{figure*}

\begin{figure*}
\includegraphics[width=500pt]{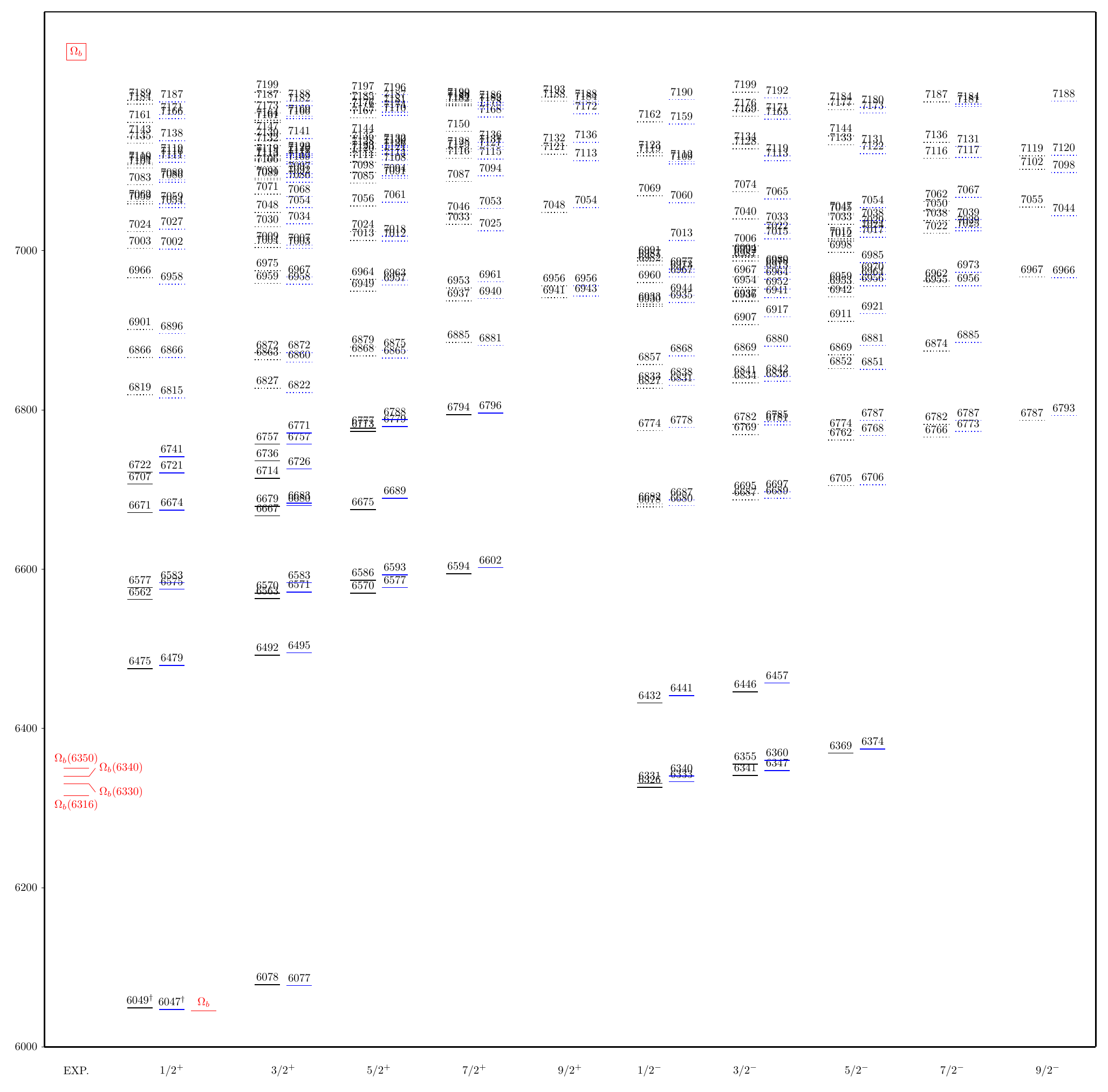}
\caption{Masses of the $\Omega_{b}$ baryons (in units of MeV). Here, the black/blue lines stand for the numerical results obtained by the GI/MGI models. The red lines represent the experimental results taken from PDG~\cite{ParticleDataGroup:2022pth}. In the first column, we list the experimental observed states whose quantum numbers are not determined yet. The states used for fittings are indicated by daggers ($\dagger$).}
\label{fig:mass:ssb}
\end{figure*}

Next we consider the $\Omega_{Q}$ baryons.
We present their masses in Table~\ref{table:mass:ssQ} and plot their relative position in Figs.~\ref{fig:mass:ssc}--\ref{fig:mass:ssb}.
The $\Omega_{Q}$ system is the strange counterpart of the $\Sigma_{Q}$ system.
So their mass spectra share a similar pattern.

For the $\Omega_{c}$ system (see Fig.~\ref{fig:mass:ssc}), the $\Omega_{c}$ and $\Omega_{c}^{*}$ are ground states, which are used for fits in this work.
Next we consider the $P$-wave excitation with negative parity, namely the seven states lying in the $3.0\sim3.15~\mathrm{GeV}$ mass region.
Among them, the $\Omega_{c}(3097/3103,1/2^{-})$ and $\Omega_{c}(3132/3141,3/2^{-})$ states are dominated by the $p_{\rho}$ excitations, while the others are $p_{\lambda}$ one.
As shown in Fig.~\ref{fig:mass:ssc}, these states may correspond to the five narrow $\Omega_{c}$ excited states observed by the LHCb collaboration~\cite{LHCb:2017uwr}.
Note that the $\Omega_{c}(3020/3025,1/2^{-})$ and $\Omega_{c}(3033/3041,1/2^{-})$ states are dominated by the mixing of
\begin{equation}
\psi_{1}=\ket{\left(1s_{\rho},1p_{\lambda}\right)_{P},1,0,1/2^{-}}\,,
\end{equation}
and
\begin{equation}
\psi_{2}=\ket{\left(1s_{\rho},1p_{\lambda}\right)_{P},1,1,1/2^{-}}
\end{equation}
bases in the notation $\ket{(l_{\rho},l_{\lambda})_{L},S_{12},j_{q}=L+S_{12},J^{P}}$.
The calculations based on $^{3}P_{0}$ model suggested that a pure $\psi_{1}$ state is very broad [$\Gamma\sim\mathcal{O}(10^{2}~\text{MeV})$] while a pure $\psi_{2}$ state is very narrow ($\Gamma\sim0~\text{MeV}$)~\cite{Zhao:2017fov}.
Since our parameters are solely fitted from the masses, we are unable to resolve the mixing angle between these two states with small mass difference.
We can only say that one of the two states is broad.
More precisely, the $\Omega_{c}(3000)$ is a $1/2^{-}$ state dominated by the $p_{\lambda}$ excitation, though we cannot determine whether it corresponds to $\Omega_{c}(3020/3025,1/2^{-})$ or $\Omega_{c}(3033/3041,1/2^{-})$.
The $\Omega_{c}(3050)$ and $\Omega_{c}(3065)$ correspond to the $\Omega_{c}(3054/3059,3/2^{-})$ and $\Omega_{c}(3073/3080,3/2^{-})$, which are also dominated by the $p_{\lambda}$ excitations.
The $\Omega_{c}(3090)$ is close to $\Omega_{c}(3097/3103,1/2^{-})$ and $\Omega_{c}(3104/3110,5/2^{-})$ states.
However, Ref.~\cite{Zhao:2017fov} suggested that the $J^{P}=1/2^{-}$ $p_{\rho}$ excitation should be a broad state ($\sim10^{2}~\text{MeV}$).
Thus we interpret $\Omega_{c}(3090)$ as the $\Omega_{c}(3104/3110,5/2^{-})$ state, which is dominated by the $p_{\lambda}$ excitation .
The next state, $\Omega_{c}(3120)$, corresponds to the $p_{\rho}$-dominated excitation $\Omega_{c}(3132/3143,3/2^{-})$.
In the higher mass region, the LHCb collaboration found a broad structure around 3188~MeV~\cite{LHCb:2017uwr}, which was later confirmed by the Belle~\cite{Belle:2017ext} and LHCb~\cite{LHCb:2023sxp} collaborations.
In Ref.~\cite{LHCb:2023sxp}, the LHCb collaboration also found the $\Omega_{c}(3327)$.
Comparing their masses with our calculated masses, we interpret the $\Omega_{c}(3188)$ as the $\Omega_{c}(3176/3178,1/2^{+})$ state dominated by the $2s_{\lambda}$ excitation, while $\Omega_{c}(3327)$ is a $d_{\lambda}$-dominated excitation with quantum numbers $3/2^{+}$ or $5/2^{+}$.

Now we consider the $\Omega_{b}$ system (see Fig.~\ref{fig:mass:ssb}).
First we consider the ground states.
Only the $1/2^{+}$ state is observed in experiment, while the $3/2^{-}$ state, $\Omega_{b}^{*}$, is not observed yet.
Our calculation predicts the mass of $\Omega_{b}^{*}$ to be 6078/6077~MeV.
Next we consider the $P$-wave excitations.
Unlike the $\Omega_{c}$ case, an explicit two semi-band structure appears among the $P$-wave $\Omega_{b}$ excited states.
The five lower states are dominated by $p_{\lambda}$ excitations while the two higher states are dominated by $p_{\rho}$ excitations.
In 2020, the LHCb collaboration observed four narrow excited $\Omega_{b}$ states, namely the $\Omega_{b}(6316)$, $\Omega_{b}(6330)$, $\Omega_{b}(6340)$, and $\Omega_{b}(6350)$ states~\cite{LHCb:2020tqd}.
As shown in Fig.~\ref{fig:mass:ssb}, the masses of these states lie in the mass region of the $p_{\lambda}$-dominated excitations.
Similar to the logic of the $\Omega_{c}$ baryons, the $\Omega_{b}(6316)$ is a $1/2^{-}$ state.
The $\Omega_{b}(6330)$ and $\Omega_{b}(6340)$ correspond to $\Omega_{b}(6341/6347,3/2^{-})$ and $\Omega_{b}(6355/6360,3/2^{-})$.
And the $\Omega_{b}(6350)$ corresponds to $\Omega_{b}(6369/6374,5/2^{-})$.
The $p_{\rho}$-dominated excitations is about 100~MeV above the $p_{\lambda}$-dominated excitations.
In particular, the $\Omega_{b}(6446/6457,3/2^{-})$, which is the bottom counterpart of the $\Omega_{c}(3120)$ state, should be a narrow state.
We hope the future experiments can search for these higher states.
%

\subsection{The $\Xi_{Q}$ baryons}
\label{sec:nsQ}

\begin{figure*}
\includegraphics[width=500pt]{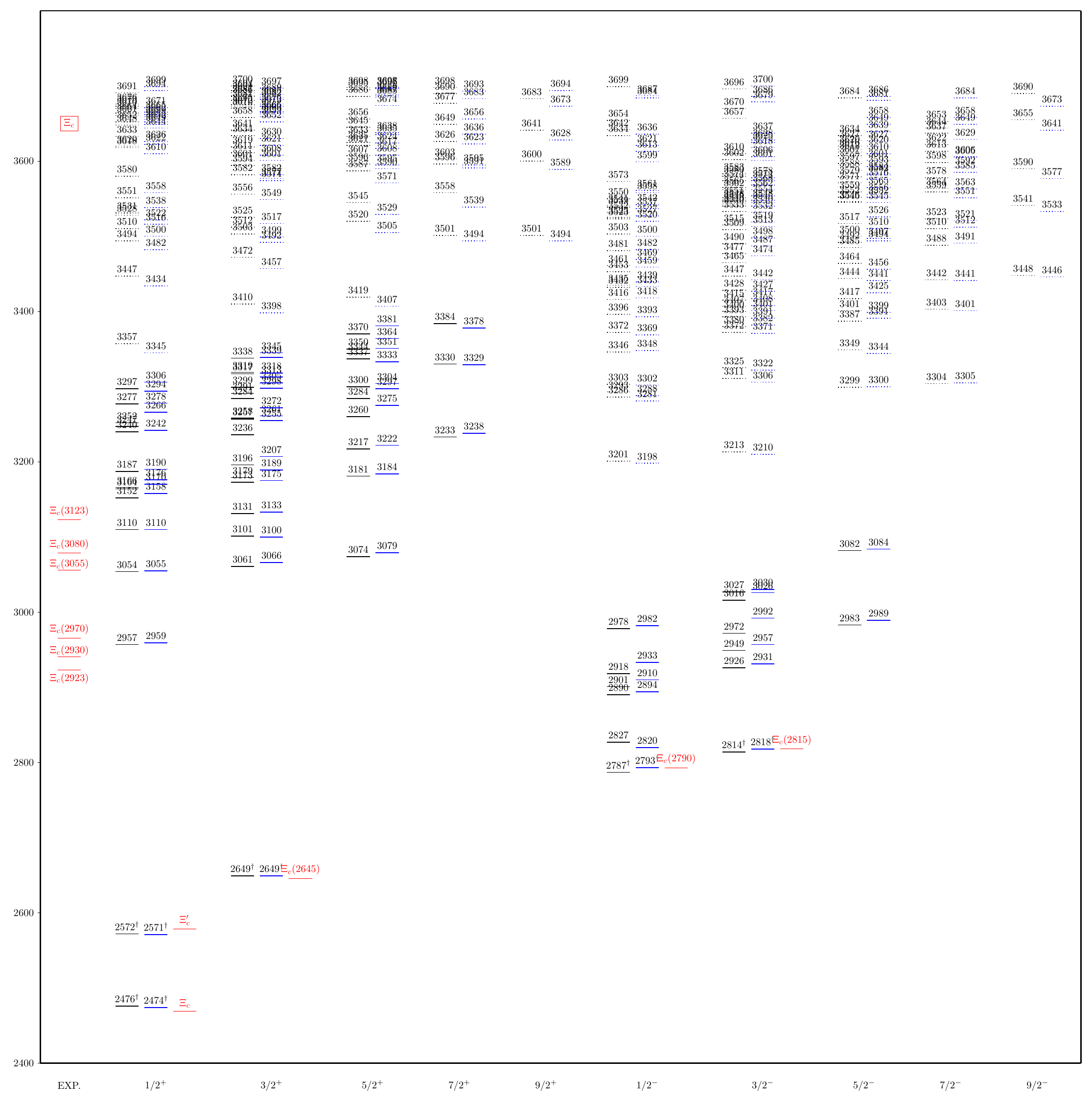}
\caption{Masses of the $\Xi_{c}$ baryons (in units of MeV). Here, the black/blue lines stand for the numerical results obtained by the GI/MGI models. The red lines represent the experimental results taken from PDG~\cite{ParticleDataGroup:2022pth}. In the first column, we list the experimental observed states whose quantum numbers are not determined yet. The states used for fittings are indicated by daggers ($\dagger$).}
\label{fig:mass:nsc}
\end{figure*}

\begin{figure*}
\includegraphics[width=500pt]{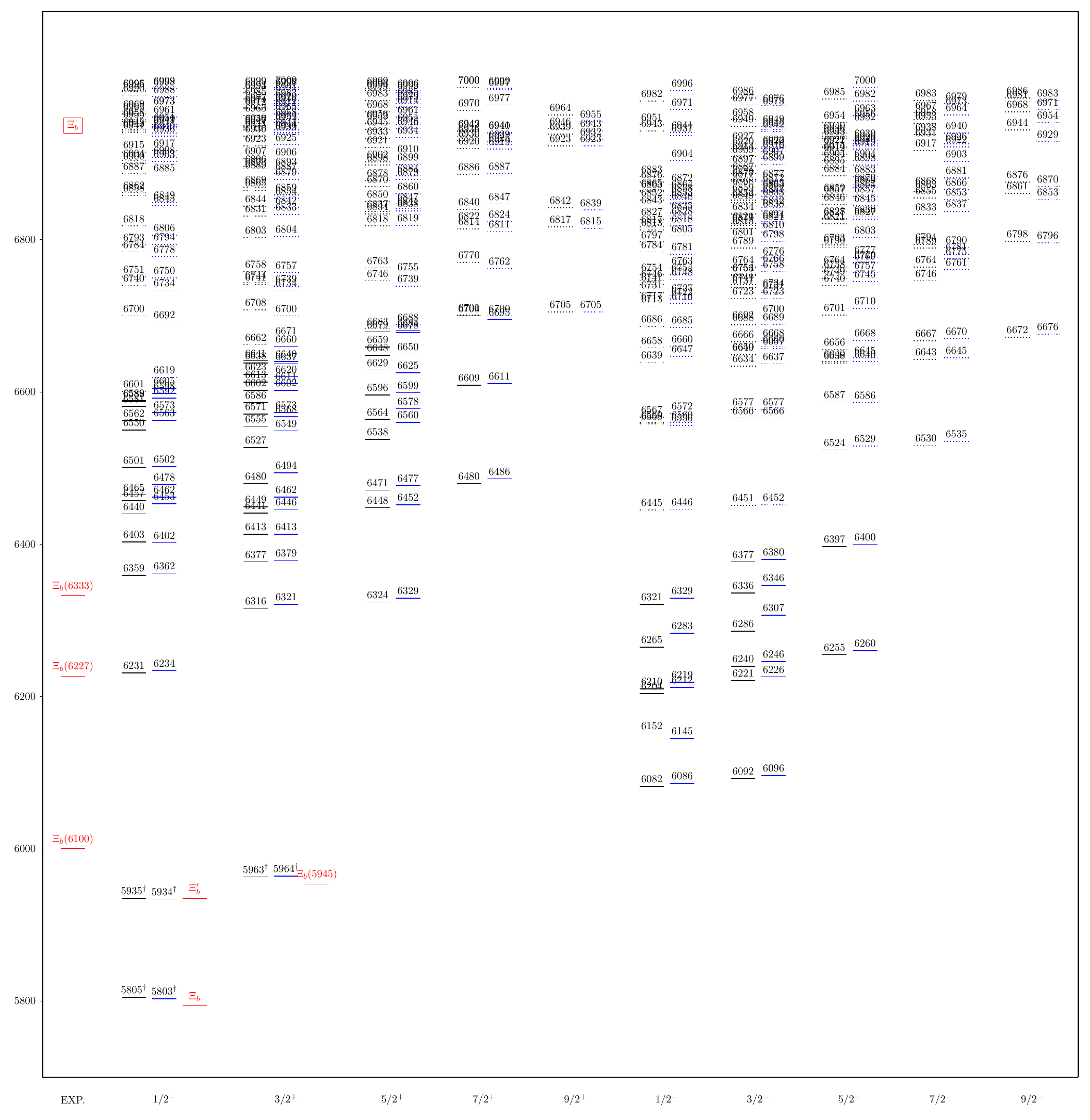}
\caption{Masses of the $\Xi_{b}$ baryons (in units of MeV). Here, the black/blue lines stand for the numerical results obtained by the GI/MGI models. The red lines represent the experimental results taken from PDG~\cite{ParticleDataGroup:2022pth}. In the first column, we list the experimental observed states whose quantum numbers are not determined yet. The states used for fittings are indicated by daggers ($\dagger$).}
\label{fig:mass:nsb}
\end{figure*}

Now we move to the $\Xi_{Q}$ system with strangeness $-1$.
Their masses are listed in Table~\ref{table:mass:nsQ}.
The $\Xi_{Q}$ baryons are composed of three different quarks, so they are not constrained by the Pauli principle.
The consequence is that more bases are allowed to be mixed.
In Figs.~\ref{fig:mass:nsc}--\ref{fig:mass:nsb}, we plot the calculated masses of the $\Xi_{Q}$ baryons.
We also plot the experimental states for comparison.
For the $\Xi_{c}$ baryons, we used the ground states and the lowest $1/2^{-}$ and $3/2^{-}$ states for parameter fits in Sec.~\ref{sec:Parameter}.
There are still six excited states whose quantum numbers are undetermined.
From Fig.~\ref{fig:mass:nsc}, we find that it is unlikely to uniquely determine their quantum numbers.
The $\Xi_{c}(2923)$/$\Xi_{c}(2930)$ might be a $1/2^{-}$ or $3/2^{-}$ state, while the $\Xi_{c}(2970)$ might be a $1/2^{-}$, $3/2^{-}$ or $5/2^{-}$ state.
In the higher mass region, the quantum numbers of $\Xi_{c}(3055)$/$\Xi_{c}(3123)$ might be $1/2^{+}$ or $3/2^{+}$, while the quantum numbers of $\Xi_{c}(3030)$ might be $5/2^{\pm}$.
The $\Xi_{b}$ baryons are similar.
We used the three ground states for fits.
For the higher states, the $\Xi_{b}(6100)$ is most likely a $3/2^{-}$ excitation, the $\Xi_{b}(6227)$ could be a $1/2^{+}$ or $3/2^{-}$ excitation, while the quantum numbers of the $\Xi_{b}(6333)$ might be $5/2^{+}$, $1/2^{-}$ or $3/2^{-}$.
We hope the future experiments can determine their quantum numbers.
A detailed study of their decay properties are also necessary.
%

\section{Summary}
\label{Sec:Summary}

In this work, we have systematically studied the masses of the heavy baryons in the relativized quark model, namely the (modified) Godfrey-Isgur [(M)GI] model.
The key ingredients of the GI model are a universal one-gluon-exchange interaction and a linear confinement potential motivated by QCD.
The relativistic effects are included semi-quantitatively.
In the MGI model, the linear confinement interaction is replaced by a screened one.
The parameters are fitted from the masses of the ground state baryons, the lowest $1/2^{-}$ and $3/2^{-}$ singly heavy baryons and the lowest $\frac{3}{2}^{+}\Lambda_{Q}$ states.
We found that all heavy baryons observed can be explained as three-quark states.
In particular,
\begin{enumerate}[(a)]
\item The $\Omega_{c}(3000)$, $\Omega_{c}(3050)$, $\Omega_{c}(3065)$ and $\Omega_{c}(3090)$ states are $p_{\lambda}$-dominated excitations with quantum numbers $1/2^{-}$, $3/2^{-}$, $3/2^{-}$ and $5/2^{-}$, while the $\Omega_{c}(3120)$ is a $3/2^{-}$ state dominated the $p_{\rho}$ excitation.
The higher state $\Omega_{c}(3188)$ is the $2s_{\lambda}$ excitation with quantum numbers $1/2^{+}$, and $\Omega_{c}(3327)$ is a $d_{\lambda}$ excitation with quantum numbers $3/2^{+}$ or $5/2^{+}$.
\item The $\Omega_{b}(6316)$, $\Omega_{b}(6330)$, $\Omega_{b}(6340)$, and $\Omega_{b}(6350)$ states are $p_{\lambda}$-dominated excitations with quantum numbers $1/2^{-}$, $3/2^{-}$, $3/2^{-}$, and $5/2^{-}$, while a narrow $p_{\rho}$-dominated excited state $\Omega_{b}(6446/6457,3/2^{-})$ is predicted to be about 100~MeV above them.
\item The $\Sigma_{c}(2800)$ is most likely a $3/2^{-}$ state dominated by the $p_{\lambda}$ excitation, while the $\Sigma_{b}(6097)$ is a $1/2^{-}$ or $3/2^{-}$ state dominated by the $p_{\lambda}$ excitation.
\item Our calculation reproduced the masses of the well-known $\Lambda_{c}(2880)$ and $\Lambda_{c}(2940)$ states, whose quantum numbers are $5/2^{+}$ and $3/2^{-}$ respectively.
The $\Lambda_{c}(2765)$ might be a $1/2^{+}$ or $1/2^{-}$ state, and the $\Lambda_{c}(2910)$ could be a $1/2^{+}$ or $3/2^{+}$ one.
Furthermore, the $\Lambda_{b}(6070)$ is a $1/2^{+}$ state dominated by the $2s_{\lambda}$ excitation, while the $\Lambda_{b}(6146)$ and $\Lambda_{b}(6152)$ states might belong to the $\{3/2^{+},5/2^{+}\}$ or $\{1/2^{-},3/2^{-}\}$ doublet.
\end{enumerate}
Hopefully, our calculations will be useful for the search of the new baryon states.
More experimental and theoretical investigations are expected to verify and understand the baryon system in the future.
%

\section*{Acknowledgments}

We are grateful to Prof.~Marek~Karliner, Prof.~Guang-Juan~Wang and Dr.~Lu~Meng for helpful comments and discussions.
We thank the High-performance Computing Platform of Peking University for providing computational resources.
This project was supported by the National Natural Science Foundation of China (NSFC) under Grant No.~11975033 and No.~12070131001;
and the NSFC-ISF under Grant No. 3423/19.
%


\begin{table*}[htbp]
\centering
\caption{Masses of $\Lambda_{c}$ and $\Sigma_{c}$ baryons (in units of MeV).}
\label{table:mass:nnc}
\begin{tabular}{ccccccccccccccccccccccc}
\toprule[1pt]
\toprule[1pt]
State&$J^{P}$&Model&Mass \\
\midrule[1pt]
$\Lambda_{c}$
&$1/2^{+}$&GI&2290&2772&2921&3007&3041&3101&3155&3185&3326&3389\\
&&MGI&2288&2772&2902&3014&3041&3135&3164&3173&3306&3379\\
&$3/2^{+}$&GI&2870&2941&3054&3094&3169&3196&3218&3238&3338\\
&&MGI&2871&2926&3066&3117&3177&3197&3208&3237&3319\\
&$5/2^{+}$&GI&2887&3120&3186&3228&3246&3263\\
&&MGI&2888&3140&3184&3217&3239&3273\\
&$7/2^{+}$&GI&3219&3322\\
&&MGI&3214&3308\\
&$9/2^{+}$&GI&3324\\
&&MGI&3311\\
&$1/2^{-}$&GI&2594&2622&2767&3022&3115&3205&3260&3298&3323&3350&3359&3390\\
&&MGI&2592&2599&2782&3016&3098&3208&3254&3292&3324&3337&3355&3383\\
&$3/2^{-}$&GI&2624&2826&2916&3036&3202&3235&3267&3287&3336&3349&3368\\
&&MGI&2627&2849&2916&3028&3192&3236&3266&3281&3335&3343&3367\\
&$5/2^{-}$&GI&2973&3119&3216&3313&3350&3367\\
&&MGI&2973&3114&3209&3312&3361&3364\\
&$7/2^{-}$&GI&3126&3355\\
&&MGI&3122&3358\\
\midrule[1pt]
$\Sigma_{c}$
&$1/2^{+}$&GI&2439&2925&3025&3035&3096&3119&3165&3323&3377&3404&3457&3488\\
&&MGI&2437&2923&3032&3040&3112&3118&3180&3305&3364&3383&3431&3480\\
&$3/2^{+}$&GI&2519&2973&3030&3044&3124&3139&3151&3176&3193&3348&3376&3389&3473&3480&3492\\
&&MGI&2518&2969&3027&3054&3135&3143&3168&3189&3217&3328&3354&3377&3453&3457&3477\\
&$5/2^{+}$&GI&3040&3091&3165&3227&3233&3384&3418&3478&3485\\
&&MGI&3036&3092&3181&3220&3248&3360&3394&3458&3476\\
&$7/2^{+}$&GI&3109&3280&3431&3460&3485\\
&&MGI&3109&3269&3406&3440&3479\\
&$9/2^{+}$&GI&3459\\
&&MGI&3440\\
&$1/2^{-}$&GI&2751&2764&2845&3158&3166&3246&3290&3304&3329&3394&3401&3423&3451\\
&&MGI&2753&2773&2852&3149&3162&3244&3287&3305&3338&3387&3410&3433&3480\\
&$3/2^{-}$&GI&2787&2820&2890&3177&3197&3268&3273&3308&3325&3366&3370&3408&3420&3442&3447&3459&3486\\
&&MGI&2788&2825&2905&3167&3189&3263&3276&3300&3314&3371&3380&3404&3422&3452&3455&3485&3489\\
&$5/2^{-}$&GI&2857&3223&3258&3278&3348&3372&3391&3431&3440&3456&3496\\
&&MGI&2861&3212&3248&3284&3334&3382&3395&3437&3445&3466&3487\\
&$7/2^{-}$&GI&3261&3317&3392&3450&3463\\
&&MGI&3251&3309&3395&3436&3472\\
&$9/2^{-}$&GI&3324&3483\\
&&MGI&3315&3464\\
\bottomrule[1pt]
\bottomrule[1pt]
\end{tabular}
\end{table*}
%
\begin{table*}[htbp]
\centering
\caption{Masses of $\Lambda_{b}$ and $\Sigma_{b}$ baryons (in units of MeV).}
\label{table:mass:nnb}
\begin{tabular}{ccccccccccccccccccccccc}
\toprule[1pt]
\toprule[1pt]
State&$J^{P}$&Model&Mass \\
\midrule[1pt]
$\Lambda_{b}$
&$1/2^{+}$&GI&5625&6047&6214&6325&6357&6406&6429&6472&6572&6659&6661\\
&         &MGI&5623&6047&6195&6339&6353&6406&6457&6483&6555&6636&6660\\
&$3/2^{+}$&GI&6130&6225&6342&6390&6450&6478&6528&6552&6579&6670&6693\\
&         &MGI&6130&6207&6357&6412&6442&6488&6524&6562&6568&6670&6672\\
&$5/2^{+}$&GI&6140&6403&6456&6493&6537&6570&6650\\
&         &MGI&6141&6425&6448&6490&6534&6586&6639&6676\\
&$7/2^{+}$&GI&6505&6525&6657\\
&         &MGI&6503&6518&6647\\
&$9/2^{+}$&GI&6530\\
&         &MGI&6523\\
&$1/2^{-}$&GI&5894&5952&6122&6271&6389&6502&6541&6578&6616&6624&6653&6679&6696\\
&         &MGI&5896&5922&6141&6268&6373&6509&6528&6571&6618&6639&6648&6674&6685\\
&$3/2^{-}$&GI&5906&6146&6274&6282&6460&6514&6545&6559&6581&6635&6641&6685\\
&         &MGI&5908&6167&6273&6279&6450&6521&6532&6570&6575&6639&6653&6689\\
&$5/2^{-}$&GI&6294&6348&6468&6570&6602&6616&6653\\
&         &MGI&6295&6347&6460&6582&6588&6629&6650\\
&$7/2^{-}$&GI&6355&6609&6620&6685\\
&         &MGI&6354&6594&6633&6677\\
&$9/2^{-}$&GI&6681&6697\\
&         &MGI&6666&6687\\
\midrule[1pt]
$\Sigma_{b}$
&$1/2^{+}$&GI&5809&6237&6313&6332&6439&6446&6479&6578&6620&6661&6743&6762&6802&6837&6844&6846&6873&6880&6883\\
&&MGI&5809&6236&6324&6337&6442&6459&6502&6565&6616&6653&6726&6760&6803&6822&6834&6847&6854&6876&6883&6898\\
&$3/2^{+}$&GI&5838&6253&6307&6322&6437&6454&6458&6491&6527&6585&6610&6627&6724&6753&6765&6777&6810&6829&6847&6856\\
&&MGI&5838&6252&6306&6334&6451&6454&6473&6519&6545&6572&6596&6622&6719&6736&6763&6771&6811&6827&6841&6852\\
&$5/2^{+}$&GI&6314&6352&6448&6547&6582&6615&6642&6704&6730&6783&6813&6839&6840&6877&6891&6897\\
&&MGI&6313&6354&6465&6567&6573&6601&6629&6707&6723&6778&6795&6836&6845&6858&6869&6879&6887\\
&$7/2^{+}$&GI&6362&6600&6650&6680&6709&6822&6825&6843&6897\\
&         &MGI&6364&6590&6637&6669&6713&6804&6826&6849&6883&6895\\
&$9/2^{+}$&GI&6684&6725&6828\\
&         &MGI&6674&6716&6829\\
&$1/2^{-}$&GI&6075&6081&6195&6430&6438&6535&6587&6595&6631&6696&6700&6729&6742&6750&6763&6830&6890&6898\\
&&MGI&6080&6090&6207&6424&6440&6537&6591&6597&6645&6697&6706&6723&6738&6739&6793&6818&6874&6880\\
&$3/2^{-}$&GI&6090&6119&6212&6436&6453&6520&6544&6603&6627&6647&6659&6701&6707&6736&6746&6748&6761&6768&6788&6802\\
&&MGI&6092&6122&6227&6431&6448&6529&6545&6601&6618&6660&6668&6703&6718&6728&6740&6748&6753&6790&6797&6806\\
&$5/2^{-}$&GI&6134&6464&6503&6527&6639&6645&6664&6711&6733&6754&6768&6784&6796&6808&6818&6846\\
&&MGI&6138&6458&6498&6536&6629&6657&6673&6728&6748&6749&6759&6763&6797&6805&6814&6850&6884&6899\\
&$7/2^{-}$&GI&6508&6548&6651&6745&6760&6789&6825&6827&6863&6879\\
&         &MGI&6504&6545&6663&6745&6760&6767&6798&6814&6868&6871&6887\\
&$9/2^{-}$&GI&6554&6772&6830&6851&6883\\
&         &MGI&6552&6757&6803&6830&6875\\
\bottomrule[1pt]
\bottomrule[1pt]
\end{tabular}
\end{table*}
%
\begin{table*}[htbp]
\centering
\caption{Masses of $\Xi_{c}$ and $\Xi_{b}$ baryons (in units of MeV).}
\label{table:mass:nsQ}
\begin{tabular}{ccccccccccccccccccccccc}
\toprule[1pt]
\toprule[1pt]
State&$J^{P}$&Model&Mass \\
\midrule[1pt]
$\Xi_{c}$
&$1/2^{+}$&GI&2476&2572&2957&3054&3110&3152&3164&3166&3187&3240&3247&3252&3277&3297&3357&3447&3494&3510&3528&3531\\
&&MGI&2474&2571&2959&3055&3110&3158&3170&3176&3190&3242&3266&3278&3294&3306&3345&3434&3482&3500&3516&3522\\
&$3/2^{+}$&GI&2649&3061&3101&3131&3173&3179&3196&3236&3257&3258&3284&3291&3299&3317&3319&3338&3410&3472&3503&3512\\
&&MGI&2649&3066&3100&3133&3175&3189&3207&3255&3261&3272&3298&3305&3313&3318&3339&3345&3398&3457&3492&3499\\
&$5/2^{+}$&GI&3074&3181&3217&3260&3284&3300&3337&3344&3350&3370&3419&3520&3545&3587&3596\\
&&MGI&3079&3184&3222&3275&3297&3304&3333&3351&3364&3381&3407&3505&3529&3571&3590&3595\\
&$7/2^{+}$&GI&3233&3330&3384&3501&3558&3596\\
&         &MGI&3238&3329&3378&3494&3539&3591&3595\\
&$9/2^{+}$&GI&3501\\
&         &MGI&3494&3589\\
&$1/2^{-}$&GI&2787&2827&2890&2901&2918&2978&3201&3286&3293&3303&3346&3372&3396&3416&3432&3435&3453&3461&3481&3503\\
&&MGI&2793&2820&2894&2910&2933&2982&3198&3281&3288&3302&3348&3369&3393&3418&3433&3439&3459&3469&3482&3500\\
&$3/2^{-}$&GI&2814&2926&2949&2972&3016&3027&3213&3311&3325&3372&3380&3393&3400&3407&3415&3428&3447&3465&3477&3490\\
&&MGI&2818&2931&2957&2992&3026&3030&3210&3306&3322&3371&3382&3391&3401&3408&3417&3427&3442&3474&3487&3498\\
&$5/2^{-}$&GI&2983&3082&3299&3349&3387&3401&3417&3444&3464&3485&3492&3500&3517&3545&3546&3552&3559&3571&3579&3588\\
&&MGI&2989&3084&3300&3344&3391&3399&3425&3441&3456&3494&3497&3510&3526&3545&3552&3556&3565&3576&3582&3584\\
&$7/2^{-}$&GI&3304&3403&3442&3488&3510&3523&3559&3564&3578&3598\\
&         &MGI&3305&3401&3441&3491&3512&3521&3551&3563&3585&3592\\
&$9/2^{-}$&GI&3448&3541&3590\\
&         &MGI&3446&3533&3577\\
\midrule[1pt]
$\Xi_{b}$
&$1/2^{+}$&GI&5805&5935&6231&6359&6403&6440&6457&6465&6501&6550&6562&6581&6588&6589&6601&6700&6740&6751&6784&6793\\
&&MGI&5803&5934&6234&6362&6402&6453&6462&6478&6502&6563&6573&6592&6598&6605&6619&6692&6734&6750&6778&6794\\
&$3/2^{+}$&GI&5963&6316&6377&6413&6441&6449&6480&6527&6555&6571&6586&6602&6613&6623&6638&6641&6662&6708&6741&6744\\
&&MGI&5964&6321&6379&6413&6446&6462&6494&6549&6568&6573&6602&6611&6620&6637&6640&6660&6671&6700&6734&6739\\
&$5/2^{+}$&GI&6324&6448&6471&6538&6564&6596&6629&6648&6659&6679&6683&6746&6763&6818&6834&6837&6850&6870&6878&6898\\
&&MGI&6329&6452&6477&6560&6578&6599&6625&6650&6678&6681&6688&6739&6755&6819&6838&6841&6847&6860&6879&6884\\
&$7/2^{+}$&GI&6480&6609&6700&6701&6770&6814&6822&6840&6886\\
&         &MGI&6486&6611&6695&6700&6762&6811&6824&6847&6887\\
&$9/2^{+}$&GI&6705&6817&6842\\
&         &MGI&6705&6815&6839\\
&$1/2^{-}$&GI&6082&6152&6204&6210&6265&6321&6445&6558&6560&6567&6639&6658&6686&6713&6717&6731&6741&6746&6754&6784\\
&&MGI&6086&6145&6212&6219&6283&6329&6446&6556&6560&6572&6647&6660&6685&6716&6722&6727&6748&6754&6763&6781\\
&$3/2^{-}$&GI&6092&6221&6240&6286&6336&6377&6451&6566&6577&6634&6649&6650&6666&6688&6692&6723&6737&6741&6753&6754\\
&&MGI&6096&6226&6246&6307&6346&6380&6452&6566&6577&6637&6657&6660&6668&6689&6700&6723&6731&6734&6758&6766\\
&$5/2^{-}$&GI&6255&6397&6524&6587&6638&6640&6656&6701&6740&6749&6758&6764&6790&6793&6821&6827&6828&6846&6857&6859\\
&&MGI&6260&6400&6529&6586&6640&6645&6668&6710&6745&6757&6769&6770&6777&6803&6827&6830&6845&6857&6864&6869\\
&$7/2^{-}$&GI&6530&6643&6667&6746&6764&6789&6794&6833&6855&6863&6868\\
&         &MGI&6535&6645&6670&6761&6775&6781&6790&6837&6853&6866&6881\\
&$9/2^{-}$&GI&6672&6798&6861&6876\\
&         &MGI&6676&6796&6853&6870\\
\bottomrule[1pt]
\bottomrule[1pt]
\end{tabular}
\end{table*}
%
\begin{table*}[htbp]
\centering
\caption{Masses of $\Omega_{c}$ and $\Omega_{b}$ baryons (in units of MeV).}
\label{table:mass:ssQ}
\begin{tabular}{ccccccccccccccccccccccc}
\toprule[1pt]
\toprule[1pt]
State&$J^{P}$&Model&Mass \\
\midrule[1pt]
$\Omega_{c}$
&$1/2^{+}$&GI&2695&3176&3290&3298&3353&3390&3412&3572&3638&3650&3692&3738&3764&3772&3784&3792\\
&&MGI&2692&3178&3296&3308&3357&3405&3426&3561&3630&3640&3677&3730&3762&3769&3777&3790\\
&$3/2^{+}$&GI&2769&3223&3304&3310&3370&3379&3417&3420&3439&3595&3642&3647&3712&3715&3735&3749&3783&3790&3798\\
&&MGI&2768&3223&3311&3322&3374&3393&3429&3430&3459&3583&3631&3640&3694&3708&3726&3742&3781&3784&3794\\
&$5/2^{+}$&GI&3312&3340&3401&3438&3462&3649&3667&3724&3746&3757&3765\\
&         &MGI&3317&3347&3413&3440&3475&3638&3657&3712&3740&3747&3753\\
&$7/2^{+}$&GI&3354&3482&3679&3735&3746&3781\\
&         &MGI&3360&3483&3667&3731&3746&3763\\
&$9/2^{+}$&GI&3733&3764\\
&         &MGI&3728&3756\\
&$1/2^{-}$&GI&3020&3033&3097&3421&3428&3492&3533&3551&3577&3637&3647&3663&3697&3765&3771\\
&&MGI&3025&3041&3103&3419&3428&3490&3535&3553&3585&3640&3657&3673&3717&3745&3753\\
&$3/2^{-}$&GI&3054&3073&3132&3440&3450&3511&3536&3539&3560&3605&3637&3649&3657&3670&3675&3705&3718&3780&3788\\
&&MGI&3059&3080&3143&3437&3449&3510&3536&3546&3559&3612&3646&3653&3664&3683&3687&3722&3734&3761&3769\\
&$5/2^{-}$&GI&3104&3472&3533&3542&3571&3620&3640&3663&3670&3680&3711&3722&3735\\
&         &MGI&3110&3468&3535&3551&3568&3629&3649&3671&3676&3694&3712&3728&3749&3779\\
&$7/2^{-}$&GI&3534&3565&3624&3664&3686&3738&3752\\
&         &MGI&3536&3568&3631&3662&3694&3738&3758\\
&$9/2^{-}$&GI&3569&3689\\
&         &MGI&3571&3684\\
\midrule[1pt]
$\Omega_{b}$
&$1/2^{+}$&GI&6049&6475&6562&6577&6671&6707&6722&6819&6866&6901&6966&7003&7024&7059&7062&7083\\
&&MGI&6047&6479&6575&6583&6674&6721&6741&6815&6866&6896&6958&7002&7027&7054&7059&7086&7089\\
&$3/2^{+}$&GI&6078&6492&6563&6570&6667&6679&6714&6736&6757&6827&6863&6872&6959&6975&7004&7009&7030&7048&7071&7089\\
&&MGI&6077&6495&6571&6583&6680&6683&6726&6757&6771&6822&6860&6872&6958&6967&7003&7007&7034&7054&7068&7086\\
&$5/2^{+}$&GI&6570&6586&6675&6773&6777&6868&6879&6949&6964&7013&7024&7056&7085&7098\\
&         &MGI&6577&6593&6689&6779&6788&6865&6875&6957&6963&7012&7018&7061&7091&7094\\
&$7/2^{+}$&GI&6594&6794&6885&6937&6953&7033&7046&7087\\
&         &MGI&6602&6796&6881&6940&6961&7025&7053&7094\\
&$9/2^{+}$&GI&6941&6956&7048\\
&         &MGI&6943&6956&7054\\
&$1/2^{-}$&GI&6326&6331&6432&6678&6682&6774&6827&6833&6857&6930&6933&6960&6982&6987&6991&7069\\
&&MGI&6333&6340&6441&6680&6687&6778&6831&6838&6868&6935&6944&6967&6973&6977&7013&7060\\
&$3/2^{-}$&GI&6341&6355&6446&6687&6695&6769&6782&6834&6841&6869&6907&6936&6937&6954&6967&6987&6992&6994&7006&7040\\
&&MGI&6347&6360&6457&6689&6697&6781&6785&6836&6842&6880&6917&6941&6952&6964&6973&6978&6980&7015&7022&7033\\
&$5/2^{-}$&GI&6369&6705&6762&6774&6852&6869&6911&6942&6953&6959&6998&7012&7015&7033&7045&7047\\
&&MGI&6374&6706&6768&6787&6851&6881&6921&6956&6964&6970&6985&7017&7024&7029&7038&7054\\
&$7/2^{-}$&GI&6766&6782&6874&6955&6962&7022&7038&7050&7062\\
&         &MGI&6773&6787&6885&6956&6973&7025&7029&7039&7067\\
&$9/2^{-}$&GI&6787&6967&7055\\
&         &MGI&6793&6966&7044&7098\\
\bottomrule[1pt]
\bottomrule[1pt]
\end{tabular}
\end{table*}
%


%
%
%
%
%
%
%
%

\bibliography{myreference}

\begin{thebibliography}{99}%
\makeatletter
\providecommand \@ifxundefined [1]{%
 \@ifx{#1\undefined}
}%
\providecommand \@ifnum [1]{%
 \ifnum #1\expandafter \@firstoftwo
 \else \expandafter \@secondoftwo
 \fi
}%
\providecommand \@ifx [1]{%
 \ifx #1\expandafter \@firstoftwo
 \else \expandafter \@secondoftwo
 \fi
}%
\providecommand \natexlab [1]{#1}%
\providecommand \enquote  [1]{``#1''}%
\providecommand \bibnamefont  [1]{#1}%
\providecommand \bibfnamefont [1]{#1}%
\providecommand \citenamefont [1]{#1}%
\providecommand \href@noop [0]{\@secondoftwo}%
\providecommand \href [0]{\begingroup \@sanitize@url \@href}%
\providecommand \@href[1]{\@@startlink{#1}\@@href}%
\providecommand \@@href[1]{\endgroup#1\@@endlink}%
\providecommand \@sanitize@url [0]{\catcode `\\12\catcode `\$12\catcode
  `\&12\catcode `\#12\catcode `\^12\catcode `\_12\catcode `\%12\relax}%
\providecommand \@@startlink[1]{}%
\providecommand \@@endlink[0]{}%
\providecommand \url  [0]{\begingroup\@sanitize@url \@url }%
\providecommand \@url [1]{\endgroup\@href {#1}{\urlprefix }}%
\providecommand \urlprefix  [0]{URL }%
\providecommand \Eprint [0]{\href }%
\providecommand \doibase [0]{https://doi.org/}%
\providecommand \selectlanguage [0]{\@gobble}%
\providecommand \bibinfo  [0]{\@secondoftwo}%
\providecommand \bibfield  [0]{\@secondoftwo}%
\providecommand \translation [1]{[#1]}%
\providecommand \BibitemOpen [0]{}%
\providecommand \bibitemStop [0]{}%
\providecommand \bibitemNoStop [0]{.\EOS\space}%
\providecommand \EOS [0]{\spacefactor3000\relax}%
\providecommand \BibitemShut  [1]{\csname bibitem#1\endcsname}%
\let\auto@bib@innerbib\@empty
\bibitem [{\citenamefont {Aaij}\ \emph
  {et~al.}(2017{\natexlab{a}})\citenamefont {Aaij} \emph
  {et~al.}}]{LHCb:2017uwr}%
  \BibitemOpen
  \bibfield  {author} {\bibinfo {author} {\bibfnamefont {R.}~\bibnamefont
  {Aaij}} \emph {et~al.} (\bibinfo {collaboration} {LHCb Collaboration}),\
  }\bibfield  {title} {\bibinfo {title} {{Observation of Five New Narrow
  ${\mathrm{\ensuremath{\Omega}}}_{c}^{0}$ States Decaying to
  ${\mathrm{\ensuremath{\Xi}}}_{c}^{+}{K}^{\ensuremath{-}}$}},\ }\href
  {https://doi.org/10.1103/PhysRevLett.118.182001} {\bibfield  {journal}
  {\bibinfo  {journal} {Phys. Rev. Lett.}\ }\textbf {\bibinfo {volume} {118}},\
  \bibinfo {pages} {182001} (\bibinfo {year} {2017}{\natexlab{a}})},\ \Eprint
  {https://arxiv.org/abs/1703.04639} {arXiv:1703.04639 [hep-ex]} \BibitemShut
  {NoStop}%
\bibitem [{\citenamefont {Yelton}\ \emph {et~al.}(2018)\citenamefont {Yelton}
  \emph {et~al.}}]{Belle:2017ext}%
  \BibitemOpen
  \bibfield  {author} {\bibinfo {author} {\bibfnamefont {J.}~\bibnamefont
  {Yelton}} \emph {et~al.} (\bibinfo {collaboration} {Belle Collaboration}),\
  }\bibfield  {title} {\bibinfo {title} {{Observation of excited
  ${\mathrm{\ensuremath{\Omega}}}_{c}$ charmed baryons in
  ${e}^{+}{e}^{\ensuremath{-}}$ collisions}},\ }\href
  {https://doi.org/10.1103/PhysRevD.97.051102} {\bibfield  {journal} {\bibinfo
  {journal} {Phys. Rev. D}\ }\textbf {\bibinfo {volume} {97}},\ \bibinfo
  {pages} {051102} (\bibinfo {year} {2018})},\ \Eprint
  {https://arxiv.org/abs/1711.07927} {arXiv:1711.07927 [hep-ex]} \BibitemShut
  {NoStop}%
\bibitem [{\citenamefont {Aaij}\ \emph
  {et~al.}(2017{\natexlab{b}})\citenamefont {Aaij} \emph
  {et~al.}}]{LHCb:2017jym}%
  \BibitemOpen
  \bibfield  {author} {\bibinfo {author} {\bibfnamefont {R.}~\bibnamefont
  {Aaij}} \emph {et~al.} (\bibinfo {collaboration} {LHCb Collaboration}),\
  }\bibfield  {title} {\bibinfo {title} {{Study of the $D^{0}p$ amplitude in
  $\Lambda_{b}^{0}{\rightarrow}D^{0}p\pi^{-}$ decays}},\ }\href
  {https://doi.org/10.1007/JHEP05(2017)030} {\bibfield  {journal} {\bibinfo
  {journal} {JHEP}\ }\textbf {\bibinfo {volume} {05}},\ \bibinfo {pages}
  {030}},\ \Eprint {https://arxiv.org/abs/1701.07873} {arXiv:1701.07873
  [hep-ex]} \BibitemShut {NoStop}%
\bibitem [{\citenamefont {Aaij}\ \emph
  {et~al.}(2020{\natexlab{a}})\citenamefont {Aaij} \emph
  {et~al.}}]{LHCb:2020iby}%
  \BibitemOpen
  \bibfield  {author} {\bibinfo {author} {\bibfnamefont {R.}~\bibnamefont
  {Aaij}} \emph {et~al.} (\bibinfo {collaboration} {LHCb Collaboration}),\
  }\bibfield  {title} {\bibinfo {title} {{Observation of New
  ${\mathrm{\ensuremath{\Xi}}}_{c}^{0}$ Baryons Decaying to
  ${\mathrm{\ensuremath{\Lambda}}}_{c}^{+}{K}^{\ensuremath{-}}$}},\ }\href
  {https://doi.org/10.1103/PhysRevLett.124.222001} {\bibfield  {journal}
  {\bibinfo  {journal} {Phys. Rev. Lett.}\ }\textbf {\bibinfo {volume} {124}},\
  \bibinfo {pages} {222001} (\bibinfo {year} {2020}{\natexlab{a}})},\ \Eprint
  {https://arxiv.org/abs/2003.13649} {arXiv:2003.13649 [hep-ex]} \BibitemShut
  {NoStop}%
\bibitem [{\citenamefont {Li}\ \emph {et~al.}(2023{\natexlab{a}})\citenamefont
  {Li} \emph {et~al.}}]{Belle:2022hnm}%
  \BibitemOpen
  \bibfield  {author} {\bibinfo {author} {\bibfnamefont {Y.~B.}\ \bibnamefont
  {Li}} \emph {et~al.} (\bibinfo {collaboration} {Belle Collaboration}),\
  }\bibfield  {title} {\bibinfo {title} {{Evidence of a New Excited Charmed
  Baryon Decaying to
  ${\mathrm{\ensuremath{\Sigma}}}_{c}(2455)^{0,++}{\ensuremath{\pi}}^{\pm}$}},\
  }\href {https://doi.org/10.1103/PhysRevLett.130.031901} {\bibfield  {journal}
  {\bibinfo  {journal} {Phys. Rev. Lett.}\ }\textbf {\bibinfo {volume} {130}},\
  \bibinfo {pages} {031901} (\bibinfo {year} {2023}{\natexlab{a}})},\ \Eprint
  {https://arxiv.org/abs/2206.08822} {arXiv:2206.08822 [hep-ex]} \BibitemShut
  {NoStop}%
\bibitem [{\citenamefont {Aaij}\ \emph {et~al.}(2023)\citenamefont {Aaij} \emph
  {et~al.}}]{LHCb:2023sxp}%
  \BibitemOpen
  \bibfield  {author} {\bibinfo {author} {\bibfnamefont {R.}~\bibnamefont
  {Aaij}} \emph {et~al.} (\bibinfo {collaboration} {LHCb Collaboration}),\
  }\bibfield  {title} {\bibinfo {title} {{Observation of New
  ${\mathrm{\ensuremath{\Omega}}}_{c}^{0}$ States Decaying to the
  ${\mathrm{\ensuremath{\Xi}}}_{c}^{+}{K}^{\ensuremath{-}}$ Final State}},\
  }\href {https://doi.org/10.1103/PhysRevLett.131.131902} {\bibfield  {journal}
  {\bibinfo  {journal} {Phys. Rev. Lett.}\ }\textbf {\bibinfo {volume} {131}},\
  \bibinfo {pages} {131902} (\bibinfo {year} {2023})},\ \Eprint
  {https://arxiv.org/abs/2302.04733} {arXiv:2302.04733 [hep-ex]} \BibitemShut
  {NoStop}%
\bibitem [{\citenamefont {Aaij}\ \emph
  {et~al.}(2019{\natexlab{a}})\citenamefont {Aaij} \emph
  {et~al.}}]{LHCb:2018haf}%
  \BibitemOpen
  \bibfield  {author} {\bibinfo {author} {\bibfnamefont {R.}~\bibnamefont
  {Aaij}} \emph {et~al.} (\bibinfo {collaboration} {LHCb Collaboration}),\
  }\bibfield  {title} {\bibinfo {title} {{Observation of Two Resonances in the
  ${\Lambda}_{b}^{0}{\pi}^{\pm}$ Systems and Precise Measurement of
  ${\Sigma}_{b}^{\pm}$ and ${\Sigma}_{b}^{*\pm}$ Properties}},\ }\href
  {https://doi.org/10.1103/PhysRevLett.122.012001} {\bibfield  {journal}
  {\bibinfo  {journal} {Phys. Rev. Lett.}\ }\textbf {\bibinfo {volume} {122}},\
  \bibinfo {pages} {012001} (\bibinfo {year} {2019}{\natexlab{a}})},\ \Eprint
  {https://arxiv.org/abs/1809.07752} {arXiv:1809.07752 [hep-ex]} \BibitemShut
  {NoStop}%
\bibitem [{\citenamefont {Aaij}\ \emph {et~al.}(2018)\citenamefont {Aaij} \emph
  {et~al.}}]{LHCb:2018vuc}%
  \BibitemOpen
  \bibfield  {author} {\bibinfo {author} {\bibfnamefont {R.}~\bibnamefont
  {Aaij}} \emph {et~al.} (\bibinfo {collaboration} {LHCb Collaboration}),\
  }\bibfield  {title} {\bibinfo {title} {{Observation of a New ${\Xi}_{b}^{-}$
  Resonance}},\ }\href {https://doi.org/10.1103/PhysRevLett.121.072002}
  {\bibfield  {journal} {\bibinfo  {journal} {Phys. Rev. Lett.}\ }\textbf
  {\bibinfo {volume} {121}},\ \bibinfo {pages} {072002} (\bibinfo {year}
  {2018})},\ \Eprint {https://arxiv.org/abs/1805.09418} {arXiv:1805.09418
  [hep-ex]} \BibitemShut {NoStop}%
\bibitem [{\citenamefont {Aaij}\ \emph {et~al.}(2021)\citenamefont {Aaij} \emph
  {et~al.}}]{LHCb:2020xpu}%
  \BibitemOpen
  \bibfield  {author} {\bibinfo {author} {\bibfnamefont {R.}~\bibnamefont
  {Aaij}} \emph {et~al.} (\bibinfo {collaboration} {LHCb Collaboration}),\
  }\bibfield  {title} {\bibinfo {title} {{Observation of a new
  ${{\ensuremath{\Xi}}}_{b}^{0}$ state}},\ }\href
  {https://doi.org/10.1103/PhysRevD.103.012004} {\bibfield  {journal} {\bibinfo
   {journal} {Phys. Rev. D}\ }\textbf {\bibinfo {volume} {103}},\ \bibinfo
  {pages} {012004} (\bibinfo {year} {2021})},\ \Eprint
  {https://arxiv.org/abs/2010.14485} {arXiv:2010.14485 [hep-ex]} \BibitemShut
  {NoStop}%
\bibitem [{\citenamefont {Aaij}\ \emph
  {et~al.}(2019{\natexlab{b}})\citenamefont {Aaij} \emph
  {et~al.}}]{LHCb:2019soc}%
  \BibitemOpen
  \bibfield  {author} {\bibinfo {author} {\bibfnamefont {R.}~\bibnamefont
  {Aaij}} \emph {et~al.} (\bibinfo {collaboration} {LHCb Collaboration}),\
  }\bibfield  {title} {\bibinfo {title} {{Observation of New Resonances in the
  ${\mathrm{\ensuremath{\Lambda}}}_{b}^{0}{\ensuremath{\pi}}^{+}{\ensuremath{\pi}}^{\ensuremath{-}}$
  System}},\ }\href {https://doi.org/10.1103/PhysRevLett.123.152001} {\bibfield
   {journal} {\bibinfo  {journal} {Phys. Rev. Lett.}\ }\textbf {\bibinfo
  {volume} {123}},\ \bibinfo {pages} {152001} (\bibinfo {year}
  {2019}{\natexlab{b}})},\ \Eprint {https://arxiv.org/abs/1907.13598}
  {arXiv:1907.13598 [hep-ex]} \BibitemShut {NoStop}%
\bibitem [{\citenamefont {Sirunyan}\ \emph {et~al.}(2020)\citenamefont
  {Sirunyan} \emph {et~al.}}]{CMS:2020zzv}%
  \BibitemOpen
  \bibfield  {author} {\bibinfo {author} {\bibfnamefont {A.~M.}\ \bibnamefont
  {Sirunyan}} \emph {et~al.} (\bibinfo {collaboration} {CMS Collaboration}),\
  }\bibfield  {title} {\bibinfo {title} {{Study of excited $\Lambda_{b}^{0}$
  states decaying to $\Lambda_{b}^{0}\pi^{+}\pi^{-}$ in proton-proton
  collisions at $\sqrt{s}=$ 13 TeV}},\ }\href
  {https://doi.org/10.1016/j.physletb.2020.135345} {\bibfield  {journal}
  {\bibinfo  {journal} {Phys. Lett. B}\ }\textbf {\bibinfo {volume} {803}},\
  \bibinfo {pages} {135345} (\bibinfo {year} {2020})},\ \Eprint
  {https://arxiv.org/abs/2001.06533} {arXiv:2001.06533 [hep-ex]} \BibitemShut
  {NoStop}%
\bibitem [{\citenamefont {Aaij}\ \emph
  {et~al.}(2020{\natexlab{b}})\citenamefont {Aaij} \emph
  {et~al.}}]{LHCb:2020tqd}%
  \BibitemOpen
  \bibfield  {author} {\bibinfo {author} {\bibfnamefont {R.}~\bibnamefont
  {Aaij}} \emph {et~al.} (\bibinfo {collaboration} {LHCb Collaboration}),\
  }\bibfield  {title} {\bibinfo {title} {{First Observation of Excited
  ${\mathrm{\ensuremath{\Omega}}}_{b}^{\ensuremath{-}}$ States}},\ }\href
  {https://doi.org/10.1103/PhysRevLett.124.082002} {\bibfield  {journal}
  {\bibinfo  {journal} {Phys. Rev. Lett.}\ }\textbf {\bibinfo {volume} {124}},\
  \bibinfo {pages} {082002} (\bibinfo {year} {2020}{\natexlab{b}})},\ \Eprint
  {https://arxiv.org/abs/2001.00851} {arXiv:2001.00851 [hep-ex]} \BibitemShut
  {NoStop}%
\bibitem [{\citenamefont {Sirunyan}\ \emph {et~al.}(2021)\citenamefont
  {Sirunyan} \emph {et~al.}}]{CMS:2021rvl}%
  \BibitemOpen
  \bibfield  {author} {\bibinfo {author} {\bibfnamefont {A.~M.}\ \bibnamefont
  {Sirunyan}} \emph {et~al.} (\bibinfo {collaboration} {CMS Collaboration}),\
  }\bibfield  {title} {\bibinfo {title} {{Observation of a New Excited Beauty
  Strange Baryon Decaying to
  ${{\ensuremath{\Xi}}}_{b}^{\ensuremath{-}}{\ensuremath{\pi}}^{+}{\ensuremath{\pi}}^{\ensuremath{-}}$}},\
  }\href {https://doi.org/10.1103/PhysRevLett.126.252003} {\bibfield  {journal}
  {\bibinfo  {journal} {Phys. Rev. Lett.}\ }\textbf {\bibinfo {volume} {126}},\
  \bibinfo {pages} {252003} (\bibinfo {year} {2021})},\ \Eprint
  {https://arxiv.org/abs/2102.04524} {arXiv:2102.04524 [hep-ex]} \BibitemShut
  {NoStop}%
\bibitem [{\citenamefont {Aaij}\ \emph {et~al.}(2022)\citenamefont {Aaij} \emph
  {et~al.}}]{LHCb:2021ssn}%
  \BibitemOpen
  \bibfield  {author} {\bibinfo {author} {\bibfnamefont {R.}~\bibnamefont
  {Aaij}} \emph {et~al.} (\bibinfo {collaboration} {LHCb Collaboration}),\
  }\bibfield  {title} {\bibinfo {title} {{Observation of Two New Excited
  ${\mathrm{\ensuremath{\Xi}}}_{b}^{0}$ States Decaying to
  ${\mathrm{\ensuremath{\Lambda}}}_{b}^{0}{K}^{\ensuremath{-}}{\ensuremath{\pi}}^{+}$}},\
  }\href {https://doi.org/10.1103/PhysRevLett.128.162001} {\bibfield  {journal}
  {\bibinfo  {journal} {Phys. Rev. Lett.}\ }\textbf {\bibinfo {volume} {128}},\
  \bibinfo {pages} {162001} (\bibinfo {year} {2022})},\ \Eprint
  {https://arxiv.org/abs/2110.04497} {arXiv:2110.04497 [hep-ex]} \BibitemShut
  {NoStop}%
\bibitem [{\citenamefont {Workman}\ \emph {et~al.}(2022)\citenamefont {Workman}
  \emph {et~al.}}]{ParticleDataGroup:2022pth}%
  \BibitemOpen
  \bibfield  {author} {\bibinfo {author} {\bibfnamefont {R.~L.}\ \bibnamefont
  {Workman}} \emph {et~al.} (\bibinfo {collaboration} {Particle Data Group}),\
  }\bibfield  {title} {\bibinfo {title} {{Review of Particle Physics}},\ }\href
  {https://doi.org/10.1093/ptep/ptac097} {\bibfield  {journal} {\bibinfo
  {journal} {PTEP}\ }\textbf {\bibinfo {volume} {2022}},\ \bibinfo {pages}
  {083C01} (\bibinfo {year} {2022})}\BibitemShut {NoStop}%
\bibitem [{\citenamefont {Isgur}\ and\ \citenamefont
  {Karl}(1977)}]{Isgur:1977ef}%
  \BibitemOpen
  \bibfield  {author} {\bibinfo {author} {\bibfnamefont {N.}~\bibnamefont
  {Isgur}}\ and\ \bibinfo {author} {\bibfnamefont {G.}~\bibnamefont {Karl}},\
  }\bibfield  {title} {\bibinfo {title} {{Hyperfine interactions in negative
  parity baryons}},\ }\href {https://doi.org/10.1016/0370-2693(77)90074-0}
  {\bibfield  {journal} {\bibinfo  {journal} {Phys. Lett.}\ }\textbf {\bibinfo
  {volume} {72B}},\ \bibinfo {pages} {109} (\bibinfo {year}
  {1977})}\BibitemShut {NoStop}%
\bibitem [{\citenamefont {Isgur}\ and\ \citenamefont
  {Karl}(1979)}]{Isgur:1978wd}%
  \BibitemOpen
  \bibfield  {author} {\bibinfo {author} {\bibfnamefont {N.}~\bibnamefont
  {Isgur}}\ and\ \bibinfo {author} {\bibfnamefont {G.}~\bibnamefont {Karl}},\
  }\bibfield  {title} {\bibinfo {title} {{Positive-parity excited baryons in a
  quark model with hyperfine interactions}},\ }\href
  {https://doi.org/10.1103/PhysRevD.19.2653} {\bibfield  {journal} {\bibinfo
  {journal} {Phys. Rev.}\ }\textbf {\bibinfo {volume} {D19}},\ \bibinfo {pages}
  {2653} (\bibinfo {year} {1979})},\ \bibinfo {note} {[Erratum: Phys.
  Rev.D23,817(1981)]}\BibitemShut {NoStop}%
\bibitem [{\citenamefont {Martin}\ and\ \citenamefont
  {Richard}(1995)}]{Martin:1995vk}%
  \BibitemOpen
  \bibfield  {author} {\bibinfo {author} {\bibfnamefont {A.}~\bibnamefont
  {Martin}}\ and\ \bibinfo {author} {\bibfnamefont {J.-M.}\ \bibnamefont
  {Richard}},\ }\bibfield  {title} {\bibinfo {title} {{$\Omega_c$ and other
  charmed baryons revisited}},\ }\href
  {https://doi.org/10.1016/0370-2693(95)00704-O} {\bibfield  {journal}
  {\bibinfo  {journal} {Phys. Lett.}\ }\textbf {\bibinfo {volume} {B355}},\
  \bibinfo {pages} {345} (\bibinfo {year} {1995})},\ \Eprint
  {https://arxiv.org/abs/hep-ph/9504276} {arXiv:hep-ph/9504276 [hep-ph]}
  \BibitemShut {NoStop}%
\bibitem [{\citenamefont {Ebert}\ \emph {et~al.}(2005)\citenamefont {Ebert},
  \citenamefont {Faustov},\ and\ \citenamefont {Galkin}}]{Ebert:2005xj}%
  \BibitemOpen
  \bibfield  {author} {\bibinfo {author} {\bibfnamefont {D.}~\bibnamefont
  {Ebert}}, \bibinfo {author} {\bibfnamefont {R.~N.}\ \bibnamefont {Faustov}},\
  and\ \bibinfo {author} {\bibfnamefont {V.~O.}\ \bibnamefont {Galkin}},\
  }\bibfield  {title} {\bibinfo {title} {{Masses of heavy baryons in the
  relativistic quark model}},\ }\href
  {https://doi.org/10.1103/PhysRevD.72.034026} {\bibfield  {journal} {\bibinfo
  {journal} {Phys. Rev.}\ }\textbf {\bibinfo {volume} {D72}},\ \bibinfo {pages}
  {034026} (\bibinfo {year} {2005})},\ \Eprint
  {https://arxiv.org/abs/hep-ph/0504112} {arXiv:hep-ph/0504112 [hep-ph]}
  \BibitemShut {NoStop}%
\bibitem [{\citenamefont {Ebert}\ \emph {et~al.}(2008)\citenamefont {Ebert},
  \citenamefont {Faustov},\ and\ \citenamefont {Galkin}}]{Ebert:2007nw}%
  \BibitemOpen
  \bibfield  {author} {\bibinfo {author} {\bibfnamefont {D.}~\bibnamefont
  {Ebert}}, \bibinfo {author} {\bibfnamefont {R.~N.}\ \bibnamefont {Faustov}},\
  and\ \bibinfo {author} {\bibfnamefont {V.~O.}\ \bibnamefont {Galkin}},\
  }\bibfield  {title} {\bibinfo {title} {{Masses of excited heavy baryons in
  the relativistic quark-diquark picture}},\ }\href
  {https://doi.org/10.1016/j.physletb.2007.11.037} {\bibfield  {journal}
  {\bibinfo  {journal} {Phys. Lett.}\ }\textbf {\bibinfo {volume} {B659}},\
  \bibinfo {pages} {612} (\bibinfo {year} {2008})},\ \Eprint
  {https://arxiv.org/abs/0705.2957} {arXiv:0705.2957 [hep-ph]} \BibitemShut
  {NoStop}%
\bibitem [{\citenamefont {Roberts}\ and\ \citenamefont
  {Pervin}(2008)}]{Roberts:2007ni}%
  \BibitemOpen
  \bibfield  {author} {\bibinfo {author} {\bibfnamefont {W.}~\bibnamefont
  {Roberts}}\ and\ \bibinfo {author} {\bibfnamefont {M.}~\bibnamefont
  {Pervin}},\ }\bibfield  {title} {\bibinfo {title} {{Heavy baryons in a quark
  model}},\ }\href {https://doi.org/10.1142/S0217751X08041219} {\bibfield
  {journal} {\bibinfo  {journal} {Int. J. Mod. Phys.}\ }\textbf {\bibinfo
  {volume} {A23}},\ \bibinfo {pages} {2817} (\bibinfo {year} {2008})},\ \Eprint
  {https://arxiv.org/abs/0711.2492} {arXiv:0711.2492 [nucl-th]} \BibitemShut
  {NoStop}%
\bibitem [{\citenamefont {Karliner}\ \emph {et~al.}(2009)\citenamefont
  {Karliner}, \citenamefont {Keren-Zur}, \citenamefont {Lipkin},\ and\
  \citenamefont {Rosner}}]{Karliner:2008sv}%
  \BibitemOpen
  \bibfield  {author} {\bibinfo {author} {\bibfnamefont {M.}~\bibnamefont
  {Karliner}}, \bibinfo {author} {\bibfnamefont {B.}~\bibnamefont {Keren-Zur}},
  \bibinfo {author} {\bibfnamefont {H.~J.}\ \bibnamefont {Lipkin}},\ and\
  \bibinfo {author} {\bibfnamefont {J.~L.}\ \bibnamefont {Rosner}},\ }\bibfield
   {title} {\bibinfo {title} {The quark model and $b$ baryons},\ }\href
  {https://doi.org/10.1016/j.aop.2008.05.003} {\bibfield  {journal} {\bibinfo
  {journal} {Annals Phys.}\ }\textbf {\bibinfo {volume} {324}},\ \bibinfo
  {pages} {2} (\bibinfo {year} {2009})},\ \Eprint
  {https://arxiv.org/abs/0804.1575} {arXiv:0804.1575 [hep-ph]} \BibitemShut
  {NoStop}%
\bibitem [{\citenamefont {Valcarce}\ \emph {et~al.}(2008)\citenamefont
  {Valcarce}, \citenamefont {Garcilazo},\ and\ \citenamefont
  {Vijande}}]{Valcarce:2008dr}%
  \BibitemOpen
  \bibfield  {author} {\bibinfo {author} {\bibfnamefont {A.}~\bibnamefont
  {Valcarce}}, \bibinfo {author} {\bibfnamefont {H.}~\bibnamefont
  {Garcilazo}},\ and\ \bibinfo {author} {\bibfnamefont {J.}~\bibnamefont
  {Vijande}},\ }\bibfield  {title} {\bibinfo {title} {{Towards an understanding
  of heavy baryon spectroscopy}},\ }\href
  {https://doi.org/10.1140/epja/i2008-10616-4} {\bibfield  {journal} {\bibinfo
  {journal} {Eur. Phys. J.}\ }\textbf {\bibinfo {volume} {A37}},\ \bibinfo
  {pages} {217} (\bibinfo {year} {2008})},\ \Eprint
  {https://arxiv.org/abs/0807.2973} {arXiv:0807.2973 [hep-ph]} \BibitemShut
  {NoStop}%
\bibitem [{\citenamefont {Yang}\ \emph {et~al.}(2008)\citenamefont {Yang},
  \citenamefont {Deng}, \citenamefont {Huang},\ and\ \citenamefont
  {Ping}}]{Yang:2008zzi}%
  \BibitemOpen
  \bibfield  {author} {\bibinfo {author} {\bibfnamefont {Y.}~\bibnamefont
  {Yang}}, \bibinfo {author} {\bibfnamefont {C.}~\bibnamefont {Deng}}, \bibinfo
  {author} {\bibfnamefont {H.}~\bibnamefont {Huang}},\ and\ \bibinfo {author}
  {\bibfnamefont {J.}~\bibnamefont {Ping}},\ }\bibfield  {title} {\bibinfo
  {title} {{Dynamical study of heavy-baryon spectroscopy}},\ }\href
  {https://doi.org/10.1142/S0217732308027102} {\bibfield  {journal} {\bibinfo
  {journal} {Mod. Phys. Lett.}\ }\textbf {\bibinfo {volume} {A23}},\ \bibinfo
  {pages} {1819} (\bibinfo {year} {2008})}\BibitemShut {NoStop}%
\bibitem [{\citenamefont {Vijande}\ \emph {et~al.}(2014)\citenamefont
  {Vijande}, \citenamefont {Valcarce},\ and\ \citenamefont
  {Garcilazo}}]{Vijande:2014uma}%
  \BibitemOpen
  \bibfield  {author} {\bibinfo {author} {\bibfnamefont {J.}~\bibnamefont
  {Vijande}}, \bibinfo {author} {\bibfnamefont {A.}~\bibnamefont {Valcarce}},\
  and\ \bibinfo {author} {\bibfnamefont {H.}~\bibnamefont {Garcilazo}},\
  }\bibfield  {title} {\bibinfo {title} {{Heavy-baryon quark model picture from
  lattice QCD}},\ }\href {https://doi.org/10.1103/PhysRevD.90.094004}
  {\bibfield  {journal} {\bibinfo  {journal} {Phys. Rev.}\ }\textbf {\bibinfo
  {volume} {D90}},\ \bibinfo {pages} {094004} (\bibinfo {year} {2014})},\
  \Eprint {https://arxiv.org/abs/1507.03736} {arXiv:1507.03736 [hep-ph]}
  \BibitemShut {NoStop}%
\bibitem [{\citenamefont {Karliner}\ and\ \citenamefont
  {Rosner}(2015)}]{Karliner:2015ema}%
  \BibitemOpen
  \bibfield  {author} {\bibinfo {author} {\bibfnamefont {M.}~\bibnamefont
  {Karliner}}\ and\ \bibinfo {author} {\bibfnamefont {J.~L.}\ \bibnamefont
  {Rosner}},\ }\bibfield  {title} {\bibinfo {title} {{Prospects for observing
  the lowest-lying odd-parity ${\Sigma}_{c}$ and ${\Sigma}_{b}$ baryons}},\
  }\href {https://doi.org/10.1103/PhysRevD.92.074026} {\bibfield  {journal}
  {\bibinfo  {journal} {Phys. Rev.}\ }\textbf {\bibinfo {volume} {D92}},\
  \bibinfo {pages} {074026} (\bibinfo {year} {2015})},\ \Eprint
  {https://arxiv.org/abs/1506.01702} {arXiv:1506.01702 [hep-ph]} \BibitemShut
  {NoStop}%
\bibitem [{\citenamefont {Karliner}\ and\ \citenamefont
  {Rosner}(2017)}]{Karliner:2017kfm}%
  \BibitemOpen
  \bibfield  {author} {\bibinfo {author} {\bibfnamefont {M.}~\bibnamefont
  {Karliner}}\ and\ \bibinfo {author} {\bibfnamefont {J.~L.}\ \bibnamefont
  {Rosner}},\ }\bibfield  {title} {\bibinfo {title} {Very narrow excited
  $\omega_c$ baryons},\ }\href {https://doi.org/10.1103/PhysRevD.95.114012}
  {\bibfield  {journal} {\bibinfo  {journal} {Phys. Rev.}\ }\textbf {\bibinfo
  {volume} {D95}},\ \bibinfo {pages} {114012} (\bibinfo {year} {2017})},\
  \Eprint {https://arxiv.org/abs/1703.07774} {arXiv:1703.07774 [hep-ph]}
  \BibitemShut {NoStop}%
\bibitem [{\citenamefont {Wang}\ \emph {et~al.}(2017)\citenamefont {Wang},
  \citenamefont {Yao}, \citenamefont {Zhong},\ and\ \citenamefont
  {Zhao}}]{Wang:2017kfr}%
  \BibitemOpen
  \bibfield  {author} {\bibinfo {author} {\bibfnamefont {K.-L.}\ \bibnamefont
  {Wang}}, \bibinfo {author} {\bibfnamefont {Y.-X.}\ \bibnamefont {Yao}},
  \bibinfo {author} {\bibfnamefont {X.-H.}\ \bibnamefont {Zhong}},\ and\
  \bibinfo {author} {\bibfnamefont {Q.}~\bibnamefont {Zhao}},\ }\bibfield
  {title} {\bibinfo {title} {{Strong and radiative decays of the low-lying $S$-
  and $P$-wave singly heavy baryons}},\ }\href
  {https://doi.org/10.1103/PhysRevD.96.116016} {\bibfield  {journal} {\bibinfo
  {journal} {Phys. Rev.}\ }\textbf {\bibinfo {volume} {D96}},\ \bibinfo {pages}
  {116016} (\bibinfo {year} {2017})},\ \Eprint
  {https://arxiv.org/abs/1709.04268} {arXiv:1709.04268 [hep-ph]} \BibitemShut
  {NoStop}%
\bibitem [{\citenamefont {Yang}\ \emph {et~al.}(2018)\citenamefont {Yang},
  \citenamefont {Ping},\ and\ \citenamefont {Segovia}}]{Yang:2017qan}%
  \BibitemOpen
  \bibfield  {author} {\bibinfo {author} {\bibfnamefont {G.}~\bibnamefont
  {Yang}}, \bibinfo {author} {\bibfnamefont {J.}~\bibnamefont {Ping}},\ and\
  \bibinfo {author} {\bibfnamefont {J.}~\bibnamefont {Segovia}},\ }\bibfield
  {title} {\bibinfo {title} {{The S- and P-Wave Low-Lying Baryons in the Chiral
  Quark Model}},\ }\href {https://doi.org/10.1007/s00601-018-1433-4} {\bibfield
   {journal} {\bibinfo  {journal} {Few Body Syst.}\ }\textbf {\bibinfo {volume}
  {59}},\ \bibinfo {pages} {113} (\bibinfo {year} {2018})},\ \Eprint
  {https://arxiv.org/abs/1709.09315} {arXiv:1709.09315 [hep-ph]} \BibitemShut
  {NoStop}%
\bibitem [{\citenamefont {Karliner}\ and\ \citenamefont
  {Rosner}(2018)}]{Karliner:2018bms}%
  \BibitemOpen
  \bibfield  {author} {\bibinfo {author} {\bibfnamefont {M.}~\bibnamefont
  {Karliner}}\ and\ \bibinfo {author} {\bibfnamefont {J.~L.}\ \bibnamefont
  {Rosner}},\ }\bibfield  {title} {\bibinfo {title} {{Scaling of P-wave
  excitation energies in heavy-quark systems}},\ }\href
  {https://doi.org/10.1103/PhysRevD.98.074026} {\bibfield  {journal} {\bibinfo
  {journal} {Phys. Rev.}\ }\textbf {\bibinfo {volume} {D98}},\ \bibinfo {pages}
  {074026} (\bibinfo {year} {2018})},\ \Eprint
  {https://arxiv.org/abs/1808.07869} {arXiv:1808.07869 [hep-ph]} \BibitemShut
  {NoStop}%
\bibitem [{\citenamefont {Shi}\ \emph {et~al.}(2020)\citenamefont {Shi},
  \citenamefont {Zhao},\ and\ \citenamefont {Zhuang}}]{Shi:2019tji}%
  \BibitemOpen
  \bibfield  {author} {\bibinfo {author} {\bibfnamefont {S.}~\bibnamefont
  {Shi}}, \bibinfo {author} {\bibfnamefont {J.}~\bibnamefont {Zhao}},\ and\
  \bibinfo {author} {\bibfnamefont {P.}~\bibnamefont {Zhuang}},\ }\bibfield
  {title} {\bibinfo {title} {{Heavy flavor dissociation in framework of
  multi-body Dirac equations}},\ }\href
  {https://doi.org/10.1088/1674-1137/44/8/084101} {\bibfield  {journal}
  {\bibinfo  {journal} {Chin. Phys.}\ }\textbf {\bibinfo {volume} {C44}},\
  \bibinfo {pages} {8} (\bibinfo {year} {2020})},\ \Eprint
  {https://arxiv.org/abs/1905.10627} {arXiv:1905.10627 [nucl-th]} \BibitemShut
  {NoStop}%
\bibitem [{\citenamefont {Karliner}\ and\ \citenamefont
  {Rosner}(2020)}]{Karliner:2020fqe}%
  \BibitemOpen
  \bibfield  {author} {\bibinfo {author} {\bibfnamefont {M.}~\bibnamefont
  {Karliner}}\ and\ \bibinfo {author} {\bibfnamefont {J.~L.}\ \bibnamefont
  {Rosner}},\ }\bibfield  {title} {\bibinfo {title} {{Interpretation of excited
  ${\Omega}_{b}$ signals}},\ }\href
  {https://doi.org/10.1103/PhysRevD.102.014027} {\bibfield  {journal} {\bibinfo
   {journal} {Phys. Rev.}\ }\textbf {\bibinfo {volume} {D102}},\ \bibinfo
  {pages} {014027} (\bibinfo {year} {2020})},\ \Eprint
  {https://arxiv.org/abs/2005.12424} {arXiv:2005.12424 [hep-ph]} \BibitemShut
  {NoStop}%
\bibitem [{\citenamefont {Wang}\ \emph {et~al.}(2020)\citenamefont {Wang},
  \citenamefont {Xiao},\ and\ \citenamefont {Zhong}}]{Wang:2020gkn}%
  \BibitemOpen
  \bibfield  {author} {\bibinfo {author} {\bibfnamefont {K.-L.}\ \bibnamefont
  {Wang}}, \bibinfo {author} {\bibfnamefont {L.-Y.}\ \bibnamefont {Xiao}},\
  and\ \bibinfo {author} {\bibfnamefont {X.-H.}\ \bibnamefont {Zhong}},\
  }\bibfield  {title} {\bibinfo {title} {{Understanding the newly observed
  ${\Xi}_{c}^{0}$ states through their decays}},\ }\href
  {https://doi.org/10.1103/PhysRevD.102.034029} {\bibfield  {journal} {\bibinfo
   {journal} {Phys. Rev.}\ }\textbf {\bibinfo {volume} {D102}},\ \bibinfo
  {pages} {034029} (\bibinfo {year} {2020})},\ \Eprint
  {https://arxiv.org/abs/2004.03221} {arXiv:2004.03221 [hep-ph]} \BibitemShut
  {NoStop}%
\bibitem [{\citenamefont {Xiao}\ and\ \citenamefont
  {Zhong}(2020)}]{Xiao:2020gjo}%
  \BibitemOpen
  \bibfield  {author} {\bibinfo {author} {\bibfnamefont {L.-Y.}\ \bibnamefont
  {Xiao}}\ and\ \bibinfo {author} {\bibfnamefont {X.-H.}\ \bibnamefont
  {Zhong}},\ }\bibfield  {title} {\bibinfo {title} {{Toward establishing the
  low-lying $P$-wave ${\Sigma}_{b}$ states}},\ }\href
  {https://doi.org/10.1103/PhysRevD.102.014009} {\bibfield  {journal} {\bibinfo
   {journal} {Phys. Rev.}\ }\textbf {\bibinfo {volume} {D102}},\ \bibinfo
  {pages} {014009} (\bibinfo {year} {2020})},\ \Eprint
  {https://arxiv.org/abs/2004.11106} {arXiv:2004.11106 [hep-ph]} \BibitemShut
  {NoStop}%
\bibitem [{\citenamefont {Chen}\ \emph {et~al.}(2021)\citenamefont {Chen},
  \citenamefont {Luo},\ and\ \citenamefont {Liu}}]{Chen:2021eyk}%
  \BibitemOpen
  \bibfield  {author} {\bibinfo {author} {\bibfnamefont {B.}~\bibnamefont
  {Chen}}, \bibinfo {author} {\bibfnamefont {S.-Q.}\ \bibnamefont {Luo}},\ and\
  \bibinfo {author} {\bibfnamefont {X.}~\bibnamefont {Liu}},\ }\bibfield
  {title} {\bibinfo {title} {{Universal behavior of mass gaps existing in the
  single heavy baryon family}},\ }\href
  {https://doi.org/10.1140/epjc/s10052-021-09234-1} {\bibfield  {journal}
  {\bibinfo  {journal} {Eur. Phys. J. C}\ }\textbf {\bibinfo {volume} {81}},\
  \bibinfo {pages} {474} (\bibinfo {year} {2021})},\ \Eprint
  {https://arxiv.org/abs/2101.10806} {arXiv:2101.10806 [hep-ph]} \BibitemShut
  {NoStop}%
\bibitem [{\citenamefont {Wang}\ and\ \citenamefont
  {Zhong}(2022)}]{Wang:2021bmz}%
  \BibitemOpen
  \bibfield  {author} {\bibinfo {author} {\bibfnamefont {K.-L.}\ \bibnamefont
  {Wang}}\ and\ \bibinfo {author} {\bibfnamefont {X.-H.}\ \bibnamefont
  {Zhong}},\ }\bibfield  {title} {\bibinfo {title} {{Toward discovering
  low-lying $P$-wave excited $\ensuremath{\Sigma}_{c}$ baryon states}},\ }\href
  {https://doi.org/10.1088/1674-1137/ac3123} {\bibfield  {journal} {\bibinfo
  {journal} {Chin. Phys. C}\ }\textbf {\bibinfo {volume} {46}},\ \bibinfo
  {pages} {023103} (\bibinfo {year} {2022})},\ \Eprint
  {https://arxiv.org/abs/2110.12443} {arXiv:2110.12443 [hep-ph]} \BibitemShut
  {NoStop}%
\bibitem [{\citenamefont {Garc\'{\i}a-Tecocoatzi}\ \emph
  {et~al.}(2023)\citenamefont {Garc\'{\i}a-Tecocoatzi}, \citenamefont
  {Giachino}, \citenamefont {Li}, \citenamefont {Ramirez-Morales},\ and\
  \citenamefont {Santopinto}}]{Garcia-Tecocoatzi:2022zrf}%
  \BibitemOpen
  \bibfield  {author} {\bibinfo {author} {\bibfnamefont {H.}~\bibnamefont
  {Garc\'{\i}a-Tecocoatzi}}, \bibinfo {author} {\bibfnamefont {A.}~\bibnamefont
  {Giachino}}, \bibinfo {author} {\bibfnamefont {J.}~\bibnamefont {Li}},
  \bibinfo {author} {\bibfnamefont {A.}~\bibnamefont {Ramirez-Morales}},\ and\
  \bibinfo {author} {\bibfnamefont {E.}~\bibnamefont {Santopinto}},\ }\bibfield
   {title} {\bibinfo {title} {{Strong decay widths and mass spectra of charmed
  baryons}},\ }\href {https://doi.org/10.1103/PhysRevD.107.034031} {\bibfield
  {journal} {\bibinfo  {journal} {Phys. Rev. D}\ }\textbf {\bibinfo {volume}
  {107}},\ \bibinfo {pages} {034031} (\bibinfo {year} {2023})},\ \Eprint
  {https://arxiv.org/abs/2205.07049} {arXiv:2205.07049 [hep-ph]} \BibitemShut
  {NoStop}%
\bibitem [{\citenamefont {Ma}\ \emph {et~al.}(2023)\citenamefont {Ma},
  \citenamefont {Meng}, \citenamefont {Chen},\ and\ \citenamefont
  {Zhu}}]{Ma:2022vqf}%
  \BibitemOpen
  \bibfield  {author} {\bibinfo {author} {\bibfnamefont {Y.}~\bibnamefont
  {Ma}}, \bibinfo {author} {\bibfnamefont {L.}~\bibnamefont {Meng}}, \bibinfo
  {author} {\bibfnamefont {Y.-K.}\ \bibnamefont {Chen}},\ and\ \bibinfo
  {author} {\bibfnamefont {S.-L.}\ \bibnamefont {Zhu}},\ }\bibfield  {title}
  {\bibinfo {title} {{Ground state baryons in the flux-tube three-body
  confinement model using diffusion Monte Carlo}},\ }\href
  {https://doi.org/10.1103/PhysRevD.107.054035} {\bibfield  {journal} {\bibinfo
   {journal} {Phys. Rev. D}\ }\textbf {\bibinfo {volume} {107}},\ \bibinfo
  {pages} {054035} (\bibinfo {year} {2023})},\ \Eprint
  {https://arxiv.org/abs/2211.09021} {arXiv:2211.09021 [hep-ph]} \BibitemShut
  {NoStop}%
\bibitem [{\citenamefont {Wang}\ \emph {et~al.}(2022)\citenamefont {Wang},
  \citenamefont {Xiao},\ and\ \citenamefont {Zhong}}]{Wang:2022dmw}%
  \BibitemOpen
  \bibfield  {author} {\bibinfo {author} {\bibfnamefont {W.-J.}\ \bibnamefont
  {Wang}}, \bibinfo {author} {\bibfnamefont {L.-Y.}\ \bibnamefont {Xiao}},\
  and\ \bibinfo {author} {\bibfnamefont {X.-H.}\ \bibnamefont {Zhong}},\
  }\bibfield  {title} {\bibinfo {title} {{Strong decays of the low-lying
  $\ensuremath{\rho}$-mode $1P$-wave singly heavy baryons}},\ }\href
  {https://doi.org/10.1103/PhysRevD.106.074020} {\bibfield  {journal} {\bibinfo
   {journal} {Phys. Rev. D}\ }\textbf {\bibinfo {volume} {106}},\ \bibinfo
  {pages} {074020} (\bibinfo {year} {2022})},\ \Eprint
  {https://arxiv.org/abs/2208.10116} {arXiv:2208.10116 [hep-ph]} \BibitemShut
  {NoStop}%
\bibitem [{\citenamefont {Karliner}\ and\ \citenamefont
  {Rosner}(2023)}]{Karliner:2023okv}%
  \BibitemOpen
  \bibfield  {author} {\bibinfo {author} {\bibfnamefont {M.}~\bibnamefont
  {Karliner}}\ and\ \bibinfo {author} {\bibfnamefont {J.~L.}\ \bibnamefont
  {Rosner}},\ }\bibfield  {title} {\bibinfo {title} {{Excited
  ${\mathrm{\ensuremath{\Omega}}}_{c}$ Baryons as $2S$ states}},\ }\href
  {https://doi.org/10.1103/PhysRevD.108.014006} {\bibfield  {journal} {\bibinfo
   {journal} {Phys. Rev. D}\ }\textbf {\bibinfo {volume} {108}},\ \bibinfo
  {pages} {014006} (\bibinfo {year} {2023})},\ \Eprint
  {https://arxiv.org/abs/2304.00407} {arXiv:2304.00407 [hep-ph]} \BibitemShut
  {NoStop}%
\bibitem [{\citenamefont {Ortiz-Pacheco}\ and\ \citenamefont
  {Bijker}(2023)}]{Ortiz-Pacheco:2023bns}%
  \BibitemOpen
  \bibfield  {author} {\bibinfo {author} {\bibfnamefont {E.}~\bibnamefont
  {Ortiz-Pacheco}}\ and\ \bibinfo {author} {\bibfnamefont {R.}~\bibnamefont
  {Bijker}},\ }\bibfield  {title} {\bibinfo {title} {{Heavy $\Xi_{c/b}$ and
  $\Xi_{c/b}'$ baryons in the quark model}},\ }\href
  {https://doi.org/10.1088/1742-6596/2619/1/012011} {\bibfield  {journal}
  {\bibinfo  {journal} {J. Phys. Conf. Ser.}\ }\textbf {\bibinfo {volume}
  {2619}},\ \bibinfo {pages} {012011} (\bibinfo {year} {2023})},\ \Eprint
  {https://arxiv.org/abs/2309.12266} {arXiv:2309.12266 [hep-ph]} \BibitemShut
  {NoStop}%
\bibitem [{\citenamefont {Liu}\ \emph {et~al.}(2010)\citenamefont {Liu},
  \citenamefont {Lin}, \citenamefont {Orginos},\ and\ \citenamefont
  {Walker-Loud}}]{Liu:2009jc}%
  \BibitemOpen
  \bibfield  {author} {\bibinfo {author} {\bibfnamefont {L.}~\bibnamefont
  {Liu}}, \bibinfo {author} {\bibfnamefont {H.-W.}\ \bibnamefont {Lin}},
  \bibinfo {author} {\bibfnamefont {K.}~\bibnamefont {Orginos}},\ and\ \bibinfo
  {author} {\bibfnamefont {A.}~\bibnamefont {Walker-Loud}},\ }\bibfield
  {title} {\bibinfo {title} {{Singly and Doubly Charmed J=1/2 Baryon Spectrum
  from Lattice QCD}},\ }\href {https://doi.org/10.1103/PhysRevD.81.094505}
  {\bibfield  {journal} {\bibinfo  {journal} {Phys. Rev. D}\ }\textbf {\bibinfo
  {volume} {81}},\ \bibinfo {pages} {094505} (\bibinfo {year} {2010})},\
  \Eprint {https://arxiv.org/abs/0909.3294} {arXiv:0909.3294 [hep-lat]}
  \BibitemShut {NoStop}%
\bibitem [{\citenamefont {Brice\~no}\ \emph {et~al.}(2012)\citenamefont
  {Brice\~no}, \citenamefont {Lin},\ and\ \citenamefont
  {Bolton}}]{Briceno:2012wt}%
  \BibitemOpen
  \bibfield  {author} {\bibinfo {author} {\bibfnamefont {R.~A.}\ \bibnamefont
  {Brice\~no}}, \bibinfo {author} {\bibfnamefont {H.-W.}\ \bibnamefont {Lin}},\
  and\ \bibinfo {author} {\bibfnamefont {D.~R.}\ \bibnamefont {Bolton}},\
  }\bibfield  {title} {\bibinfo {title} {{Charmed-baryon spectroscopy from
  lattice QCD with ${N}_{f}\mathbf{=}2\mathbf{+}1\mathbf{+}1$ flavors}},\
  }\href {https://doi.org/10.1103/PhysRevD.86.094504} {\bibfield  {journal}
  {\bibinfo  {journal} {Phys. Rev.}\ }\textbf {\bibinfo {volume} {D86}},\
  \bibinfo {pages} {094504} (\bibinfo {year} {2012})},\ \Eprint
  {https://arxiv.org/abs/1207.3536} {arXiv:1207.3536 [hep-lat]} \BibitemShut
  {NoStop}%
\bibitem [{\citenamefont {Namekawa}\ \emph {et~al.}(2013)\citenamefont
  {Namekawa}, \citenamefont {Aoki}, \citenamefont {Ishikawa}, \citenamefont
  {Ishizuka}, \citenamefont {Kanaya}, \citenamefont {Kuramashi}, \citenamefont
  {Okawa}, \citenamefont {Taniguchi}, \citenamefont {Ukawa}, \citenamefont
  {Ukita},\ and\ \citenamefont {Yoshi\'e}}]{Namekawa:2013vu}%
  \BibitemOpen
  \bibfield  {author} {\bibinfo {author} {\bibfnamefont {Y.}~\bibnamefont
  {Namekawa}}, \bibinfo {author} {\bibfnamefont {S.}~\bibnamefont {Aoki}},
  \bibinfo {author} {\bibfnamefont {K.-I.}\ \bibnamefont {Ishikawa}}, \bibinfo
  {author} {\bibfnamefont {N.}~\bibnamefont {Ishizuka}}, \bibinfo {author}
  {\bibfnamefont {K.}~\bibnamefont {Kanaya}}, \bibinfo {author} {\bibfnamefont
  {Y.}~\bibnamefont {Kuramashi}}, \bibinfo {author} {\bibfnamefont
  {M.}~\bibnamefont {Okawa}}, \bibinfo {author} {\bibfnamefont
  {Y.}~\bibnamefont {Taniguchi}}, \bibinfo {author} {\bibfnamefont
  {A.}~\bibnamefont {Ukawa}}, \bibinfo {author} {\bibfnamefont
  {N.}~\bibnamefont {Ukita}},\ and\ \bibinfo {author} {\bibfnamefont
  {T.}~\bibnamefont {Yoshi\'e}} (\bibinfo {collaboration} {PACS-CS
  Collaboration}),\ }\bibfield  {title} {\bibinfo {title} {{Charmed baryons at
  the physical point in $2\mathbf{+}1$ flavor lattice QCD}},\ }\href
  {https://doi.org/10.1103/PhysRevD.87.094512} {\bibfield  {journal} {\bibinfo
  {journal} {Phys. Rev.}\ }\textbf {\bibinfo {volume} {D87}},\ \bibinfo {pages}
  {094512} (\bibinfo {year} {2013})},\ \Eprint
  {https://arxiv.org/abs/1301.4743} {arXiv:1301.4743 [hep-lat]} \BibitemShut
  {NoStop}%
\bibitem [{\citenamefont {Brown}\ \emph {et~al.}(2014)\citenamefont {Brown},
  \citenamefont {Detmold}, \citenamefont {Meinel},\ and\ \citenamefont
  {Orginos}}]{Brown:2014ena}%
  \BibitemOpen
  \bibfield  {author} {\bibinfo {author} {\bibfnamefont {Z.~S.}\ \bibnamefont
  {Brown}}, \bibinfo {author} {\bibfnamefont {W.}~\bibnamefont {Detmold}},
  \bibinfo {author} {\bibfnamefont {S.}~\bibnamefont {Meinel}},\ and\ \bibinfo
  {author} {\bibfnamefont {K.}~\bibnamefont {Orginos}},\ }\bibfield  {title}
  {\bibinfo {title} {{Charmed bottom baryon spectroscopy from lattice QCD}},\
  }\href {https://doi.org/10.1103/PhysRevD.90.094507} {\bibfield  {journal}
  {\bibinfo  {journal} {Phys. Rev.}\ }\textbf {\bibinfo {volume} {D90}},\
  \bibinfo {pages} {094507} (\bibinfo {year} {2014})},\ \Eprint
  {https://arxiv.org/abs/1409.0497} {arXiv:1409.0497 [hep-lat]} \BibitemShut
  {NoStop}%
\bibitem [{\citenamefont {P\'erez-Rubio}\ \emph {et~al.}(2015)\citenamefont
  {P\'erez-Rubio}, \citenamefont {Collins},\ and\ \citenamefont
  {Bali}}]{Bali:2015lka}%
  \BibitemOpen
  \bibfield  {author} {\bibinfo {author} {\bibfnamefont {P.}~\bibnamefont
  {P\'erez-Rubio}}, \bibinfo {author} {\bibfnamefont {S.}~\bibnamefont
  {Collins}},\ and\ \bibinfo {author} {\bibfnamefont {G.~S.}\ \bibnamefont
  {Bali}},\ }\bibfield  {title} {\bibinfo {title} {{Charmed baryon spectroscopy
  and light flavor symmetry from lattice QCD}},\ }\href
  {https://doi.org/10.1103/PhysRevD.92.034504} {\bibfield  {journal} {\bibinfo
  {journal} {Phys. Rev.}\ }\textbf {\bibinfo {volume} {D92}},\ \bibinfo {pages}
  {034504} (\bibinfo {year} {2015})},\ \Eprint
  {https://arxiv.org/abs/1503.08440} {arXiv:1503.08440 [hep-lat]} \BibitemShut
  {NoStop}%
\bibitem [{\citenamefont {Bahtiyar}\ \emph {et~al.}(2020)\citenamefont
  {Bahtiyar}, \citenamefont {Can}, \citenamefont {Erkol}, \citenamefont
  {Gubler}, \citenamefont {Oka},\ and\ \citenamefont
  {Takahashi}}]{Bahtiyar:2020uuj}%
  \BibitemOpen
  \bibfield  {author} {\bibinfo {author} {\bibfnamefont {H.}~\bibnamefont
  {Bahtiyar}}, \bibinfo {author} {\bibfnamefont {K.~U.}\ \bibnamefont {Can}},
  \bibinfo {author} {\bibfnamefont {G.}~\bibnamefont {Erkol}}, \bibinfo
  {author} {\bibfnamefont {P.}~\bibnamefont {Gubler}}, \bibinfo {author}
  {\bibfnamefont {M.}~\bibnamefont {Oka}},\ and\ \bibinfo {author}
  {\bibfnamefont {T.~T.}\ \bibnamefont {Takahashi}} (\bibinfo {collaboration}
  {TRJQCD Collaboration}),\ }\bibfield  {title} {\bibinfo {title} {{Charmed
  baryon spectrum from lattice QCD near the physical point}},\ }\href
  {https://doi.org/10.1103/PhysRevD.102.054513} {\bibfield  {journal} {\bibinfo
   {journal} {Phys. Rev.}\ }\textbf {\bibinfo {volume} {D102}},\ \bibinfo
  {pages} {054513} (\bibinfo {year} {2020})},\ \Eprint
  {https://arxiv.org/abs/2004.08999} {arXiv:2004.08999 [hep-lat]} \BibitemShut
  {NoStop}%
\bibitem [{\citenamefont {Zhang}\ \emph {et~al.}(2022)\citenamefont {Zhang},
  \citenamefont {Hua}, \citenamefont {Huang}, \citenamefont {Li}, \citenamefont
  {Li}, \citenamefont {Lu}, \citenamefont {Sun}, \citenamefont {Sun},
  \citenamefont {Wang},\ and\ \citenamefont {Yang}}]{Zhang:2021oja}%
  \BibitemOpen
  \bibfield  {author} {\bibinfo {author} {\bibfnamefont {Q.-A.}\ \bibnamefont
  {Zhang}}, \bibinfo {author} {\bibfnamefont {J.}~\bibnamefont {Hua}}, \bibinfo
  {author} {\bibfnamefont {F.}~\bibnamefont {Huang}}, \bibinfo {author}
  {\bibfnamefont {R.}~\bibnamefont {Li}}, \bibinfo {author} {\bibfnamefont
  {Y.}~\bibnamefont {Li}}, \bibinfo {author} {\bibfnamefont {C.-D.}\
  \bibnamefont {Lu}}, \bibinfo {author} {\bibfnamefont {P.}~\bibnamefont
  {Sun}}, \bibinfo {author} {\bibfnamefont {W.}~\bibnamefont {Sun}}, \bibinfo
  {author} {\bibfnamefont {W.}~\bibnamefont {Wang}},\ and\ \bibinfo {author}
  {\bibfnamefont {Y.-B.}\ \bibnamefont {Yang}},\ }\bibfield  {title} {\bibinfo
  {title} {{First lattice QCD calculation of semileptonic decays of
  charmed-strange baryons $\ensuremath{\Xi}_{c}$}},\ }\href
  {https://doi.org/10.1088/1674-1137/ac2b12} {\bibfield  {journal} {\bibinfo
  {journal} {Chin. Phys. C}\ }\textbf {\bibinfo {volume} {46}},\ \bibinfo
  {pages} {011002} (\bibinfo {year} {2022})},\ \Eprint
  {https://arxiv.org/abs/2103.07064} {arXiv:2103.07064 [hep-lat]} \BibitemShut
  {NoStop}%
\bibitem [{\citenamefont {Bagan}\ \emph {et~al.}(1992)\citenamefont {Bagan},
  \citenamefont {Chabab}, \citenamefont {Dosch},\ and\ \citenamefont
  {Narison}}]{Bagan:1992tp}%
  \BibitemOpen
  \bibfield  {author} {\bibinfo {author} {\bibfnamefont {E.}~\bibnamefont
  {Bagan}}, \bibinfo {author} {\bibfnamefont {M.}~\bibnamefont {Chabab}},
  \bibinfo {author} {\bibfnamefont {H.~G.}\ \bibnamefont {Dosch}},\ and\
  \bibinfo {author} {\bibfnamefont {S.}~\bibnamefont {Narison}},\ }\bibfield
  {title} {\bibinfo {title} {{Spectra of heavy baryons from QCD spectral sum
  rules}},\ }\href {https://doi.org/10.1016/0370-2693(92)91896-H} {\bibfield
  {journal} {\bibinfo  {journal} {Phys. Lett. B}\ }\textbf {\bibinfo {volume}
  {287}},\ \bibinfo {pages} {176} (\bibinfo {year} {1992})}\BibitemShut
  {NoStop}%
\bibitem [{\citenamefont {Wang}(2010)}]{Wang:2009cr}%
  \BibitemOpen
  \bibfield  {author} {\bibinfo {author} {\bibfnamefont {Z.-G.}\ \bibnamefont
  {Wang}},\ }\bibfield  {title} {\bibinfo {title} {{Reanalysis of the heavy
  baryon states Omega(b), Omega(c), Xi'(b), Xi'(c), Sigma(b) and Sigma(c) with
  QCD sum rules}},\ }\href {https://doi.org/10.1016/j.physletb.2010.01.039}
  {\bibfield  {journal} {\bibinfo  {journal} {Phys. Lett. B}\ }\textbf
  {\bibinfo {volume} {685}},\ \bibinfo {pages} {59} (\bibinfo {year} {2010})},\
  \Eprint {https://arxiv.org/abs/0912.1648} {arXiv:0912.1648 [hep-ph]}
  \BibitemShut {NoStop}%
\bibitem [{\citenamefont {Zhang}\ and\ \citenamefont
  {Huang}(2009)}]{Zhang:2009iya}%
  \BibitemOpen
  \bibfield  {author} {\bibinfo {author} {\bibfnamefont {J.-R.}\ \bibnamefont
  {Zhang}}\ and\ \bibinfo {author} {\bibfnamefont {M.-Q.}\ \bibnamefont
  {Huang}},\ }\bibfield  {title} {\bibinfo {title} {{Heavy flavor baryon
  spectra via QCD sum rules}},\ }\href
  {https://doi.org/10.1088/1674-1137/33/12/061} {\bibfield  {journal} {\bibinfo
   {journal} {Chin. Phys. C}\ }\textbf {\bibinfo {volume} {33}},\ \bibinfo
  {pages} {1385} (\bibinfo {year} {2009})},\ \Eprint
  {https://arxiv.org/abs/0904.3391} {arXiv:0904.3391 [hep-ph]} \BibitemShut
  {NoStop}%
\bibitem [{\citenamefont {Chen}\ \emph {et~al.}(2015)\citenamefont {Chen},
  \citenamefont {Chen}, \citenamefont {Mao}, \citenamefont {Hosaka},
  \citenamefont {Liu},\ and\ \citenamefont {Zhu}}]{Chen:2015kpa}%
  \BibitemOpen
  \bibfield  {author} {\bibinfo {author} {\bibfnamefont {H.-X.}\ \bibnamefont
  {Chen}}, \bibinfo {author} {\bibfnamefont {W.}~\bibnamefont {Chen}}, \bibinfo
  {author} {\bibfnamefont {Q.}~\bibnamefont {Mao}}, \bibinfo {author}
  {\bibfnamefont {A.}~\bibnamefont {Hosaka}}, \bibinfo {author} {\bibfnamefont
  {X.}~\bibnamefont {Liu}},\ and\ \bibinfo {author} {\bibfnamefont {S.-L.}\
  \bibnamefont {Zhu}},\ }\bibfield  {title} {\bibinfo {title} {{$P$-wave
  charmed baryons from QCD sum rules}},\ }\href
  {https://doi.org/10.1103/PhysRevD.91.054034} {\bibfield  {journal} {\bibinfo
  {journal} {Phys. Rev.}\ }\textbf {\bibinfo {volume} {D91}},\ \bibinfo {pages}
  {054034} (\bibinfo {year} {2015})},\ \Eprint
  {https://arxiv.org/abs/1502.01103} {arXiv:1502.01103 [hep-ph]} \BibitemShut
  {NoStop}%
\bibitem [{\citenamefont {Mao}\ \emph {et~al.}(2015)\citenamefont {Mao},
  \citenamefont {Chen}, \citenamefont {Chen}, \citenamefont {Hosaka},
  \citenamefont {Liu},\ and\ \citenamefont {Zhu}}]{Mao:2015gya}%
  \BibitemOpen
  \bibfield  {author} {\bibinfo {author} {\bibfnamefont {Q.}~\bibnamefont
  {Mao}}, \bibinfo {author} {\bibfnamefont {H.-X.}\ \bibnamefont {Chen}},
  \bibinfo {author} {\bibfnamefont {W.}~\bibnamefont {Chen}}, \bibinfo {author}
  {\bibfnamefont {A.}~\bibnamefont {Hosaka}}, \bibinfo {author} {\bibfnamefont
  {X.}~\bibnamefont {Liu}},\ and\ \bibinfo {author} {\bibfnamefont {S.-L.}\
  \bibnamefont {Zhu}},\ }\bibfield  {title} {\bibinfo {title} {{QCD sum rule
  calculation for $P$-wave bottom baryons}},\ }\href
  {https://doi.org/10.1103/PhysRevD.92.114007} {\bibfield  {journal} {\bibinfo
  {journal} {Phys. Rev.}\ }\textbf {\bibinfo {volume} {D92}},\ \bibinfo {pages}
  {114007} (\bibinfo {year} {2015})},\ \Eprint
  {https://arxiv.org/abs/1510.05267} {arXiv:1510.05267 [hep-ph]} \BibitemShut
  {NoStop}%
\bibitem [{\citenamefont {Agaev}\ \emph {et~al.}(2017)\citenamefont {Agaev},
  \citenamefont {Azizi},\ and\ \citenamefont {Sundu}}]{Agaev:2017lip}%
  \BibitemOpen
  \bibfield  {author} {\bibinfo {author} {\bibfnamefont {S.~S.}\ \bibnamefont
  {Agaev}}, \bibinfo {author} {\bibfnamefont {K.}~\bibnamefont {Azizi}},\ and\
  \bibinfo {author} {\bibfnamefont {H.}~\bibnamefont {Sundu}},\ }\bibfield
  {title} {\bibinfo {title} {Interpretation of the new $\omega_c^{0}$ states
  via their mass and width},\ }\href
  {https://doi.org/10.1140/epjc/s10052-017-4953-z} {\bibfield  {journal}
  {\bibinfo  {journal} {Eur. Phys. J.}\ }\textbf {\bibinfo {volume} {C77}},\
  \bibinfo {pages} {395} (\bibinfo {year} {2017})},\ \Eprint
  {https://arxiv.org/abs/1704.04928} {arXiv:1704.04928 [hep-ph]} \BibitemShut
  {NoStop}%
\bibitem [{\citenamefont {Yang}\ \emph
  {et~al.}(2022{\natexlab{a}})\citenamefont {Yang}, \citenamefont {Chen},
  \citenamefont {Cui},\ and\ \citenamefont {Mao}}]{Yang:2022oog}%
  \BibitemOpen
  \bibfield  {author} {\bibinfo {author} {\bibfnamefont {H.-M.}\ \bibnamefont
  {Yang}}, \bibinfo {author} {\bibfnamefont {H.-X.}\ \bibnamefont {Chen}},
  \bibinfo {author} {\bibfnamefont {E.-L.}\ \bibnamefont {Cui}},\ and\ \bibinfo
  {author} {\bibfnamefont {Q.}~\bibnamefont {Mao}},\ }\bibfield  {title}
  {\bibinfo {title} {{Identifying the ${\mathrm{\ensuremath{\Xi}}}_{b}(6100)$
  as the $P$-wave bottom baryon of ${J}^{P}=3/{2}^{\ensuremath{-}}$}},\ }\href
  {https://doi.org/10.1103/PhysRevD.106.036018} {\bibfield  {journal} {\bibinfo
   {journal} {Phys. Rev. D}\ }\textbf {\bibinfo {volume} {106}},\ \bibinfo
  {pages} {036018} (\bibinfo {year} {2022}{\natexlab{a}})},\ \Eprint
  {https://arxiv.org/abs/2205.07224} {arXiv:2205.07224 [hep-ph]} \BibitemShut
  {NoStop}%
\bibitem [{\citenamefont {Vishwakarma}\ and\ \citenamefont
  {Upadhyay}(2022)}]{Vishwakarma:2022vzy}%
  \BibitemOpen
  \bibfield  {author} {\bibinfo {author} {\bibfnamefont {K.~K.}\ \bibnamefont
  {Vishwakarma}}\ and\ \bibinfo {author} {\bibfnamefont {A.}~\bibnamefont
  {Upadhyay}},\ }\bibfield  {title} {\bibinfo {title} {{Masses of 2S single
  heavy baryons using non-perturbative parameters in HQET}},\ }\href
  {https://arxiv.org/abs/2208.02536} {\  (\bibinfo {year} {2022})},\ \Eprint
  {https://arxiv.org/abs/2208.02536} {arXiv:2208.02536 [hep-ph]} \BibitemShut
  {NoStop}%
\bibitem [{\citenamefont {Wei}\ \emph {et~al.}(2017)\citenamefont {Wei},
  \citenamefont {Chen}, \citenamefont {Liu}, \citenamefont {Wang},\ and\
  \citenamefont {Guo}}]{Wei:2016jyk}%
  \BibitemOpen
  \bibfield  {author} {\bibinfo {author} {\bibfnamefont {K.-W.}\ \bibnamefont
  {Wei}}, \bibinfo {author} {\bibfnamefont {B.}~\bibnamefont {Chen}}, \bibinfo
  {author} {\bibfnamefont {N.}~\bibnamefont {Liu}}, \bibinfo {author}
  {\bibfnamefont {Q.-Q.}\ \bibnamefont {Wang}},\ and\ \bibinfo {author}
  {\bibfnamefont {X.-H.}\ \bibnamefont {Guo}},\ }\bibfield  {title} {\bibinfo
  {title} {{Spectroscopy of singly, doubly, and triply bottom baryons}},\
  }\href {https://doi.org/10.1103/PhysRevD.95.116005} {\bibfield  {journal}
  {\bibinfo  {journal} {Phys. Rev.}\ }\textbf {\bibinfo {volume} {D95}},\
  \bibinfo {pages} {116005} (\bibinfo {year} {2017})},\ \Eprint
  {https://arxiv.org/abs/1609.02512} {arXiv:1609.02512 [hep-ph]} \BibitemShut
  {NoStop}%
\bibitem [{\citenamefont {Jia}\ \emph {et~al.}(2020)\citenamefont {Jia},
  \citenamefont {Liu},\ and\ \citenamefont {Hosaka}}]{Jia:2019bkr}%
  \BibitemOpen
  \bibfield  {author} {\bibinfo {author} {\bibfnamefont {D.}~\bibnamefont
  {Jia}}, \bibinfo {author} {\bibfnamefont {W.-N.}\ \bibnamefont {Liu}},\ and\
  \bibinfo {author} {\bibfnamefont {A.}~\bibnamefont {Hosaka}},\ }\bibfield
  {title} {\bibinfo {title} {{Regge behaviors in orbitally excited spectroscopy
  of charmed and bottom baryons}},\ }\href
  {https://doi.org/10.1103/PhysRevD.101.034016} {\bibfield  {journal} {\bibinfo
   {journal} {Phys. Rev.}\ }\textbf {\bibinfo {volume} {D101}},\ \bibinfo
  {pages} {034016} (\bibinfo {year} {2020})},\ \Eprint
  {https://arxiv.org/abs/1907.04958} {arXiv:1907.04958 [hep-ph]} \BibitemShut
  {NoStop}%
\bibitem [{\citenamefont {Oudichhya}\ and\ \citenamefont
  {Rai}(2023)}]{Oudichhya:2023awb}%
  \BibitemOpen
  \bibfield  {author} {\bibinfo {author} {\bibfnamefont {J.}~\bibnamefont
  {Oudichhya}}\ and\ \bibinfo {author} {\bibfnamefont {A.~K.}\ \bibnamefont
  {Rai}},\ }\bibfield  {title} {\bibinfo {title} {{Spin-parity identification
  of newly observed singly charmed baryons in Regge phenomenology}},\ }\href
  {https://doi.org/10.1140/epja/s10050-023-01024-5} {\bibfield  {journal}
  {\bibinfo  {journal} {Eur. Phys. J. A}\ }\textbf {\bibinfo {volume} {59}},\
  \bibinfo {pages} {123} (\bibinfo {year} {2023})}\BibitemShut {NoStop}%
\bibitem [{\citenamefont {Klempt}\ and\ \citenamefont
  {Richard}(2010)}]{Klempt:2009pi}%
  \BibitemOpen
  \bibfield  {author} {\bibinfo {author} {\bibfnamefont {E.}~\bibnamefont
  {Klempt}}\ and\ \bibinfo {author} {\bibfnamefont {J.-M.}\ \bibnamefont
  {Richard}},\ }\bibfield  {title} {\bibinfo {title} {{Baryon spectroscopy}},\
  }\href {https://doi.org/10.1103/RevModPhys.82.1095} {\bibfield  {journal}
  {\bibinfo  {journal} {Rev. Mod. Phys.}\ }\textbf {\bibinfo {volume} {82}},\
  \bibinfo {pages} {1095} (\bibinfo {year} {2010})},\ \Eprint
  {https://arxiv.org/abs/0901.2055} {arXiv:0901.2055 [hep-ph]} \BibitemShut
  {NoStop}%
\bibitem [{\citenamefont {Crede}\ and\ \citenamefont
  {Roberts}(2013)}]{Crede:2013sze}%
  \BibitemOpen
  \bibfield  {author} {\bibinfo {author} {\bibfnamefont {V.}~\bibnamefont
  {Crede}}\ and\ \bibinfo {author} {\bibfnamefont {W.}~\bibnamefont
  {Roberts}},\ }\bibfield  {title} {\bibinfo {title} {{Progress towards
  understanding baryon resonances}},\ }\href
  {https://doi.org/10.1088/0034-4885/76/7/076301} {\bibfield  {journal}
  {\bibinfo  {journal} {Rept. Prog. Phys.}\ }\textbf {\bibinfo {volume} {76}},\
  \bibinfo {pages} {076301} (\bibinfo {year} {2013})},\ \Eprint
  {https://arxiv.org/abs/1302.7299} {arXiv:1302.7299 [nucl-ex]} \BibitemShut
  {NoStop}%
\bibitem [{\citenamefont {Chen}\ \emph {et~al.}(2017)\citenamefont {Chen},
  \citenamefont {Chen}, \citenamefont {Liu}, \citenamefont {Liu},\ and\
  \citenamefont {Zhu}}]{Chen:2016spr}%
  \BibitemOpen
  \bibfield  {author} {\bibinfo {author} {\bibfnamefont {H.-X.}\ \bibnamefont
  {Chen}}, \bibinfo {author} {\bibfnamefont {W.}~\bibnamefont {Chen}}, \bibinfo
  {author} {\bibfnamefont {X.}~\bibnamefont {Liu}}, \bibinfo {author}
  {\bibfnamefont {Y.-R.}\ \bibnamefont {Liu}},\ and\ \bibinfo {author}
  {\bibfnamefont {S.-L.}\ \bibnamefont {Zhu}},\ }\bibfield  {title} {\bibinfo
  {title} {{A review of the open charm and open bottom systems}},\ }\href
  {https://doi.org/10.1088/1361-6633/aa6420} {\bibfield  {journal} {\bibinfo
  {journal} {Rept. Prog. Phys.}\ }\textbf {\bibinfo {volume} {80}},\ \bibinfo
  {pages} {076201} (\bibinfo {year} {2017})},\ \Eprint
  {https://arxiv.org/abs/1609.08928} {arXiv:1609.08928 [hep-ph]} \BibitemShut
  {NoStop}%
\bibitem [{\citenamefont {Eichmann}\ \emph {et~al.}(2016)\citenamefont
  {Eichmann}, \citenamefont {Sanchis-Alepuz}, \citenamefont {Williams},
  \citenamefont {Alkofer},\ and\ \citenamefont {Fischer}}]{Eichmann:2016yit}%
  \BibitemOpen
  \bibfield  {author} {\bibinfo {author} {\bibfnamefont {G.}~\bibnamefont
  {Eichmann}}, \bibinfo {author} {\bibfnamefont {H.}~\bibnamefont
  {Sanchis-Alepuz}}, \bibinfo {author} {\bibfnamefont {R.}~\bibnamefont
  {Williams}}, \bibinfo {author} {\bibfnamefont {R.}~\bibnamefont {Alkofer}},\
  and\ \bibinfo {author} {\bibfnamefont {C.~S.}\ \bibnamefont {Fischer}},\
  }\bibfield  {title} {\bibinfo {title} {{Baryons as relativistic three-quark
  bound states}},\ }\href {https://doi.org/10.1016/j.ppnp.2016.07.001}
  {\bibfield  {journal} {\bibinfo  {journal} {Prog. Part. Nucl. Phys.}\
  }\textbf {\bibinfo {volume} {91}},\ \bibinfo {pages} {1} (\bibinfo {year}
  {2016})},\ \Eprint {https://arxiv.org/abs/1606.09602} {arXiv:1606.09602
  [hep-ph]} \BibitemShut {NoStop}%
\bibitem [{\citenamefont {Cheng}(2022)}]{Cheng:2021qpd}%
  \BibitemOpen
  \bibfield  {author} {\bibinfo {author} {\bibfnamefont {H.-Y.}\ \bibnamefont
  {Cheng}},\ }\bibfield  {title} {\bibinfo {title} {{Charmed baryon physics
  circa 2021}},\ }\href {https://doi.org/10.1016/j.cjph.2022.06.021} {\bibfield
   {journal} {\bibinfo  {journal} {Chin. J. Phys.}\ }\textbf {\bibinfo {volume}
  {78}},\ \bibinfo {pages} {324} (\bibinfo {year} {2022})},\ \Eprint
  {https://arxiv.org/abs/2109.01216} {arXiv:2109.01216 [hep-ph]} \BibitemShut
  {NoStop}%
\bibitem [{\citenamefont {Chen}\ \emph {et~al.}(2023)\citenamefont {Chen},
  \citenamefont {Chen}, \citenamefont {Liu}, \citenamefont {Liu},\ and\
  \citenamefont {Zhu}}]{Chen:2022asf}%
  \BibitemOpen
  \bibfield  {author} {\bibinfo {author} {\bibfnamefont {H.-X.}\ \bibnamefont
  {Chen}}, \bibinfo {author} {\bibfnamefont {W.}~\bibnamefont {Chen}}, \bibinfo
  {author} {\bibfnamefont {X.}~\bibnamefont {Liu}}, \bibinfo {author}
  {\bibfnamefont {Y.-R.}\ \bibnamefont {Liu}},\ and\ \bibinfo {author}
  {\bibfnamefont {S.-L.}\ \bibnamefont {Zhu}},\ }\bibfield  {title} {\bibinfo
  {title} {{An updated review of the new hadron states}},\ }\href
  {https://doi.org/10.1088/1361-6633/aca3b6} {\bibfield  {journal} {\bibinfo
  {journal} {Rept. Prog. Phys.}\ }\textbf {\bibinfo {volume} {86}},\ \bibinfo
  {pages} {026201} (\bibinfo {year} {2023})},\ \Eprint
  {https://arxiv.org/abs/2204.02649} {arXiv:2204.02649 [hep-ph]} \BibitemShut
  {NoStop}%
\bibitem [{\citenamefont {Godfrey}\ and\ \citenamefont
  {Isgur}(1985)}]{Godfrey:1985xj}%
  \BibitemOpen
  \bibfield  {author} {\bibinfo {author} {\bibfnamefont {S.}~\bibnamefont
  {Godfrey}}\ and\ \bibinfo {author} {\bibfnamefont {N.}~\bibnamefont
  {Isgur}},\ }\bibfield  {title} {\bibinfo {title} {{Mesons in a relativized
  quark model with chromodynamics}},\ }\href
  {https://doi.org/10.1103/PhysRevD.32.189} {\bibfield  {journal} {\bibinfo
  {journal} {Phys. Rev. D}\ }\textbf {\bibinfo {volume} {32}},\ \bibinfo
  {pages} {189} (\bibinfo {year} {1985})}\BibitemShut {NoStop}%
\bibitem [{\citenamefont {Godfrey}\ and\ \citenamefont
  {Kokoski}(1991)}]{Godfrey:1986wj}%
  \BibitemOpen
  \bibfield  {author} {\bibinfo {author} {\bibfnamefont {S.}~\bibnamefont
  {Godfrey}}\ and\ \bibinfo {author} {\bibfnamefont {R.}~\bibnamefont
  {Kokoski}},\ }\bibfield  {title} {\bibinfo {title} {{Properties of $P$-wave
  mesons with one heavy quark}},\ }\href
  {https://doi.org/10.1103/PhysRevD.43.1679} {\bibfield  {journal} {\bibinfo
  {journal} {Phys. Rev.}\ }\textbf {\bibinfo {volume} {D43}},\ \bibinfo {pages}
  {1679} (\bibinfo {year} {1991})}\BibitemShut {NoStop}%
\bibitem [{\citenamefont {Godfrey}(2004)}]{Godfrey:2004ya}%
  \BibitemOpen
  \bibfield  {author} {\bibinfo {author} {\bibfnamefont {S.}~\bibnamefont
  {Godfrey}},\ }\bibfield  {title} {\bibinfo {title} {{Spectroscopy of $B_c$
  mesons in the relativized quark model}},\ }\href
  {https://doi.org/10.1103/PhysRevD.70.054017} {\bibfield  {journal} {\bibinfo
  {journal} {Phys. Rev.}\ }\textbf {\bibinfo {volume} {D70}},\ \bibinfo {pages}
  {054017} (\bibinfo {year} {2004})},\ \Eprint
  {https://arxiv.org/abs/hep-ph/0406228} {arXiv:hep-ph/0406228 [hep-ph]}
  \BibitemShut {NoStop}%
\bibitem [{\citenamefont {Barnes}\ \emph {et~al.}(2005)\citenamefont {Barnes},
  \citenamefont {Godfrey},\ and\ \citenamefont {Swanson}}]{Barnes:2005pb}%
  \BibitemOpen
  \bibfield  {author} {\bibinfo {author} {\bibfnamefont {T.}~\bibnamefont
  {Barnes}}, \bibinfo {author} {\bibfnamefont {S.}~\bibnamefont {Godfrey}},\
  and\ \bibinfo {author} {\bibfnamefont {E.~S.}\ \bibnamefont {Swanson}},\
  }\bibfield  {title} {\bibinfo {title} {{Higher charmonia}},\ }\href
  {https://doi.org/10.1103/PhysRevD.72.054026} {\bibfield  {journal} {\bibinfo
  {journal} {Phys. Rev.}\ }\textbf {\bibinfo {volume} {D72}},\ \bibinfo {pages}
  {054026} (\bibinfo {year} {2005})},\ \Eprint
  {https://arxiv.org/abs/hep-ph/0505002} {arXiv:hep-ph/0505002 [hep-ph]}
  \BibitemShut {NoStop}%
\bibitem [{\citenamefont {Godfrey}\ and\ \citenamefont
  {Moats}(2015)}]{Godfrey:2015dia}%
  \BibitemOpen
  \bibfield  {author} {\bibinfo {author} {\bibfnamefont {S.}~\bibnamefont
  {Godfrey}}\ and\ \bibinfo {author} {\bibfnamefont {K.}~\bibnamefont
  {Moats}},\ }\bibfield  {title} {\bibinfo {title} {{Bottomonium mesons and
  strategies for their observation}},\ }\href
  {https://doi.org/10.1103/PhysRevD.92.054034} {\bibfield  {journal} {\bibinfo
  {journal} {Phys. Rev. D}\ }\textbf {\bibinfo {volume} {92}},\ \bibinfo
  {pages} {054034} (\bibinfo {year} {2015})},\ \Eprint
  {https://arxiv.org/abs/1507.00024} {arXiv:1507.00024 [hep-ph]} \BibitemShut
  {NoStop}%
\bibitem [{\citenamefont {Godfrey}\ \emph {et~al.}(2016)\citenamefont
  {Godfrey}, \citenamefont {Moats},\ and\ \citenamefont
  {Swanson}}]{Godfrey:2016nwn}%
  \BibitemOpen
  \bibfield  {author} {\bibinfo {author} {\bibfnamefont {S.}~\bibnamefont
  {Godfrey}}, \bibinfo {author} {\bibfnamefont {K.}~\bibnamefont {Moats}},\
  and\ \bibinfo {author} {\bibfnamefont {E.~S.}\ \bibnamefont {Swanson}},\
  }\bibfield  {title} {\bibinfo {title} {{$B$ and ${B}_{s}$ meson
  spectroscopy}},\ }\href {https://doi.org/10.1103/PhysRevD.94.054025}
  {\bibfield  {journal} {\bibinfo  {journal} {Phys. Rev. D}\ }\textbf {\bibinfo
  {volume} {94}},\ \bibinfo {pages} {054025} (\bibinfo {year} {2016})},\
  \Eprint {https://arxiv.org/abs/1607.02169} {arXiv:1607.02169 [hep-ph]}
  \BibitemShut {NoStop}%
\bibitem [{\citenamefont {Godfrey}\ and\ \citenamefont
  {Moats}(2016)}]{Godfrey:2015dva}%
  \BibitemOpen
  \bibfield  {author} {\bibinfo {author} {\bibfnamefont {S.}~\bibnamefont
  {Godfrey}}\ and\ \bibinfo {author} {\bibfnamefont {K.}~\bibnamefont
  {Moats}},\ }\bibfield  {title} {\bibinfo {title} {{Properties of excited
  charm and charm-strange mesons}},\ }\href
  {https://doi.org/10.1103/PhysRevD.93.034035} {\bibfield  {journal} {\bibinfo
  {journal} {Phys. Rev. D}\ }\textbf {\bibinfo {volume} {93}},\ \bibinfo
  {pages} {034035} (\bibinfo {year} {2016})},\ \Eprint
  {https://arxiv.org/abs/1510.08305} {arXiv:1510.08305 [hep-ph]} \BibitemShut
  {NoStop}%
\bibitem [{\citenamefont {Capstick}\ and\ \citenamefont
  {Isgur}(1986)}]{Capstick:1986ter}%
  \BibitemOpen
  \bibfield  {author} {\bibinfo {author} {\bibfnamefont {S.}~\bibnamefont
  {Capstick}}\ and\ \bibinfo {author} {\bibfnamefont {N.}~\bibnamefont
  {Isgur}},\ }\bibfield  {title} {\bibinfo {title} {{Baryons in a relativized
  quark model with chromodynamics}},\ }\href
  {https://doi.org/10.1103/PhysRevD.34.2809} {\bibfield  {journal} {\bibinfo
  {journal} {Phys. Rev. D}\ }\textbf {\bibinfo {volume} {34}},\ \bibinfo
  {pages} {2809} (\bibinfo {year} {1986})}\BibitemShut {NoStop}%
\bibitem [{\citenamefont {L\"u}\ \emph {et~al.}(2018)\citenamefont {L\"u},
  \citenamefont {Dong}, \citenamefont {Liu},\ and\ \citenamefont
  {Matsuki}}]{Lu:2016ctt}%
  \BibitemOpen
  \bibfield  {author} {\bibinfo {author} {\bibfnamefont {Q.-F.}\ \bibnamefont
  {L\"u}}, \bibinfo {author} {\bibfnamefont {Y.}~\bibnamefont {Dong}}, \bibinfo
  {author} {\bibfnamefont {X.}~\bibnamefont {Liu}},\ and\ \bibinfo {author}
  {\bibfnamefont {T.}~\bibnamefont {Matsuki}},\ }\bibfield  {title} {\bibinfo
  {title} {{Puzzle of the $\Lambda_{c}$ Spectrum}},\ }\href
  {https://doi.org/10.11804/NuclPhysRev.35.01.001} {\bibfield  {journal}
  {\bibinfo  {journal} {Nucl. Phys. Rev.}\ }\textbf {\bibinfo {volume} {35}},\
  \bibinfo {pages} {1} (\bibinfo {year} {2018})},\ \Eprint
  {https://arxiv.org/abs/1610.09605} {arXiv:1610.09605 [hep-ph]} \BibitemShut
  {NoStop}%
\bibitem [{\citenamefont {L\"u}\ \emph {et~al.}(2017)\citenamefont {L\"u},
  \citenamefont {Wang}, \citenamefont {Xiao},\ and\ \citenamefont
  {Zhong}}]{Lu:2017meb}%
  \BibitemOpen
  \bibfield  {author} {\bibinfo {author} {\bibfnamefont {Q.-F.}\ \bibnamefont
  {L\"u}}, \bibinfo {author} {\bibfnamefont {K.-L.}\ \bibnamefont {Wang}},
  \bibinfo {author} {\bibfnamefont {L.-Y.}\ \bibnamefont {Xiao}},\ and\
  \bibinfo {author} {\bibfnamefont {X.-H.}\ \bibnamefont {Zhong}},\ }\bibfield
  {title} {\bibinfo {title} {{Mass spectra and radiative transitions of doubly
  heavy baryons in a relativized quark model}},\ }\href
  {https://doi.org/10.1103/PhysRevD.96.114006} {\bibfield  {journal} {\bibinfo
  {journal} {Phys. Rev. D}\ }\textbf {\bibinfo {volume} {96}},\ \bibinfo
  {pages} {114006} (\bibinfo {year} {2017})},\ \Eprint
  {https://arxiv.org/abs/1708.04468} {arXiv:1708.04468 [hep-ph]} \BibitemShut
  {NoStop}%
\bibitem [{\citenamefont {Yu}\ \emph {et~al.}(2023{\natexlab{a}})\citenamefont
  {Yu}, \citenamefont {Li}, \citenamefont {Wang}, \citenamefont {Lu},\ and\
  \citenamefont {Yan}}]{Yu:2022ymb}%
  \BibitemOpen
  \bibfield  {author} {\bibinfo {author} {\bibfnamefont {G.-L.}\ \bibnamefont
  {Yu}}, \bibinfo {author} {\bibfnamefont {Z.-Y.}\ \bibnamefont {Li}}, \bibinfo
  {author} {\bibfnamefont {Z.-G.}\ \bibnamefont {Wang}}, \bibinfo {author}
  {\bibfnamefont {J.}~\bibnamefont {Lu}},\ and\ \bibinfo {author}
  {\bibfnamefont {M.}~\bibnamefont {Yan}},\ }\bibfield  {title} {\bibinfo
  {title} {{Systematic analysis of single heavy baryons
  $\ensuremath{\Lambda}_{Q}$, $\ensuremath{\Sigma}_{Q}$ and
  $\ensuremath{\Omega}_{Q}$}},\ }\href
  {https://doi.org/10.1016/j.nuclphysb.2023.116183} {\bibfield  {journal}
  {\bibinfo  {journal} {Nucl. Phys. B}\ }\textbf {\bibinfo {volume} {990}},\
  \bibinfo {pages} {116183} (\bibinfo {year} {2023}{\natexlab{a}})},\ \Eprint
  {https://arxiv.org/abs/2206.08128} {arXiv:2206.08128 [hep-ph]} \BibitemShut
  {NoStop}%
\bibitem [{\citenamefont {Li}\ \emph {et~al.}(2023{\natexlab{b}})\citenamefont
  {Li}, \citenamefont {Yu}, \citenamefont {Wang}, \citenamefont {Gu},\ and\
  \citenamefont {Lu}}]{Li:2022xtj}%
  \BibitemOpen
  \bibfield  {author} {\bibinfo {author} {\bibfnamefont {Z.-Y.}\ \bibnamefont
  {Li}}, \bibinfo {author} {\bibfnamefont {G.-L.}\ \bibnamefont {Yu}}, \bibinfo
  {author} {\bibfnamefont {Z.-G.}\ \bibnamefont {Wang}}, \bibinfo {author}
  {\bibfnamefont {J.-Z.}\ \bibnamefont {Gu}},\ and\ \bibinfo {author}
  {\bibfnamefont {J.}~\bibnamefont {Lu}},\ }\bibfield  {title} {\bibinfo
  {title} {{Systematic analysis of strange single heavy baryons $\Xi_{c}$ and
  $\Xi_{b}$}},\ }\href {https://doi.org/10.1088/1674-1137/acd365} {\bibfield
  {journal} {\bibinfo  {journal} {Chin. Phys. C}\ }\textbf {\bibinfo {volume}
  {47}},\ \bibinfo {pages} {7} (\bibinfo {year} {2023}{\natexlab{b}})},\
  \Eprint {https://arxiv.org/abs/2207.04167} {arXiv:2207.04167 [hep-ph]}
  \BibitemShut {NoStop}%
\bibitem [{\citenamefont {Li}\ \emph {et~al.}(2023{\natexlab{c}})\citenamefont
  {Li}, \citenamefont {Yu}, \citenamefont {Wang}, \citenamefont {Gu},\ and\
  \citenamefont {Shen}}]{Li:2022oth}%
  \BibitemOpen
  \bibfield  {author} {\bibinfo {author} {\bibfnamefont {Z.-Y.}\ \bibnamefont
  {Li}}, \bibinfo {author} {\bibfnamefont {G.-L.}\ \bibnamefont {Yu}}, \bibinfo
  {author} {\bibfnamefont {Z.-G.}\ \bibnamefont {Wang}}, \bibinfo {author}
  {\bibfnamefont {J.-Z.}\ \bibnamefont {Gu}},\ and\ \bibinfo {author}
  {\bibfnamefont {H.-T.}\ \bibnamefont {Shen}},\ }\bibfield  {title} {\bibinfo
  {title} {{Mass spectra of double-bottom baryons}},\ }\href
  {https://doi.org/10.1142/S0217732323500529} {\bibfield  {journal} {\bibinfo
  {journal} {Mod. Phys. Lett. A}\ }\textbf {\bibinfo {volume} {38}},\ \bibinfo
  {pages} {2350052} (\bibinfo {year} {2023}{\natexlab{c}})},\ \Eprint
  {https://arxiv.org/abs/2210.13085} {arXiv:2210.13085 [hep-ph]} \BibitemShut
  {NoStop}%
\bibitem [{\citenamefont {Yu}\ \emph {et~al.}(2023{\natexlab{b}})\citenamefont
  {Yu}, \citenamefont {Li}, \citenamefont {Wang}, \citenamefont {Lu},\ and\
  \citenamefont {Yan}}]{Yu:2022lel}%
  \BibitemOpen
  \bibfield  {author} {\bibinfo {author} {\bibfnamefont {G.-L.}\ \bibnamefont
  {Yu}}, \bibinfo {author} {\bibfnamefont {Z.-Y.}\ \bibnamefont {Li}}, \bibinfo
  {author} {\bibfnamefont {Z.-G.}\ \bibnamefont {Wang}}, \bibinfo {author}
  {\bibfnamefont {J.}~\bibnamefont {Lu}},\ and\ \bibinfo {author}
  {\bibfnamefont {M.}~\bibnamefont {Yan}},\ }\bibfield  {title} {\bibinfo
  {title} {{Systematic analysis of doubly charmed baryons $\Xi_{cc}$ and
  $\Omega_{cc}$}},\ }\href {https://doi.org/10.1140/epja/s10050-023-01044-1}
  {\bibfield  {journal} {\bibinfo  {journal} {Eur. Phys. J. A}\ }\textbf
  {\bibinfo {volume} {59}},\ \bibinfo {pages} {126} (\bibinfo {year}
  {2023}{\natexlab{b}})},\ \Eprint {https://arxiv.org/abs/2211.00510}
  {arXiv:2211.00510 [hep-ph]} \BibitemShut {NoStop}%
\bibitem [{\citenamefont {Li}\ \emph {et~al.}(2023{\natexlab{d}})\citenamefont
  {Li}, \citenamefont {Yu}, \citenamefont {Wang}, \citenamefont {Gu},\ and\
  \citenamefont {Shen}}]{Li:2022ywz}%
  \BibitemOpen
  \bibfield  {author} {\bibinfo {author} {\bibfnamefont {Z.-Y.}\ \bibnamefont
  {Li}}, \bibinfo {author} {\bibfnamefont {G.-L.}\ \bibnamefont {Yu}}, \bibinfo
  {author} {\bibfnamefont {Z.-G.}\ \bibnamefont {Wang}}, \bibinfo {author}
  {\bibfnamefont {J.-Z.}\ \bibnamefont {Gu}},\ and\ \bibinfo {author}
  {\bibfnamefont {H.-T.}\ \bibnamefont {Shen}},\ }\bibfield  {title} {\bibinfo
  {title} {{Mass spectra of bottom-charm baryons}},\ }\href
  {https://doi.org/10.1142/S0217751X23500951} {\bibfield  {journal} {\bibinfo
  {journal} {Int. J. Mod. Phys. A}\ }\textbf {\bibinfo {volume} {38}},\
  \bibinfo {pages} {2350095} (\bibinfo {year} {2023}{\natexlab{d}})},\ \Eprint
  {https://arxiv.org/abs/2211.15111} {arXiv:2211.15111 [hep-ph]} \BibitemShut
  {NoStop}%
\bibitem [{\citenamefont {Yang}\ \emph
  {et~al.}(2022{\natexlab{b}})\citenamefont {Yang}, \citenamefont {Wang},
  \citenamefont {Wu}, \citenamefont {Oka},\ and\ \citenamefont
  {Zhu}}]{Yang:2021tvc}%
  \BibitemOpen
  \bibfield  {author} {\bibinfo {author} {\bibfnamefont {Z.}~\bibnamefont
  {Yang}}, \bibinfo {author} {\bibfnamefont {G.-J.}\ \bibnamefont {Wang}},
  \bibinfo {author} {\bibfnamefont {J.-J.}\ \bibnamefont {Wu}}, \bibinfo
  {author} {\bibfnamefont {M.}~\bibnamefont {Oka}},\ and\ \bibinfo {author}
  {\bibfnamefont {S.-L.}\ \bibnamefont {Zhu}},\ }\bibfield  {title} {\bibinfo
  {title} {{Novel Coupled Channel Framework Connecting the Quark Model and
  Lattice QCD for the Near-threshold ${D}_{s}$ States}},\ }\href
  {https://doi.org/10.1103/PhysRevLett.128.112001} {\bibfield  {journal}
  {\bibinfo  {journal} {Phys. Rev. Lett.}\ }\textbf {\bibinfo {volume} {128}},\
  \bibinfo {pages} {112001} (\bibinfo {year} {2022}{\natexlab{b}})},\ \Eprint
  {https://arxiv.org/abs/2107.04860} {arXiv:2107.04860 [hep-ph]} \BibitemShut
  {NoStop}%
\bibitem [{\citenamefont {Kalashnikova}(2005)}]{Kalashnikova:2005ui}%
  \BibitemOpen
  \bibfield  {author} {\bibinfo {author} {\bibfnamefont {{\relax Yu}.~S.}\
  \bibnamefont {Kalashnikova}},\ }\bibfield  {title} {\bibinfo {title}
  {{Coupled-channel model for charmonium levels and an option for
  ${X}(3872)$}},\ }\href {https://doi.org/10.1103/PhysRevD.72.034010}
  {\bibfield  {journal} {\bibinfo  {journal} {Phys. Rev.}\ }\textbf {\bibinfo
  {volume} {D72}},\ \bibinfo {pages} {034010} (\bibinfo {year} {2005})},\
  \Eprint {https://arxiv.org/abs/hep-ph/0506270} {arXiv:hep-ph/0506270
  [hep-ph]} \BibitemShut {NoStop}%
\bibitem [{\citenamefont {Ortega}\ \emph {et~al.}(2010)\citenamefont {Ortega},
  \citenamefont {Segovia}, \citenamefont {Entem},\ and\ \citenamefont
  {Fern\'andez}}]{Ortega:2009hj}%
  \BibitemOpen
  \bibfield  {author} {\bibinfo {author} {\bibfnamefont {P.~G.}\ \bibnamefont
  {Ortega}}, \bibinfo {author} {\bibfnamefont {J.}~\bibnamefont {Segovia}},
  \bibinfo {author} {\bibfnamefont {D.~R.}\ \bibnamefont {Entem}},\ and\
  \bibinfo {author} {\bibfnamefont {F.}~\bibnamefont {Fern\'andez}},\
  }\bibfield  {title} {\bibinfo {title} {{Coupled channel approach to the
  structure of the X(3872)}},\ }\href
  {https://doi.org/10.1103/PhysRevD.81.054023} {\bibfield  {journal} {\bibinfo
  {journal} {Phys. Rev.}\ }\textbf {\bibinfo {volume} {D81}},\ \bibinfo {pages}
  {054023} (\bibinfo {year} {2010})},\ \Eprint
  {https://arxiv.org/abs/0907.3997} {arXiv:0907.3997 [hep-ph]} \BibitemShut
  {NoStop}%
\bibitem [{\citenamefont {Danilkin}\ and\ \citenamefont
  {Simonov}(2010)}]{Danilkin:2010cc}%
  \BibitemOpen
  \bibfield  {author} {\bibinfo {author} {\bibfnamefont {I.~V.}\ \bibnamefont
  {Danilkin}}\ and\ \bibinfo {author} {\bibfnamefont {{\relax Yu}.~A.}\
  \bibnamefont {Simonov}},\ }\bibfield  {title} {\bibinfo {title} {{Dynamical
  Origin and the Pole Structure of $X(3872)$}},\ }\href
  {https://doi.org/10.1103/PhysRevLett.105.102002} {\bibfield  {journal}
  {\bibinfo  {journal} {Phys. Rev. Lett.}\ }\textbf {\bibinfo {volume} {105}},\
  \bibinfo {pages} {102002} (\bibinfo {year} {2010})},\ \Eprint
  {https://arxiv.org/abs/1006.0211} {arXiv:1006.0211 [hep-ph]} \BibitemShut
  {NoStop}%
\bibitem [{\citenamefont {Padmanath}\ \emph {et~al.}(2015)\citenamefont
  {Padmanath}, \citenamefont {Lang},\ and\ \citenamefont
  {Prelovsek}}]{Padmanath:2015era}%
  \BibitemOpen
  \bibfield  {author} {\bibinfo {author} {\bibfnamefont {M.}~\bibnamefont
  {Padmanath}}, \bibinfo {author} {\bibfnamefont {C.~B.}\ \bibnamefont
  {Lang}},\ and\ \bibinfo {author} {\bibfnamefont {S.}~\bibnamefont
  {Prelovsek}},\ }\bibfield  {title} {\bibinfo {title} {{$X(3872)$ and
  $Y(4140)$ using diquark-antidiquark operators with lattice QCD}},\ }\href
  {https://doi.org/10.1103/PhysRevD.92.034501} {\bibfield  {journal} {\bibinfo
  {journal} {Phys. Rev. D}\ }\textbf {\bibinfo {volume} {92}},\ \bibinfo
  {pages} {034501} (\bibinfo {year} {2015})},\ \Eprint
  {https://arxiv.org/abs/1503.03257} {arXiv:1503.03257 [hep-lat]} \BibitemShut
  {NoStop}%
\bibitem [{\citenamefont {Luo}\ \emph {et~al.}(2020)\citenamefont {Luo},
  \citenamefont {Chen}, \citenamefont {Liu},\ and\ \citenamefont
  {Liu}}]{Luo:2019qkm}%
  \BibitemOpen
  \bibfield  {author} {\bibinfo {author} {\bibfnamefont {S.-Q.}\ \bibnamefont
  {Luo}}, \bibinfo {author} {\bibfnamefont {B.}~\bibnamefont {Chen}}, \bibinfo
  {author} {\bibfnamefont {Z.-W.}\ \bibnamefont {Liu}},\ and\ \bibinfo {author}
  {\bibfnamefont {X.}~\bibnamefont {Liu}},\ }\bibfield  {title} {\bibinfo
  {title} {{Resolving the low mass puzzle of $\Lambda_c(2940)^+$}},\ }\href
  {https://doi.org/10.1140/epjc/s10052-020-7874-1} {\bibfield  {journal}
  {\bibinfo  {journal} {Eur. Phys. J.}\ }\textbf {\bibinfo {volume} {C80}},\
  \bibinfo {pages} {301} (\bibinfo {year} {2020})},\ \Eprint
  {https://arxiv.org/abs/1910.14545} {arXiv:1910.14545 [hep-ph]} \BibitemShut
  {NoStop}%
\bibitem [{\citenamefont {Zhang}\ \emph {et~al.}(2023)\citenamefont {Zhang},
  \citenamefont {Liu}, \citenamefont {Luo}, \citenamefont {Wang}, \citenamefont
  {Wang},\ and\ \citenamefont {Xu}}]{Zhang:2022pxc}%
  \BibitemOpen
  \bibfield  {author} {\bibinfo {author} {\bibfnamefont {Z.-L.}\ \bibnamefont
  {Zhang}}, \bibinfo {author} {\bibfnamefont {Z.-W.}\ \bibnamefont {Liu}},
  \bibinfo {author} {\bibfnamefont {S.-Q.}\ \bibnamefont {Luo}}, \bibinfo
  {author} {\bibfnamefont {F.-L.}\ \bibnamefont {Wang}}, \bibinfo {author}
  {\bibfnamefont {B.}~\bibnamefont {Wang}},\ and\ \bibinfo {author}
  {\bibfnamefont {H.}~\bibnamefont {Xu}},\ }\bibfield  {title} {\bibinfo
  {title} {{${\mathrm{\ensuremath{\Lambda}}}_{c}(2910)$ and
  ${\mathrm{\ensuremath{\Lambda}}}_{c}(2940)$ as conventional baryons dressed
  with the ${D}^{*}N$ channel}},\ }\href
  {https://doi.org/10.1103/PhysRevD.107.034036} {\bibfield  {journal} {\bibinfo
   {journal} {Phys. Rev. D}\ }\textbf {\bibinfo {volume} {107}},\ \bibinfo
  {pages} {034036} (\bibinfo {year} {2023})},\ \Eprint
  {https://arxiv.org/abs/2210.17188} {arXiv:2210.17188 [hep-ph]} \BibitemShut
  {NoStop}%
\bibitem [{\citenamefont {Li}\ \emph {et~al.}(2009)\citenamefont {Li},
  \citenamefont {Meng},\ and\ \citenamefont {Chao}}]{Li:2009ad}%
  \BibitemOpen
  \bibfield  {author} {\bibinfo {author} {\bibfnamefont {B.-Q.}\ \bibnamefont
  {Li}}, \bibinfo {author} {\bibfnamefont {C.}~\bibnamefont {Meng}},\ and\
  \bibinfo {author} {\bibfnamefont {K.-T.}\ \bibnamefont {Chao}},\ }\bibfield
  {title} {\bibinfo {title} {{Coupled-channel and screening effects in
  charmonium spectrum}},\ }\href {https://doi.org/10.1103/PhysRevD.80.014012}
  {\bibfield  {journal} {\bibinfo  {journal} {Phys. Rev.}\ }\textbf {\bibinfo
  {volume} {D80}},\ \bibinfo {pages} {014012} (\bibinfo {year} {2009})},\
  \Eprint {https://arxiv.org/abs/0904.4068} {arXiv:0904.4068 [hep-ph]}
  \BibitemShut {NoStop}%
\bibitem [{\citenamefont {Song}\ \emph
  {et~al.}(2015{\natexlab{a}})\citenamefont {Song}, \citenamefont {Chen},
  \citenamefont {Liu},\ and\ \citenamefont {Matsuki}}]{Song:2015nia}%
  \BibitemOpen
  \bibfield  {author} {\bibinfo {author} {\bibfnamefont {Q.-T.}\ \bibnamefont
  {Song}}, \bibinfo {author} {\bibfnamefont {D.-Y.}\ \bibnamefont {Chen}},
  \bibinfo {author} {\bibfnamefont {X.}~\bibnamefont {Liu}},\ and\ \bibinfo
  {author} {\bibfnamefont {T.}~\bibnamefont {Matsuki}},\ }\bibfield  {title}
  {\bibinfo {title} {{Charmed-strange mesons revisited: Mass spectra and strong
  decays}},\ }\href {https://doi.org/10.1103/PhysRevD.91.054031} {\bibfield
  {journal} {\bibinfo  {journal} {Phys. Rev. D}\ }\textbf {\bibinfo {volume}
  {91}},\ \bibinfo {pages} {054031} (\bibinfo {year} {2015}{\natexlab{a}})},\
  \Eprint {https://arxiv.org/abs/1501.03575} {arXiv:1501.03575 [hep-ph]}
  \BibitemShut {NoStop}%
\bibitem [{\citenamefont {Song}\ \emph
  {et~al.}(2015{\natexlab{b}})\citenamefont {Song}, \citenamefont {Chen},
  \citenamefont {Liu},\ and\ \citenamefont {Matsuki}}]{Song:2015fha}%
  \BibitemOpen
  \bibfield  {author} {\bibinfo {author} {\bibfnamefont {Q.-T.}\ \bibnamefont
  {Song}}, \bibinfo {author} {\bibfnamefont {D.-Y.}\ \bibnamefont {Chen}},
  \bibinfo {author} {\bibfnamefont {X.}~\bibnamefont {Liu}},\ and\ \bibinfo
  {author} {\bibfnamefont {T.}~\bibnamefont {Matsuki}},\ }\bibfield  {title}
  {\bibinfo {title} {{Higher radial and orbital excitations in the charmed
  meson family}},\ }\href {https://doi.org/10.1103/PhysRevD.92.074011}
  {\bibfield  {journal} {\bibinfo  {journal} {Phys. Rev. D}\ }\textbf {\bibinfo
  {volume} {92}},\ \bibinfo {pages} {074011} (\bibinfo {year}
  {2015}{\natexlab{b}})},\ \Eprint {https://arxiv.org/abs/1503.05728}
  {arXiv:1503.05728 [hep-ph]} \BibitemShut {NoStop}%
\bibitem [{\citenamefont {Pang}\ \emph {et~al.}(2017)\citenamefont {Pang},
  \citenamefont {Wang}, \citenamefont {Liu},\ and\ \citenamefont
  {Matsuki}}]{Pang:2017dlw}%
  \BibitemOpen
  \bibfield  {author} {\bibinfo {author} {\bibfnamefont {C.-Q.}\ \bibnamefont
  {Pang}}, \bibinfo {author} {\bibfnamefont {J.-Z.}\ \bibnamefont {Wang}},
  \bibinfo {author} {\bibfnamefont {X.}~\bibnamefont {Liu}},\ and\ \bibinfo
  {author} {\bibfnamefont {T.}~\bibnamefont {Matsuki}},\ }\bibfield  {title}
  {\bibinfo {title} {{A systematic study of mass spectra and strong decay of
  strange mesons}},\ }\href {https://doi.org/10.1140/epjc/s10052-017-5434-0}
  {\bibfield  {journal} {\bibinfo  {journal} {Eur. Phys. J. C}\ }\textbf
  {\bibinfo {volume} {77}},\ \bibinfo {pages} {861} (\bibinfo {year} {2017})},\
  \Eprint {https://arxiv.org/abs/1705.03144} {arXiv:1705.03144 [hep-ph]}
  \BibitemShut {NoStop}%
\bibitem [{\citenamefont {Wang}\ \emph {et~al.}(2018)\citenamefont {Wang},
  \citenamefont {Sun}, \citenamefont {Liu},\ and\ \citenamefont
  {Matsuki}}]{Wang:2018rjg}%
  \BibitemOpen
  \bibfield  {author} {\bibinfo {author} {\bibfnamefont {J.-Z.}\ \bibnamefont
  {Wang}}, \bibinfo {author} {\bibfnamefont {Z.-F.}\ \bibnamefont {Sun}},
  \bibinfo {author} {\bibfnamefont {X.}~\bibnamefont {Liu}},\ and\ \bibinfo
  {author} {\bibfnamefont {T.}~\bibnamefont {Matsuki}},\ }\bibfield  {title}
  {\bibinfo {title} {{Higher bottomonium zoo}},\ }\href
  {https://doi.org/10.1140/epjc/s10052-018-6372-1} {\bibfield  {journal}
  {\bibinfo  {journal} {Eur. Phys. J. C}\ }\textbf {\bibinfo {volume} {78}},\
  \bibinfo {pages} {915} (\bibinfo {year} {2018})},\ \Eprint
  {https://arxiv.org/abs/1802.04938} {arXiv:1802.04938 [hep-ph]} \BibitemShut
  {NoStop}%
\bibitem [{\citenamefont {Dosch}\ and\ \citenamefont
  {M\"uller}(1976)}]{Dosch:1975gf}%
  \BibitemOpen
  \bibfield  {author} {\bibinfo {author} {\bibfnamefont {H.~G.}\ \bibnamefont
  {Dosch}}\ and\ \bibinfo {author} {\bibfnamefont {V.~F.}\ \bibnamefont
  {M\"uller}},\ }\bibfield  {title} {\bibinfo {title} {{Composite hadrons in
  non-Abelian lattice gauge theories}},\ }\href
  {https://doi.org/10.1016/0550-3213(76)90359-X} {\bibfield  {journal}
  {\bibinfo  {journal} {Nucl. Phys. B}\ }\textbf {\bibinfo {volume} {116}},\
  \bibinfo {pages} {470} (\bibinfo {year} {1976})}\BibitemShut {NoStop}%
\bibitem [{\citenamefont {Carlson}\ \emph {et~al.}(1983)\citenamefont
  {Carlson}, \citenamefont {Kogut},\ and\ \citenamefont
  {Pandharipande}}]{Carlson:1982xi}%
  \BibitemOpen
  \bibfield  {author} {\bibinfo {author} {\bibfnamefont {J.}~\bibnamefont
  {Carlson}}, \bibinfo {author} {\bibfnamefont {J.}~\bibnamefont {Kogut}},\
  and\ \bibinfo {author} {\bibfnamefont {V.~R.}\ \bibnamefont
  {Pandharipande}},\ }\bibfield  {title} {\bibinfo {title} {{Quark model for
  baryons based on quantum chromodynamics}},\ }\href
  {https://doi.org/10.1103/PhysRevD.27.233} {\bibfield  {journal} {\bibinfo
  {journal} {Phys. Rev. D}\ }\textbf {\bibinfo {volume} {27}},\ \bibinfo
  {pages} {233} (\bibinfo {year} {1983})}\BibitemShut {NoStop}%
\bibitem [{\citenamefont {Mizuk}\ \emph {et~al.}(2005)\citenamefont {Mizuk}
  \emph {et~al.}}]{Belle:2004zjl}%
  \BibitemOpen
  \bibfield  {author} {\bibinfo {author} {\bibfnamefont {R.}~\bibnamefont
  {Mizuk}} \emph {et~al.} (\bibinfo {collaboration} {Belle Collaboration}),\
  }\bibfield  {title} {\bibinfo {title} {{Observation of an Isotriplet of
  Excited Charmed Baryons Decaying to
  ${\ensuremath{\Lambda}}_{c}^{+}\ensuremath{\pi}$}},\ }\href
  {https://doi.org/10.1103/PhysRevLett.94.122002} {\bibfield  {journal}
  {\bibinfo  {journal} {Phys. Rev. Lett.}\ }\textbf {\bibinfo {volume} {94}},\
  \bibinfo {pages} {122002} (\bibinfo {year} {2005})},\ \Eprint
  {https://arxiv.org/abs/hep-ex/0412069} {arXiv:hep-ex/0412069} \BibitemShut
  {NoStop}%
\bibitem [{\citenamefont {Artuso}\ \emph {et~al.}(2001)\citenamefont {Artuso}
  \emph {et~al.}}]{CLEO:2000mbh}%
  \BibitemOpen
  \bibfield  {author} {\bibinfo {author} {\bibfnamefont {M.}~\bibnamefont
  {Artuso}} \emph {et~al.} (\bibinfo {collaboration} {CLEO Collaboration}),\
  }\bibfield  {title} {\bibinfo {title} {{Observation of New States Decaying
  into ${\Lambda}_{c}^{+}{\pi}^{-}{\pi}^{+}$}},\ }\href
  {https://doi.org/10.1103/PhysRevLett.86.4479} {\bibfield  {journal} {\bibinfo
   {journal} {Phys. Rev. Lett.}\ }\textbf {\bibinfo {volume} {86}},\ \bibinfo
  {pages} {4479} (\bibinfo {year} {2001})},\ \Eprint
  {https://arxiv.org/abs/hep-ex/0010080} {arXiv:hep-ex/0010080} \BibitemShut
  {NoStop}%
\bibitem [{\citenamefont {Abe}\ \emph {et~al.}(2007)\citenamefont {Abe} \emph
  {et~al.}}]{Belle:2006xni}%
  \BibitemOpen
  \bibfield  {author} {\bibinfo {author} {\bibfnamefont {K.}~\bibnamefont
  {Abe}} \emph {et~al.} (\bibinfo {collaboration} {Belle Collaboration}),\
  }\bibfield  {title} {\bibinfo {title} {{Experimental Constraints on the Spin
  and Parity of the ${\Lambda}_{c}(2880)^{+}$}},\ }\href
  {https://doi.org/10.1103/PhysRevLett.98.262001} {\bibfield  {journal}
  {\bibinfo  {journal} {Phys. Rev. Lett.}\ }\textbf {\bibinfo {volume} {98}},\
  \bibinfo {pages} {262001} (\bibinfo {year} {2007})},\ \Eprint
  {https://arxiv.org/abs/hep-ex/0608043} {arXiv:hep-ex/0608043} \BibitemShut
  {NoStop}%
\bibitem [{\citenamefont {Aubert}\ \emph {et~al.}(2007)\citenamefont {Aubert}
  \emph {et~al.}}]{BaBar:2006itc}%
  \BibitemOpen
  \bibfield  {author} {\bibinfo {author} {\bibfnamefont {B.}~\bibnamefont
  {Aubert}} \emph {et~al.} (\bibinfo {collaboration} {BABAR Collaboration}),\
  }\bibfield  {title} {\bibinfo {title} {{Observation of a Charmed Baryon
  Decaying to ${D}^{0}p$ at a Mass Near $2.94~\mathrm{GeV}/{c}^{2}$}},\ }\href
  {https://doi.org/10.1103/PhysRevLett.98.012001} {\bibfield  {journal}
  {\bibinfo  {journal} {Phys. Rev. Lett.}\ }\textbf {\bibinfo {volume} {98}},\
  \bibinfo {pages} {012001} (\bibinfo {year} {2007})},\ \Eprint
  {https://arxiv.org/abs/hep-ex/0603052} {arXiv:hep-ex/0603052} \BibitemShut
  {NoStop}%
\bibitem [{\citenamefont {Zhao}\ \emph {et~al.}(2017)\citenamefont {Zhao},
  \citenamefont {Ye},\ and\ \citenamefont {Zhang}}]{Zhao:2017fov}%
  \BibitemOpen
  \bibfield  {author} {\bibinfo {author} {\bibfnamefont {Z.}~\bibnamefont
  {Zhao}}, \bibinfo {author} {\bibfnamefont {D.-D.}\ \bibnamefont {Ye}},\ and\
  \bibinfo {author} {\bibfnamefont {A.}~\bibnamefont {Zhang}},\ }\bibfield
  {title} {\bibinfo {title} {Hadronic decay properties of newly observed
  $\omega_c$ baryons},\ }\href {https://doi.org/10.1103/PhysRevD.95.114024}
  {\bibfield  {journal} {\bibinfo  {journal} {Phys. Rev.}\ }\textbf {\bibinfo
  {volume} {D95}},\ \bibinfo {pages} {114024} (\bibinfo {year} {2017})},\
  \Eprint {https://arxiv.org/abs/1704.02688} {arXiv:1704.02688 [hep-ph]}
  \BibitemShut {NoStop}%
\end{thebibliography}%
\end{document}